\pdfoutput=1
\documentclass[12pt,a4paper]{article}

\usepackage{lineno}  % for line numbering during review
\usepackage{graphicx}  % to include figures (can also use other packages)
\usepackage{color}
\usepackage{colortbl}
\graphicspath{{./figs/}} % Make Latex search fig subdir for figures
\usepackage{rotating}
 
\usepackage{xspace} % To avoid problems with missing or double spaces after
\usepackage{color}
\usepackage{colortbl}
\usepackage{amsmath} % Adds a large collection of math symbols
\usepackage{ifthen} % for conditional statements
\usepackage{multirow}
\newboolean{pdflatex}
\setboolean{pdflatex}{true} % use this if using non-eps figures
\newboolean{articletitles}
\setboolean{articletitles}{true} % False removes titles in references

\newboolean{uprightparticles}
\setboolean{uprightparticles}{false} %Set to true to get roman particle symbols

\usepackage{amssymb}
\usepackage{amsfonts}
\usepackage{upgreek} % Adds in support for greek letters in roman typeset

\usepackage{hyperref}    % Hyperlinks in references

\usepackage[all]{hypcap} % Internal hyperlinks to floats.
\usepackage{afterpage}

\graphicspath{{fig/};{figures/}}
\newcommand*\patchAmsMathEnvironmentForLineno[1]{%
\expandafter\let\csname old#1\expandafter\endcsname\csname #1\endcsname
\expandafter\let\csname oldend#1\expandafter\endcsname\csname
end#1\endcsname
 \renewenvironment{#1}%
   {\linenomath\csname old#1\endcsname}%
   {\csname oldend#1\endcsname\endlinenomath}%
}
\newcommand*\patchBothAmsMathEnvironmentsForLineno[1]{%
  \patchAmsMathEnvironmentForLineno{#1}%
  \patchAmsMathEnvironmentForLineno{#1*}%
}
\AtBeginDocument{%
\patchBothAmsMathEnvironmentsForLineno{equation}%
\patchBothAmsMathEnvironmentsForLineno{align}%
\patchBothAmsMathEnvironmentsForLineno{flalign}%
\patchBothAmsMathEnvironmentsForLineno{alignat}%
\patchBothAmsMathEnvironmentsForLineno{gather}%
\patchBothAmsMathEnvironmentsForLineno{multline}%
\patchBothAmsMathEnvironmentsForLineno{eqnarray}%
}

\def\lhcb {\mbox{LHCb}\xspace}

\ifthenelse{\boolean{uprightparticles}}%
{

 \def\Pmu         {\ensuremath{\upmu}\xspace}

 \def\Ppi         {\ensuremath{\uppi}\xspace}

 \def\Ppsi        {\ensuremath{\uppsi}\xspace}

 \def\PDelta      {\ensuremath{\Delta}\xspace}                 
 \def\PXi      {\ensuremath{\Xi}\xspace}                 
 \def\PLambda      {\ensuremath{\Lambda}\xspace}                 
 \def\PSigma      {\ensuremath{\Sigma}\xspace}                 
 \def\POmega      {\ensuremath{\Omega}\xspace}                 
 \def\PUpsilon      {\ensuremath{\Upsilon}\xspace}                 
 
 \def\PB      {\ensuremath{\mathrm{B}}\xspace}                 
                  
 \def\PD      {\ensuremath{\mathrm{D}}\xspace}

 \def\PJ      {\ensuremath{\mathrm{J}}\xspace}                 
 \def\PK      {\ensuremath{\mathrm{K}}\xspace}

 \def\Pb      {\ensuremath{\mathrm{b}}\xspace}                 
 \def\Pc      {\ensuremath{\mathrm{c}}\xspace}

 \def\Pi      {\ensuremath{\mathrm{i}}\xspace}

 \def\Pp      {\ensuremath{\mathrm{p}}\xspace}

 \def\Ps      {\ensuremath{\mathrm{s}}\xspace}

}
{

 \def\Pmu         {\ensuremath{\mu}\xspace}

 \def\Ppi         {\ensuremath{\pi}\xspace}

 \def\Ppsi        {\ensuremath{\psi}\xspace}                 
                  
 \mathchardef\PDelta="7101
 \mathchardef\PXi="7104
 \mathchardef\PLambda="7103
 \mathchardef\PSigma="7106
 \mathchardef\POmega="710A
 \mathchardef\PUpsilon="7107
                  
 \def\PB      {\ensuremath{B}\xspace}                 
                  
 \def\PD      {\ensuremath{D}\xspace}

 \def\PJ      {\ensuremath{J}\xspace}                 
 \def\PK      {\ensuremath{K}\xspace}

 \def\Pb      {\ensuremath{b}\xspace}                 
 \def\Pc      {\ensuremath{c}\xspace}

 \def\Pi      {\ensuremath{i}\xspace}

 \def\Pp      {\ensuremath{p}\xspace}

 \def\Ps      {\ensuremath{s}\xspace}

}

\def\mup        {{\ensuremath{\Pmu^+}}\xspace}
\def\mun        {{\ensuremath{\Pmu^-}}\xspace} % muon negative (\mum is taken)

\def\squark    {{\ensuremath{\Ps}}\xspace}

\def\cquark    {{\ensuremath{\Pc}}\xspace}

\def\bquark    {{\ensuremath{\Pb}}\xspace}

\def\pion   {{\ensuremath{\Ppi}}\xspace}

\def\pip    {{\ensuremath{\pion^+}}\xspace}
\def\pim    {{\ensuremath{\pion^-}}\xspace}

\def\kaon    {{\ensuremath{\PK}}\xspace}
  \def\Kbar    {{\kern 0.2em\overline{\kern -0.2em \PK}{}}\xspace}

\def\Km      {{\ensuremath{\kaon^-}}\xspace}

\def\KS      {{\ensuremath{\kaon^0_{\rm\scriptscriptstyle S}}}\xspace}

\def\Kstarzb {{\ensuremath{\Kbar^{*0}}}\xspace}

\def\Kstarb  {{\ensuremath{\Kbar^*}}\xspace}

  \def\Dbar    {{\kern 0.2em\overline{\kern -0.2em \PD}{}}\xspace}

\def\B       {{\ensuremath{\PB}}\xspace}
\def\Bbar    {{\ensuremath{\kern 0.18em\overline{\kern -0.18em \PB}{}}}\xspace}

\def\Bz      {{\ensuremath{\B^0}}\xspace}
\def\Bzb     {{\ensuremath{\Bbar^0}}\xspace}

\def\Bs      {{\ensuremath{\B^0_\squark}}\xspace}
\def\Bsb     {{\ensuremath{\Bbar^0_\squark}}\xspace}
\def\Bdb     {{\ensuremath{\Bbar^0}}\xspace}

\def\jpsi     {{\ensuremath{{\PJ\mskip -3mu/\mskip -2mu\Ppsi\mskip 2mu}}}\xspace}

  \def\Y#1S{\ensuremath{\PUpsilon{(#1S)}}\xspace}% no space before {...}!

\def\proton      {{\ensuremath{\Pp}}\xspace}

\def\Lz          {{\ensuremath{\PLambda}}\xspace}
\def\Lbar        {{\ensuremath{\kern 0.1em\overline{\kern -0.1em\PLambda}}}\xspace}

\def\Lb      {{\ensuremath{\Lz^0_\bquark}}\xspace}

\def\to                 {\ensuremath{\rightarrow}\xspace}

\def\CP                {{\ensuremath{C\!P}}\xspace}

\def\AT#1     {\ensuremath{A_{\mathrm{T}}^{#1}}\xspace}           % 2

\def\C#1      {\ensuremath{\mathcal{C}_{#1}}\xspace}                       % 9
\def\Cp#1     {\ensuremath{\mathcal{C}_{#1}^{'}}\xspace}                    % 7
\def\Ceff#1   {\ensuremath{\mathcal{C}_{#1}^{\mathrm{(eff)}}}\xspace}        % 9  
\def\Cpeff#1  {\ensuremath{\mathcal{C}_{#1}^{'\mathrm{(eff)}}}\xspace}       % 7
\def\Ope#1    {\ensuremath{\mathcal{O}_{#1}}\xspace}                       % 2
\def\Opep#1   {\ensuremath{\mathcal{O}_{#1}^{'}}\xspace}                    % 7

\newcommand{\ket}[1]{\ensuremath{|#1\rangle}}              % {b}
\newcommand{\tev}{\ifthenelse{\boolean{inbibliography}}{\ensuremath{~T\kern -0.05em eV}\xspace}{\ensuremath{\mathrm{\,Te\kern -0.1em V}}}\xspace}
\newcommand{\gev}{\ensuremath{\mathrm{\,Ge\kern -0.1em V}}\xspace}
\newcommand{\mev}{\ensuremath{\mathrm{\,Me\kern -0.1em V}}\xspace}
\newcommand{\kev}{\ensuremath{\mathrm{\,ke\kern -0.1em V}}\xspace}
\newcommand{\ev}{\ensuremath{\mathrm{\,e\kern -0.1em V}}\xspace}
\newcommand{\gevc}{\ensuremath{{\mathrm{\,Ge\kern -0.1em V\!/}c}}\xspace}
\newcommand{\mevc}{\ensuremath{{\mathrm{\,Me\kern -0.1em V\!/}c}}\xspace}
\newcommand{\gevcc}{\ensuremath{{\mathrm{\,Ge\kern -0.1em V\!/}c^2}}\xspace}
\newcommand{\gevgevcccc}{\ensuremath{{\mathrm{\,Ge\kern -0.1em V^2\!/}c^4}}\xspace}
\newcommand{\mevcc}{\ensuremath{{\mathrm{\,Me\kern -0.1em V\!/}c^2}}\xspace}

\def\m    {\ensuremath{\rm \,m}\xspace}

\def\mm   {\ensuremath{\rm \,mm}\xspace}

\def\mum  {\ensuremath{{\,\upmu\rm m}}\xspace}

\def\invfb   {\ensuremath{\mbox{\,fb}^{-1}}\xspace}

\def\gsim{{~\raise.15em\hbox{$>$}\kern-.85em
          \lower.35em\hbox{$\sim$}~}\xspace}
\def\lsim{{~\raise.15em\hbox{$<$}\kern-.85em
          \lower.35em\hbox{$\sim$}~}\xspace}

\newcommand{\Real}{\ensuremath{\mathcal{R}e}\xspace}
\newcommand{\Imag}{\ensuremath{\mathcal{I}m}\xspace}

\def\pt         {\mbox{$p_{\rm T}$}\xspace}

\def\evtgen     {\mbox{\textsc{EvtGen}}\xspace}

\def\gauss      {\mbox{\textsc{Gauss}}\xspace}
\def\geant      {\mbox{\textsc{Geant4}}\xspace}

\def\photos     {\mbox{\textsc{Photos}}\xspace}

\def\pythia     {\mbox{\textsc{Pythia}}\xspace}

\def\tell1  {TELL1\xspace}
\def\ukl1   {UKL1\xspace}

\newcommand{\ie}{\mbox{\itshape i.e.}\xspace}

 \renewcommand{\tev}{\ensuremath{\mathrm{\,Te\kern -0.1em V}}\xspace}
\renewcommand{\gev}{\ensuremath{\mathrm{\,Ge\kern -0.1em V}}\xspace}
\renewcommand{\mev}{\ensuremath{\mathrm{\,Me\kern -0.1em V}}\xspace}
\renewcommand{\kev}{\ensuremath{\mathrm{\,ke\kern -0.1em V}}\xspace}
\renewcommand{\ev}{\ensuremath{\mathrm{\,e\kern -0.1em V}}\xspace}
\renewcommand{\gevc}{\ensuremath{{\mathrm{\,Ge\kern -0.1em V\!/}c}}\xspace}
\renewcommand{\mevc}{\ensuremath{{\mathrm{\,Me\kern -0.1em V\!/}c}}\xspace}
\renewcommand{\gevcc}{\ensuremath{{\mathrm{\,Ge\kern -0.1em V\!/}c^2}}\xspace}
\renewcommand{\gevgevcccc}{\ensuremath{{\mathrm{\,Ge\kern -0.1em V^2\!/}c^4}}\xspace}
\renewcommand{\mevcc}{\ensuremath{{\mathrm{\,Me\kern -0.1em V\!/}c^2}}\xspace}

\def \m {m_{hh}}
\def \angmu {\theta_{\jpsi}}
\def \angpi {\theta_{hh}}

\def \dv {{\rm d}}
\def \G {\Gamma}
\def \ch {\cosh \frac{\Delta \G t}{2}}
\def \sh {\sinh \frac{\Delta \G t}{2}}
\def \Ab {\overline{A}}
\def \cs {\cos(\Delta m t)}
\def \sn {\sin(\Delta m t)}

\usepackage{accents}
\newcommand*{\fancybar}{\scalebox{.4}{(}\raisebox{-1.7pt}{--}\scalebox{.4}{)}}
\newcommand*{\brabar}[1]{\accentset{\fancybar}{#1}}
\usepackage{mathrsfs}
\usepackage{cite}
\usepackage{mciteplus}

\begin{document}
\renewcommand{\thefootnote}{\fnsymbol{footnote}}
\setcounter{footnote}{1}
\begin{titlepage}

% Header ---------------------------------------------------
\belowpdfbookmark{Title page}{title}

\vspace*{-1.5cm}
\centerline{\large EUROPEAN ORGANIZATION FOR NUCLEAR RESEARCH (CERN)}
\vspace*{1.5cm}
\hspace*{-5mm}\begin{tabular*}{16cm}{lc@{\extracolsep{\fill}}r}
\vspace*{-12mm}\mbox{\!\!\!\includegraphics[width=.12\textwidth]{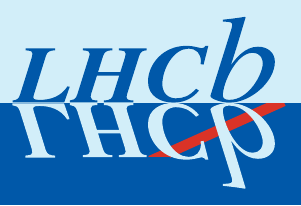}}& & \\
 & & CERN-PH-EP-2014-069\\
 & & LHCb-PAPER-2014-012\\  % ID
 & & 22 April 2014 \\ % Date - Can also hardwire e.g.: 23 March 2010
 %& & Version 4\\
 & & \\

\end{tabular*}

%\vspace*{2.0cm}

% Title --------------------------------------------------
{\bf\boldmath\huge
\begin{center}
Measurement of the resonant and \CP components in $\Bdb\rightarrow J/\psi \pi^+\pi^-$ decays
\end{center}
}
\vspace*{1.0cm}
\begin{center}
\normalsize {The LHCb collaboration\footnote{Authors are listed on the following pages.}}
%Sheldon Stone,  Liming Zhang and Zhou Xing
%\bigskip\\
%{\it\footnotesize
%Syracuse University, Syracuse, USA\\

\end{center}

\vspace{\fill}

% Abstract -----------------------------------------------
\begin{abstract}
  \noindent

The resonant structure of the reaction $\Bdb\rightarrow J/\psi \pi^+\pi^-$ is studied using data from 3\,\invfb  of integrated luminosity collected by the LHCb experiment, one-third at 7\tev center-of-mass energy  and the remainder at 8\tev. The invariant mass of the $\pip\pim$ pair and three decay angular distributions are used to determine the fractions of the resonant and non-resonant components. Six interfering $\pip\pim$ states: $\rho(770)$, $f_0(500)$, $f_2(1270)$, $\rho(1450)$, $\omega(782)$ and $\rho(1700)$ are required to give a good description of invariant mass spectra and decay angular distributions. The positive and negative  \CP fractions of each of the resonant final states are determined.  The $f_0(980)$ meson is not seen and the upper limit on its presence, compared with the observed $f_0(500)$ rate, is inconsistent with a model where these scalar mesons are formed from two quarks and two anti-quarks (tetraquarks)  at the 
 eight standard deviation level. In the $q\overline{q}$ model, the  absolute value of the mixing angle between the $f_0(980)$ and the $f_0(500)$ scalar mesons is limited to be less than $17^{\circ}$ at 90\% confidence level.

\end{abstract}

\vspace*{2.0cm}
\vspace{\fill}

%\vspace*{1.0cm}
%\vspace{\fill}

\vspace*{1.0cm}
%{\it Keywords:} LHC, \CP violation, Hadronic $B$ decays, $\Bsb$ meson, S-wave\\
%\hspace*{6mm}{\it PACS:} 13.25.Hw, 14.40.Nd, 11.30.Er\\
\begin{center}
\hspace*{6mm}Submitted to Phys. Rev. D\\
\end{center}
\vspace{\fill}

{\footnotesize
\centerline{\copyright~CERN on behalf of the \lhcb collaboration, license \href{http://creativecommons.org/licenses/by/3.0/}{CC-BY-3.0}.}}
\vspace*{2mm}

\end{titlepage}
\newpage
\setcounter{page}{2}
\mbox{~}
%\newpage

% Author List ----------------------------
%  You need to get a new author list!
%%%%%%%%%%%%%%%%%%%%%%%%%%%%%%%%%%%%%%%%%%
\centerline{\large\bf LHCb collaboration}
\begin{flushleft}
\small
R.~Aaij$^{41}$, 
B.~Adeva$^{37}$, 
M.~Adinolfi$^{46}$, 
A.~Affolder$^{52}$, 
Z.~Ajaltouni$^{5}$, 
J.~Albrecht$^{9}$, 
F.~Alessio$^{38}$, 
M.~Alexander$^{51}$, 
S.~Ali$^{41}$, 
G.~Alkhazov$^{30}$
% ...

{\footnotesize \it
% ...
$ ^{5}$Clermont Universit\'{e}, Universit\'{e} Blaise Pascal, CNRS/IN2P3, LPC, Clermont-Ferrand, France\\
% ...
$ ^{9}$Fakult\"{a}t Physik, Technische Universit\"{a}t Dortmund, Dortmund, Germany\\
% ...
$ ^{30}$Petersburg Nuclear Physics Institute (PNPI), Gatchina, Russia\\
% ...
$ ^{37}$Universidad de Santiago de Compostela, Santiago de Compostela, Spain\\
$ ^{38}$European Organization for Nuclear Research (CERN), Geneva, Switzerland\\
% ...
$ ^{41}$Nikhef National Institute for Subatomic Physics, Amsterdam, The Netherlands\\
% ...
$ ^{46}$H.H. Wills Physics Laboratory, University of Bristol, Bristol, United Kingdom\\
% ...
$ ^{51}$School of Physics and Astronomy, University of Glasgow, Glasgow, United Kingdom\\
$ ^{52}$Oliver Lodge Laboratory, University of Liverpool, Liverpool, United Kingdom\\
% ...
$ ^{q}$Universit\`{a} di Padova, Padova, Italy\\
% ...
}
\end{flushleft}
%%%%%%%%%%%%%%%%%%%%%%%%%%%%%%%%%%%%%%%%%%

\cleardoublepage

\renewcommand{\thefootnote}{\arabic{footnote}}
\setcounter{footnote}{0}

%%%%%%%%%%%%%%%%%%%%%%%%%
%%%%% Main text %%%%%%%%%
%%%%%%%%%%%%%%%%%%%%%%%%%

\pagestyle{plain} % restore page numbers for the main text
\setcounter{page}{1}
\pagenumbering{arabic}

% %%%%%%% CHOOSE --------
%% ----------------------------------
%% Line numbering on the left margin
%% ----------------------------------
%% Uncomment during review phase.
%% Comment it out before a final submission.

%% --------------------------------
% %%%%%%%%%%%%% ---------
\newpage

% You can include short sections directly in the main tex file.
% However, for larger papers it is desirable to split the text into
% several semiautonomous files, which can be revised independently.
% This is especially useful when developing a document in
% collaboration with several people, since then different parts can be
% edited independently.  This type of file organization is shown here.
%
\section{Introduction}
\label{sec:Introduction}

\begin{figure}[h]
\vskip -.4cm
\begin{center}
\includegraphics[width=6in]{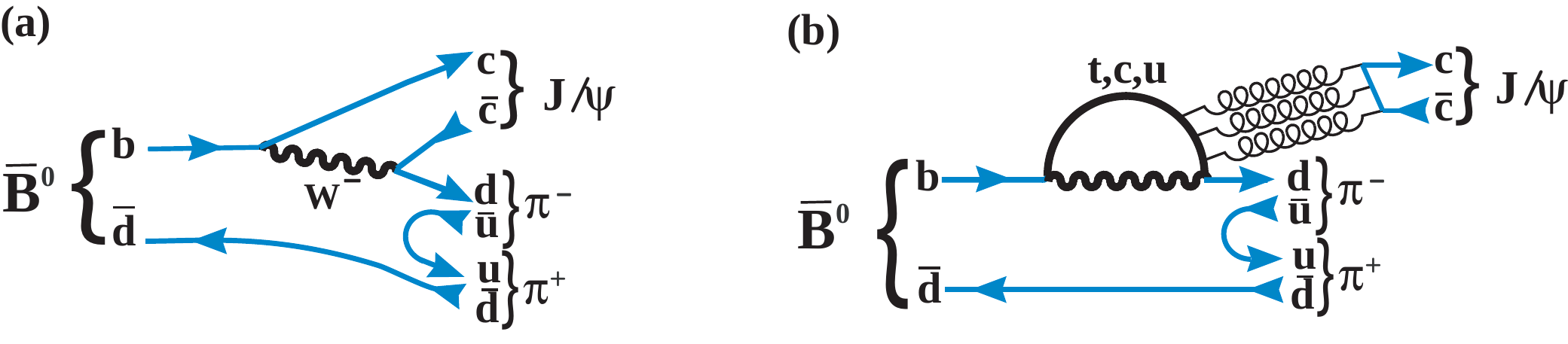}
\end{center}\label{feyn3}
\vskip -.5cm
\caption{\small (a) Tree level and (b) penguin diagram for $\Bdb$ decays into $J/\psi \pi^+\pi^-$.}
\end{figure}

The decay mode $\Bzb \to J/\psi \pi^+ \pi^-$ is of particular interest in the study of \CP violation in the $B$ system.\footnote{In this paper, mention of a particular decay mode implies the use of the charge conjugate decay as well, unless stated otherwise.} The decay can proceed either via a tree level process, shown in Fig.~\ref{feyn3}(a), or via the penguin mechanisms shown in Fig.~\ref{feyn3}(b). The ratio of penguin to tree amplitudes is enhanced in this decay relative to $\Bzb\to\jpsi \KS$ \cite{Fleischer:1999sj,*Faller:2008zc}. Thus the effects of penguin topologies can be investigated  by using the $J/\psi \pi^+ \pi^-$ decay  and comparing different measurements of the \CP violating phase, $\beta$, in $\jpsi \KS$, and individual channels such as $\Bzb \to J/\psi \rho^0$. 

The $\Bzb \to J/\psi \pi^+ \pi^-$ decay is also useful for the study of the substructure of light mesons that decay into $\pi^+\pi^-$. Tests have been proposed to ascertain if the scalar $f_0(500)$ and $f_0(980)$ mesons are formed of $q\overline{q}$ or tetraquarks.  In the model of Ref.~\cite{Stone:2013eaa}, if these  scalar states are tetraquarks, the ratio of decay widths is predicted to be 1/2.
If instead these are $q\overline{q}$ states, they can be mixtures of two base states; in this scenario the width ratio can be any value and is determined principally by the mixing angle between the base states.

The $\Bzb \to J/\psi \pi^+ \pi^-$ decay was first observed by the BaBar collaboration \cite{Aubert:2002vb,*Aubert:2007xw}. It has been previously studied by LHCb using data from 1\,\invfb of integrated luminosity \cite{Aaij:2013zpt}. The branching fraction was measured to be $(3.97\pm 0.22)\times 10^{-5}$. The mass and angular distributions were used to measure the resonant substructure. That analysis, however, did not use the angle between the $\jpsi$ and $\pi^+\pi^-$ decay planes, due to limited statistics. A new theoretical approach \cite{Zhang:2012zk} now allows us to include all the angular information and measure the fraction of \CP-even and \CP-odd states. This information is vital to any subsequent \CP violation measurements.

\section{Data sample and detector}
In this paper, we measure the resonant substructure and \CP content of the $\Bzb \to J/\psi \pi^+ \pi^-$ decay from data corresponding to 3\,\invfb of integrated luminosity collected with the \lhcb detector~\cite{LHCb-det} using $pp$ collisions. One-third of the data was acquired at a center-of-mass energy of 7\tev, and the remainder at 8\tev. The detector is a single-arm forward
spectrometer covering the \mbox{pseudorapidity} range $2<\eta <5$,
designed for the study of particles containing \bquark or \cquark
quarks. The detector includes a high-precision tracking system
consisting of a silicon-strip vertex detector surrounding the $pp$
interaction region, a large-area silicon-strip detector located
upstream of a dipole magnet with a bending power of about
$4{\rm\,Tm}$, and three stations of silicon-strip detectors and straw
drift tubes~\cite{LHCb-DP-2013-003} placed downstream.
The combined tracking system provides a momentum measurement with
relative uncertainty that varies from 0.4\% at 5\gev to 0.6\% at 100\gev,\footnote{We work in units where $c = 1$.} 
and impact parameter (IP) resolution of 20\mum for
tracks with large transverse momentum ($\pt$). Different types of charged hadrons are distinguished by information
from two ring-imaging Cherenkov (RICH) detectors\cite{LHCb-DP-2012-003}. Photon, electron and
hadron candidates are identified by a calorimeter system consisting of
scintillating-pad and preshower detectors, an electromagnetic
calorimeter and a hadronic calorimeter. Muons are identified by a
system composed of alternating layers of iron and multiwire
proportional chambers~\cite{LHCb-DP-2012-002}.

The trigger consists of a hardware stage, based
on information from the calorimeter and muon systems, followed by a
software stage that applies a full event reconstruction~\cite{Aaij:2012me}. Events selected for this analysis are triggered by a $J/\psi\to\mu^+\mu^-$  decay, where the
$J/\psi$ meson is required at the software level to be consistent with coming from the decay of a $\Bzb$ meson by use either of IP requirements or detachment of the $J/\psi$ meson decay vertex  from the primary vertex (PV). In the simulation,  $pp$ collisions are generated using
\pythia~\cite{Sjostrand:2006za,*Sjostrand:2007gs}
 with a specific \lhcb
configuration~\cite{LHCb-PROC-2010-056}.  Decays of hadronic particles
are described by \evtgen~\cite{Lange:2001uf}, in which final state
radiation is generated using \photos~\cite{Golonka:2005pn}. The
interaction of the generated particles with the detector and its
response are implemented using the \geant
toolkit~\cite{Allison:2006ve, Agostinelli:2002hh} as described in
Ref.~\cite{LHCb-PROC-2011-006}.

\section{Decay amplitude formalism}

\subsection{Observables used in the analysis}

\begin{figure}[h]
\vskip -.4cm
\begin{center}
\includegraphics[width=6in]{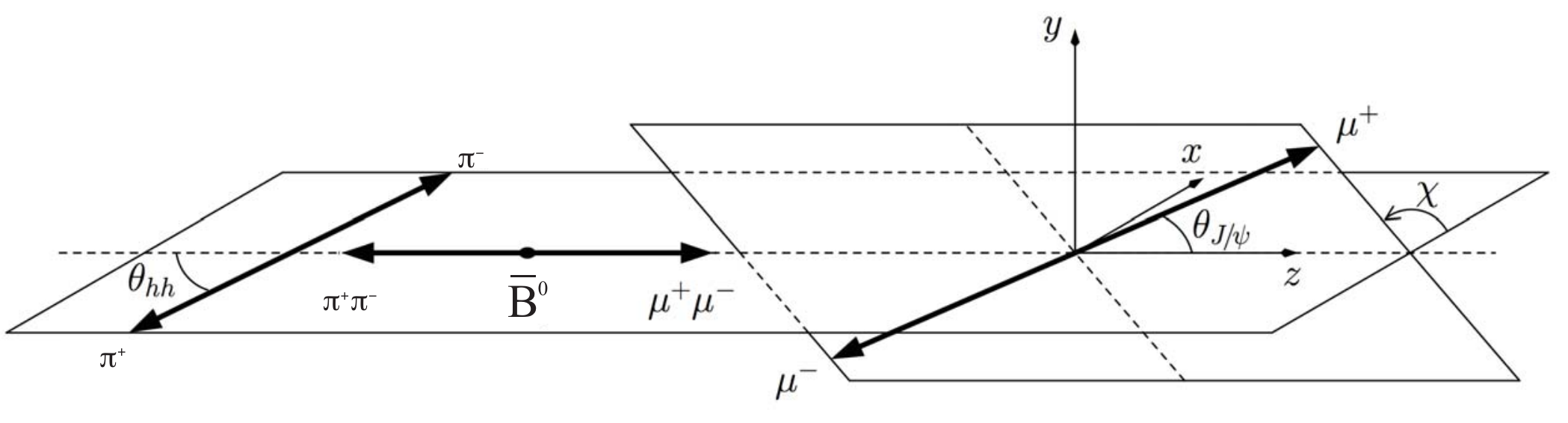}
\end{center}\label{4variables}
\vskip -.5cm
\caption{\small Illustration of the three  angles used in this analysis.}
\end{figure}

The $\Bzb \to J/\psi \pi^+ \pi^-$ decay with $\jpsi \to \mup\mun$ can be described by the invariant mass of  the $\pip\pim$ ($m_{hh}$) pair, and three  angles: (i) the angle between the $\mup$ direction in the $\jpsi$ rest frame with respect to the $\jpsi$ direction in the $\Bzb$ rest frame, $\theta_{\jpsi}$; (ii) the angle between the $\pip$ direction and the opposite direction of the \Bzb candidate momentum in the $\pip\pim$ rest frame, $\theta_{hh}$; and (iii) the angle between the $\jpsi$ and $\pip\pim$ decay planes in the $\Bzb$ rest frame, $\chi$. The angular variables are illustrated in Fig.~\ref{4variables}.

In our previous study \cite{Aaij:2013zpt}, we used the ``Dalitz-plot" variables: the invariant mass squared of $\jpsi \pip$, $s_{12} = m^2(\jpsi \pip)$, and the invariant mass squared of the $\pip\pim$ pair, $s_{23} = m^2(\pip\pim)$. Due to the $\jpsi$ spin,  the event distributions in the $s_{12}$ and $s_{23}$ plane do not  directly show the effect of the matrix-element squared. Since the probability density functions (PDFs) expressed as functions of  $m_{hh}$ and $\theta_{hh}$ are easier to normalize, we use them instead.  In this paper, the notation $hh$ is equivalent to $\pi^+\pi^-$. 
 The Dalitz-plot variables can be translated into ($m_{hh}$, $\theta_{hh}$), and vice versa.
The formalism described below is for the decay sequence $\Bzb\to\jpsi R$, $R\to\pi^+\pi^-$.

\subsection{Amplitude formalism}

The decay rate of $\brabar{B}^0\to \jpsi \pip \pim$ has been described in detail in  Ref.~\cite{Zhang:2012zk}.
The differential decay width can be written in terms of the decay time $t$ and the
four other variables $m_{hh},\angmu,\angpi,$ and $\chi$ as \cite{Nierste:2009wg,*Bigi:2000yz}
\begin{align}\label{Eq-t}
&\frac{d^5 {\Gamma}}{dt\,dm_{hh}\,d\cos\angmu\,d\cos\angpi\,d\chi } = \quad\quad\quad\quad\quad\quad\quad\quad\quad\quad\quad\quad\quad\quad\quad\quad\quad\quad\quad\quad\quad\quad\quad\quad
\nonumber\\&\quad\quad\quad\quad\quad\quad
 {\cal N} e^{-\G t}\left\{\frac{|A|^2+|(q/p)\Ab|^2}{2}\ch  + \frac{|A|^2-|(q/p)\Ab|^2}{2}\cs\right.\quad\quad\nonumber\\
&\quad\quad\quad\quad\quad\quad\quad\quad\quad\quad\quad- \left.\Real\left((q/p)A^*\Ab\right)\sh  - \Imag\left((q/p)A^*\Ab\right)\sn\right\},
\end{align}
\begin{align}
&\frac{d^5 \overline{\Gamma}}{dt\,dm_{hh}\,d\cos\angmu\,d\cos\angpi\,d\chi }  =
\quad\quad\quad\quad\quad\quad\quad\quad\quad\quad\quad\quad\quad\quad\quad\quad\quad\quad\quad\quad\quad\quad\quad\quad
\nonumber\\&\quad\quad\quad\quad
\left|\frac{p}{q}\right|^2{\cal N}  e^{-\G t}\left\{\frac{|A|^2+|(q/p)\Ab|^2}{2}\ch  - \frac{|A|^2-|(q/p)\Ab|^2}{2}\cs\right.\quad\quad\nonumber\\
&\quad\quad\quad\quad\quad\quad\quad\quad\quad\quad\quad - \left.\Real\left((q/p)A^*\Ab\right)\sh  + \Imag\left((q/p)A^*\Ab\right)\sn\right\},\label{Eqbar-t}
\end{align}
where $\cal N$ is a constant; $\brabar{A}$ is the amplitude of $\brabar{B}^0\to \jpsi \pip \pim$ at the decay time $t=0$, which is itself a function of $m_{hh},\angmu,\angpi,$ and $\chi$, summed over all resonant (and possibly non-resonant) components; $\Delta m$ is the mass difference between the heavy and light \Bz mass eigenstates, and $\Delta \Gamma$ the width difference;\footnote{We use the conventions that $\Delta m     =      m_H -      m_L$ and
$\Delta\Gamma = \Gamma_L - \Gamma_H$, where $L$ and $H$ correspond to the light and heavy mass eigenstates, respectively.}
$q$ and $p$ are complex parameters that describe the relation between mass and flavor eigenstates. In this analysis we take $|p/q|$ to be equal to unity.

Forming the sum of  $\Bz$ and $\Bzb$ decay widths and integrating over decay time, yields the time-integrated and flavor-averaged decay width
\begin{align}\label{Eq:SPDF}
S(m_{hh}, \angpi, \angmu, \chi)=&|A(m_{hh}, \angpi, \angmu, \chi)|^2+|\Ab(m_{hh}, \angpi, \angmu, \chi)|^2\nonumber\\
&-2{\cal D}\,\Real\left(\frac{q}{p}A^*(m_{hh}, \angpi, \angmu, \chi)\Ab(m_{hh}, \angpi, \angmu, \chi)\right)\nonumber\\
\approx &|A(m_{hh}, \angpi, \angmu, \chi)|^2+|\Ab(m_{hh}, \angpi, \angmu, \chi)|^2,
\end{align}
where we drop the term arising from quantum interference of the amplitudes in the last line. This results from the fact that the ${\cal D}$ factor  is 
negligibly small for $\Bzb$ meson decays. Specifically,
\begin{equation}
{\cal D} = \frac{\int_{0}^{\infty}\alpha(t) e^{-\G t}\sh \dv t}{\int_{0}^{\infty}\alpha(t)e^{-\G t}\ch \dv t} ,
\end{equation}
where $\alpha(t)$ is the decay time dependent detection efficiency.\footnote{For uniform acceptance, ${\cal D}=\Delta\Gamma/(2\Gamma)$.} Since $\Delta\Gamma/\Gamma$ is of the order of $1\%$  for 
$\Bzb$ meson decays \cite{PDG}, the ${\cal D} $ term is about the same size.

We define $A_R(m_{hh})$ to be the mass line shape of the resonance $R$, which in most cases is a Breit-Wigner function. It is combined with the decay properties of the
$\Bzb$ and resonance to form the expression for the decay amplitude. For each resonance $R$:
\begin{equation}\label{eq:DP}
{\mathscr A}_R(\m)=\sqrt{2J_R+1} \sqrt{P_R P_B}\,  F_B^{(L_B)}\left(\frac{P_B}{m_B}\right)^{L_B} F_R^{(L_R)}\left(\frac{P_R}{m_{hh}}\right)^{L_R} A_R(m_{hh}).
\end{equation}
Here  $P_R$ ($P_B$)
%, which depends only on $m(\pip\pim)$, 
is the scalar momentum of one of the two daughters of the resonance $R$
(or the $\Bzb$ meson) in the $R$ (or $\Bzb$) rest frame, $J_R$ is the spin of $R$, $L_B$ is the orbital angular momentum between the $J/\psi$ and $h^+h^-$ system, and $L_R$ the orbital angular momentum in the $h^+h^-$ decay, and thus is the same as the spin of the $h^+h^-$ resonance. $F_B^{(L_B)}$ and $F_R^{(L_R)}$ are the centrifugal barrier factors for the $\Bzb$ and the $R$ resonance, respectively~\cite{LHCb:2012ae}. The factor $\sqrt{P_RP_B}$ results from converting the phase space of  the Dalitz-plot variables $m^2_{hh}$ and  $m^2_{\jpsi h^+}$ to that of $m_{hh}$ and $\cos\angpi$. The function defined in Eq.~(\ref{eq:DP}) is based on previous amplitude analyses \cite{Mizuk:2008me, LHCb:2012ae}.

We must sum over all final states, $R$, so for each $\jpsi$ helicity,  denoted by $\lambda$, equal to $0$, $+1$ and $-1$ we have
the overall decay amplitudes:
\begin{equation}
\label{eq:heart}
{\cal \brabar{H}}_{\lambda}(m_{hh},\angpi) = \sum_R  \brabar{\bf h}_\lambda^R {\mathscr A}_R(\m) d_{-\lambda,0}^{J_R}(\angpi),
\end{equation}
where the Wigner-$d$ functions are defined in Ref.~\cite{PDG} and $\brabar{\textbf{h}}_\lambda^R$ are complex helicity coefficients. We note that the $\lambda$ value of the \jpsi is equal to that of the $R$ resonance.
Finally, the total decay rate of $\brabar{B}^0\to \jpsi \pi^+\pi^-$ at $t=0$ is given by
\begin{align}\label{Eq:Ab}
|\brabar{A}(m_{hh},\angpi,&\angmu,\chi)|^2=\nonumber\\&|{\cal \brabar{H}}_0(m_{hh},\angpi)|^2\sin^2\angmu+\frac{1}{2}\left(|{\cal \brabar{H}}_+(m_{hh},\angpi)|^2+|{\cal \brabar{H}}_-(m_{hh},\angpi)|^2\right) \nonumber\\
&\times (1+\cos^2\angmu)+\Real\left[{\cal \brabar{H}}_+(m_{hh},\angpi){\cal \brabar{H}}_-^*(m_{hh},\angpi)e^{2i\chi}\right]\sin^2\angmu \nonumber\\
&+\sqrt{2}\Real\left[\left({\cal \brabar{H}}_0(m_{hh},\angpi){\cal \brabar{H}}_+^*(m_{hh},\angpi)-{\cal \brabar{H}}^*_0(m_{hh},\angpi){\cal \brabar{H}}_-(m_{hh},\angpi)\right)e^{-i\chi}\right] \nonumber\\
&\times \sin\angmu\cos\angmu~.
\end{align}

In order to determine the  \CP components, it is convenient to replace the complex helicity coefficients $\brabar{\textbf{h}}_\lambda^R$ by the 
complex transversity coefficients $\brabar{\textbf{a}}_{\tau}^R$ using the relations
\begin{eqnarray}
\brabar{\textbf{h}}_0^R &=& \brabar{\textbf{a}}_{0}^R,\nonumber\\
%\brabar{\textbf{a}}_\parallel^R &=& \frac{1}{\sqrt{2}}(\brabar{\textbf{h}}_+^R+\brabar{\textbf{h}}_-^R),\nonumber\\
%\brabar{\textbf{a}}_\perp^R &=& \frac{1}{\sqrt{2}}(\brabar{\textbf{h}}_+^R-\brabar{\textbf{h}}_-^R).
\brabar{\textbf{h}}_+^R&=& \frac{1}{\sqrt{2}}(\brabar{\textbf{a}}_\parallel^R+\brabar{\textbf{a}}_\perp^R),\nonumber\\
\brabar{\textbf{h}}_-^R&=& \frac{1}{\sqrt{2}}(\brabar{\textbf{a}}_\parallel^R-\brabar{\textbf{a}}_\perp^R).
\end{eqnarray}
Here $\brabar{\textbf{a}}_{0}^R$ corresponds to the longitudinal polarization of the $\jpsi$ meson, and the other two coefficients correspond to polarizations of the $\jpsi$ meson and $h^+h^-$ system transverse to the decay axis: $\brabar{\textbf{a}}_\parallel^R$ for parallel polarization of the $\jpsi$ and $h^+h^-$, and $\brabar{\textbf{a}}_\perp^R$ for  perpendicular polarization.

Assuming the absence of direct \CP violation, the relation between the $\overline{B}^0$ and $B^0$ transversity coefficients is  $\bar{\textbf{a}}^R_\tau=\eta^R_\tau\textbf{a}^R_\tau$, where $\eta^R_\tau$ is the \CP eigenvalue of the $\tau$  transversity component for the intermediate state $R$, and $\tau$ denotes the $0,~\parallel,$ or $\perp$ components.
Note that for the $h^+h^-$ system both $C$ and $P$ are given by
$(-1)^{L_R}$, so the  \CP of  the $h^+h^-$ system is always even. The total \CP of the final state is $(-1)^{L_B}$, since the \CP of the \jpsi is also even. The final state \CP parities, for S, P,  and D-waves, are listed in Table~\ref{CPPart}.

In this analysis a fit determines the amplitude modulus $a_\tau^R$ and the phase $\phi_\tau^R$ of the amplitude
 \begin{equation}\label{eq:amp}
\textbf{a}^R_\tau=a_\tau^R e^{i\phi_\tau^R}
\end{equation}
for each resonance $R$, and each transversity component $\tau$.
For the $\tau=\perp$ amplitude, the $L_B$ value of spin-1 (or spin-2) resonances is 1 (or 2). While the other transversity components, $\tau=$ 0 or $\parallel$, have two possible $L_B$ values of 0 and 2 (or 1 and 3) for spin-1 (or -2) resonances. We use only the smaller values for each. Studies show that our results for fractions of different interfering components are not sensitive to these $L_B$ choices.

\begin{table}[t]
\centering
\caption{\small \CP parity of the full final state for different spin resonances. Note that spin-0 only has the 0 transversity component.}\label{CPPart}
\begin{tabular}{c|ccc}\hline
Spin& $\eta_0$& $\eta_\parallel$& $\eta_\perp$\\\hline
0 & $-1$ &  & \\
1 & ~~1 & ~~1 & $-1$\\
2 & $-1$ & $-1$ & ~~1\\ \hline
\end{tabular}
\end{table}

\subsection{Dalitz fit fractions}
A complete description of the decay is given in terms of the fitted complex amplitudes. Knowledge of the contribution of each component can be summarized by defining a fit fraction for each transversity $\tau$, ${\cal{F}}_\tau^R$.  To determine ${\cal{F}}_\tau^R$ one needs to integrate over all the four variables: $m_{hh},\angpi, \angmu,\chi$.  The interference terms between different helicity components vanish after integrating Eq.~(\ref{Eq:Ab}) over the two variables of $\cos\angmu$ and $\chi$, \ie
\begin{align}\label{eq:SumH}
\int |\brabar{A}(m_{hh},\angpi,&\angmu,\chi)|^2 d\cos\angmu\,d\chi\nonumber\\
&=\frac{4}{3}\left(|{\cal \brabar{H}}_0(m_{hh},\angpi)|^2+|{\cal \brabar{H}}_+(m_{hh},\angpi)|^2+|{\cal \brabar{H}}_-(m_{hh},\angpi)|^2\right).
\end{align}
The decay rate is the sum of the contributions from the three helicity terms.
To define the transversity fractions, we need to write Eq.~(\ref{eq:SumH}) in terms of transversity amplitudes. Since $d_{-1,0}^{J_R}=-d_{1,0}^{J_R}$, the sum of the three helicity terms is equal to the sum of three transversities, given as
\begin{align}
&|{\cal H}_0(m_{hh},\angpi)|^2+|{\cal H}_+(m_{hh},\angpi)|^2+|{\cal H}_-(m_{hh},\angpi)|^2 =\nonumber\\
&\left|\sum_R \textbf{a}_0^R {\mathscr A}_R(\m) d_{0,0}^{J_R}(\angpi)\right|^2+\left|\sum_R \textbf{a}_\|^R {\mathscr A}_R(\m) d_{1,0}^{J_R}(\angpi)\right|^2+\left|\sum_R \textbf{a}_\perp^R {\mathscr A}_R(\m) d_{1,0}^{J_R}(\angpi)\right|^2.
\end{align}
Thus, we define the transversity fit fractions as
\begin{equation}\label{eq:ff}
{\cal{F}}^R_{\tau}=\frac{\int\left| a^R_\tau e^{i\phi^R_\tau} {\mathscr A}_R(\m) d_{\lambda,0}^{J_R}(\angpi)\right|^2 d\m\;d\cos\angpi}{\int \left(|{\cal {H}}_0(m_{hh},\angpi)|^2+|{\cal {H}}_+(m_{hh},\angpi)|^2+|{\cal {H}}_-(m_{hh},\angpi)|^2\right)d\m\;d\cos\angpi},
\end{equation}
where $\lambda=0$ for $\tau=0$, and $\lambda=1$ for $\tau=\perp$ or $\parallel$.

The sum of the fit fractions is not necessarily unity due to the potential presence of interference between two resonances. Interference term fractions are given by
\begin{equation}
\label{eq:inter}
{\cal{F}}_{\tau}^{RR^\prime}=2\mathcal{R}e\left(\frac{\int a^R_\tau\; a^{R'}_\tau e^{i(\phi^R_\tau-\phi^{R'}_\tau)} {\mathscr A}_R(\m)  {\mathscr A}^{*}_{R'}(\m) d_{\lambda,0}^{J_R}(\angpi) d_{\lambda,0}^{J_{R'}}(\angpi) d\m\;d\cos\angpi}{\int \left(|{\cal {H}}_0(m_{hh},\angpi)|^2+|{\cal {H}}_+(m_{hh},\angpi)|^2+|{\cal {H}}_-(m_{hh},\angpi)|^2\right)d\m\;d\cos\angpi}\right),
\end{equation}
and
\begin{equation}
\sum_{R,\tau} {\cal{F}}_\tau^R+\sum^{R > R'}_{RR',\tau} {\cal{F}}_\tau^{RR'} =1.
\end{equation}
Interference between different spin-$J$ states vanishes when integrated over angle, because the $d^J_{\lambda0}$ angular functions are orthogonal.

\section{Selection requirements}

In this analysis we adopt a two step selection. The first step, preselection, is followed by  a multivariate selection based on a boosted decision tree (BDT)~\cite{Breiman}.
Preselection criteria are implemented to preserve a large fraction of the signal events, yet reject easily eliminated backgrounds, and are identical to those used  in Ref.~\cite{Aaij:2013zpt}. A $\Bzb \to \jpsi \pi^+\pi^-$ candidate is reconstructed by combining a $\jpsi \to \mu^+\mu^-$ candidate with two pions of opposite charge. To ensure good track reconstruction, each of the four particles in the $\Bzb$ candidate is required to have the track fit $\chi^2$/ndf to be less than 4, where ndf is the number of degrees of freedom of the fit. The $\jpsi\to \mu^+\mu^-$ candidate is formed by two identified muons of opposite charge having $\pt$ greater than 500\,\mev, and with a geometrical fit vertex $\chi^2$ less than 16. Only candidates with dimuon invariant mass between $-48$\,\mev and $+43$\,\mev from  the  observed $\jpsi$ mass peak are selected, and are then constrained to the $\jpsi$ mass~\cite{PDG} for subsequent use.

Each pion candidate is required to have $\pt$ greater than 250\,\mev, and that the scalar sum, $\pt(\pi^+)+\pt(\pi^-)$  is required to be larger than 900\,\mev. Both pions must have $\chi^2_{\rm IP}$ greater than 9 to reject particles produced from the PV. The $\chi^2_{\rm IP}$ is
computed as the difference between the $\chi^2$ of the PV
reconstructed with and without the considered track. Both pions
must also come from a common vertex with $\chi^2{\rm /ndf}<16$, and form a vertex with the $\jpsi$ with a $\chi^2$/ndf less than 10 (here ndf equals five). Pion candidates are identified using the RICH and muon systems. The particle
identification makes use of the logarithm of the likelihood
ratio comparing two particle hypotheses (DLL). For the pion
selection we require DLL$(\pi-K)>-10$ and DLL$(\pi-\mu)>-10$.
The $\Bzb$ candidate must have a flight distance of more than 1.5\,\mm. The angle between the combined momentum vector of the decay products and the vector formed from the positions of the PV and the decay vertex (pointing angle) is required to be less than $2.5^\circ$.

The BDT uses eight variables that
are chosen to provide separation between signal and background.
These are the minimum of DLL($\mu-\pi$) of the $\mu^+$ and $\mu^-$, $\pt(\pi^+)+\pt(\pi^-)$, the minimum of $\chi^2_{\rm IP}$ of the $\pi^+$ and $\pi^-$, and the $\Bzb$ properties of vertex $\chi^2$, pointing angle, flight distance, $\pt$ and $\chi^2_{\rm IP}$. The BDT is trained on a simulated sample of two million $\Bzb\to \jpsi \pi^+\pi^-$ signal events generated uniformly in phase space with unpolarized $J/\psi \rightarrow \mu^+\mu^-$ decays, and a background data sample from the sideband $5566<m(J/\psi \pi^+\pi^-)< 5616$\,\mev. Then the BDT is tested on independent samples from the same sources. The BDT can take any value from -1 to 1. The distributions of signal and background are approximately Gaussian shaped with r.m.s. of about 0.13. Signal peaks at BDT of 0.27 and background at -0.22. 
 To minimize possible bias on the signal acceptance due to the BDT, we choose a loose requirement of BDT$>0$, which has about a $95\%$ signal efficiency and a $90\%$ background rejection rate.

\section{Fit model}\label{Formalism}

\begin{figure}[b]
\begin{center}
\includegraphics[width=0.8\textwidth]{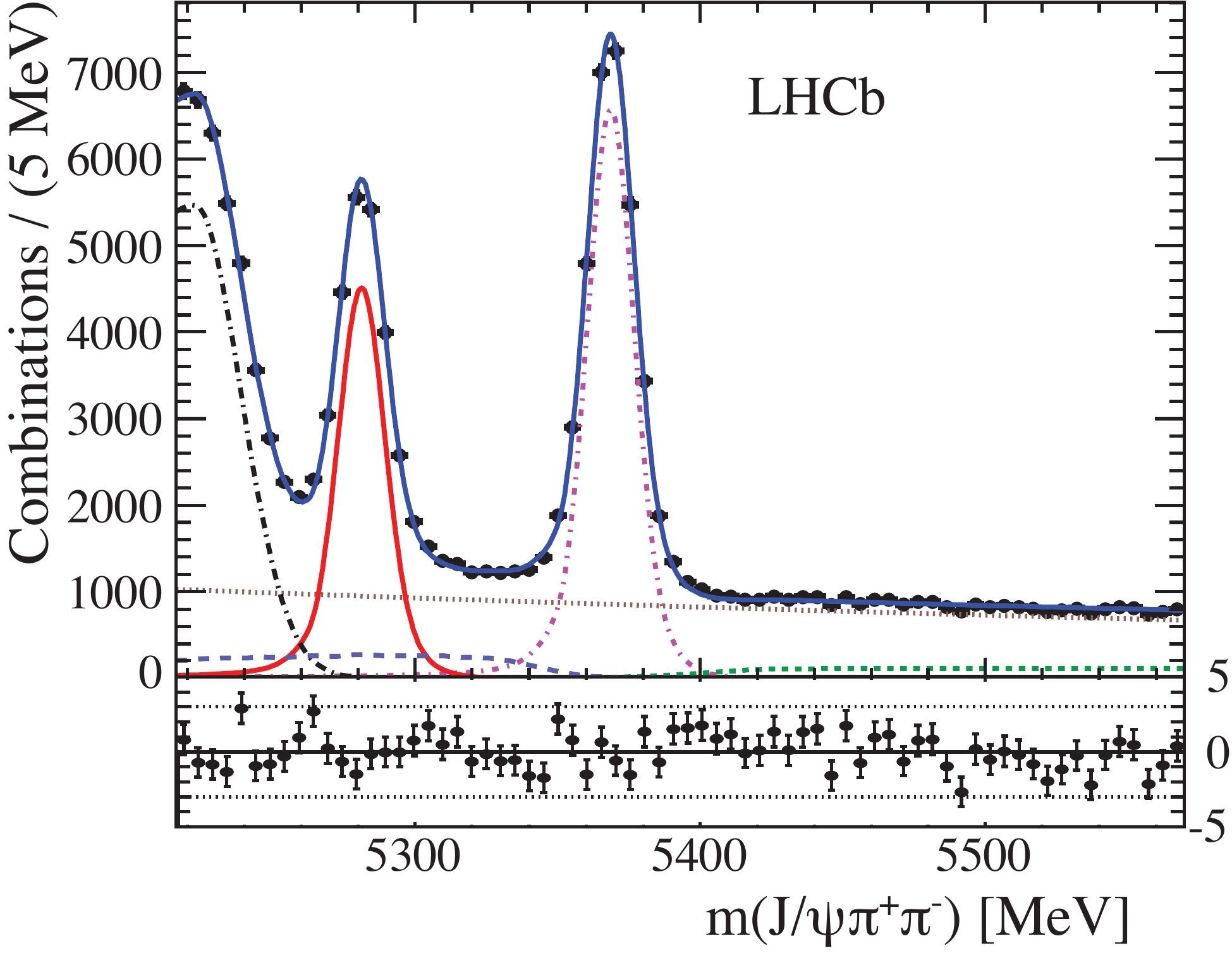}
\end{center}\label{fitmass}
\vskip -0.8cm
\caption{\small Invariant mass of $J/\psi \pi^+\pi^-$ combinations together with the data fit. The (red) solid curve shows the $\Bzb$ signal, the (brown) dotted line shows the combinatorial background, the (green) short-dashed line shows the $B^-$ background, the (purple) dot-dashed curve is $\Bsb\rightarrow J/\psi \pi^+\pi^-$, the (light blue) long-dashed line is the sum of $\Bsb\rightarrow J/\psi\eta^{(')}$, $\Bsb\rightarrow J/\psi\phi$ with  $\phi\to\pi^+\pi^-\pi^0$ backgrounds and the $\Lz_b^0 \to \jpsi K^- p$ reflection, the (black) dot-long dashed curve is the $\Bdb\rightarrow J/\psi K^- \pi^+$ reflection and the (blue) solid curve is the total. The points at the bottom show the difference between the data points and the total fit divided by the statistical uncertainty on the data.}
\end{figure}

We first select events based on their $\jpsi\pi^+\pi^-$ invariant mass and then perform a full fit to the decay variables.
The invariant mass of the selected $J/\psi\pi^+\pi^-$ combinations is shown in Fig.~\ref{fitmass}. 
There is a large peak at the $\Bsb$ mass and a smaller one at the $\Bzb$ mass on top of the background. A double Crystal Ball function with common means models the radiative tails and is used to fit each of the  signals \cite{Skwarnicki:1986xj}. The known $\Bsb - \Bzb$ mass difference \cite{PDG} is used to constrain the difference in mean values. 
 Other components in the fit model take into account background contributions from  $B^-\rightarrow J/\psi K^-$ and $B^-\rightarrow J/\psi \pi^-$ decays combined with a random $\pi^+$, $\Bsb\rightarrow J/\psi\eta(')$ with $\eta(')\rightarrow \pi^+\pi^- \gamma$, $\Bsb\rightarrow J/\psi\phi$  with $\phi\rightarrow \pi^+\pi^-\pi^0$,  $\Bdb\rightarrow J/\psi K^- \pi^+$ and $\Lb \to \jpsi \Km \proton$ reflections, and combinatorial backgrounds. The exponential  combinatorial background shape is taken from like-sign combinations, that are the sum of $\pi^+\pi^+$ and $\pi^-\pi^-$ candidates, and found to be a good description in previous studies \cite{LHCb:2012ae,Stone:2009hd}.

The shapes of the other components are taken from the Monte Carlo simulation with their normalizations allowed to vary. The mass fit gives
$18\,841\pm204$ signal and $10\,207\pm178$ background candidates
within $\pm20$ MeV of the $\Bdb$ mass peak.
Only candidates within $\pm20 \mev$ of the $\Bzb$ mass peak are retained for further analysis.
To improve the resolution of the mass and angular variables used in the amplitude analysis,  we perform a kinematic fit constraining the $\Bdb$ and $J/\psi$ masses to their PDG mass values \cite{PDG}, and recompute the final state momenta~\cite{Hulsbergen:2005pu}.

One of the main challenges in performing a mass and angular analysis is to construct a realistic probability density function, where both the kinematic and dynamical properties are modeled accurately. The PDF is  given by the sum of signal, $S$, and background, $B$, functions. The \Bzb signal includes events from the reaction $\Bzb\to \jpsi\KS$. These are described by a separate term in the PDF. The total PDF is 

\begin{eqnarray}\label{eq:pdf}
F(m_{hh}, \theta_{hh}, \theta_{J/\psi}, \chi)&=&f_{\rm sig}\times\left[\frac{1-f_{\KS}}{{\cal{N}}_{\rm sig}}\varepsilon(m_{hh}, \theta_{hh}, \theta_{J/\psi}, \chi) S(m_{hh}, \theta_{hh}, \theta_{J/\psi}, \chi) \vphantom{\frac{f_{\KS}}{{\cal N}_{\KS}}}\right. \nonumber \\
&& + \left.\frac{f_{\KS}}{{\cal N}_{\KS}}\varepsilon(m_{hh}, \theta_{hh}, \theta_{J/\psi}, \chi) G(m_{hh};m_{\KS},\sigma_{\KS}) \sin^2\theta_{J/\psi}\right]\nonumber \\
 &&+(1-f_{\rm sig})B(m_{hh}, \theta_{hh}, \theta_{J/\psi}, \chi),
\end{eqnarray}
where $f_{\text{sig}}$ is the fraction of the signal in the fitted region ($f_{\text{sig}}=(64.9\pm1.2)\%$ obtained from the mass fit in Fig.~\ref{fitmass}), $\varepsilon$ is the detection efficiency described in Sec.~\ref{sec:mc}, and $B$ is the background PDF described in Sec.~\ref{sec:background modeling}. The $\KS$  component is modeled by a Gaussian function, $G$, with mean $m_{\KS}$ and width $\sigma_{\KS}$. The Gaussian parameters together with the $\KS$ fraction in the $\Bzb$ peak, $f_{\KS}$, are determined in the fit. The normalization factors ${\cal N}_{\rm sig}$ for the signal and ${\cal N}_{\KS}$ for the $\KS$ candidates are efficiency-multiplied theoretical functions integrated over the four analysis variables, $m_{hh}$, $\theta_{hh}$, $\theta_{J/\psi}$, and $\chi$, given by
\begin{eqnarray}
{\cal{N}}_{\rm sig}&=&\int \! \varepsilon(m_{hh}, \angpi, \angmu, \chi) S(m_{hh}, \angpi, \angmu, \chi) \,
\dv \,\m \, \dv\cos\angpi \, \dv\cos\angmu \,\dv\chi.
\end{eqnarray}

\begin{figure}[t]
\begin{center}
\includegraphics[width =0.6\textwidth]{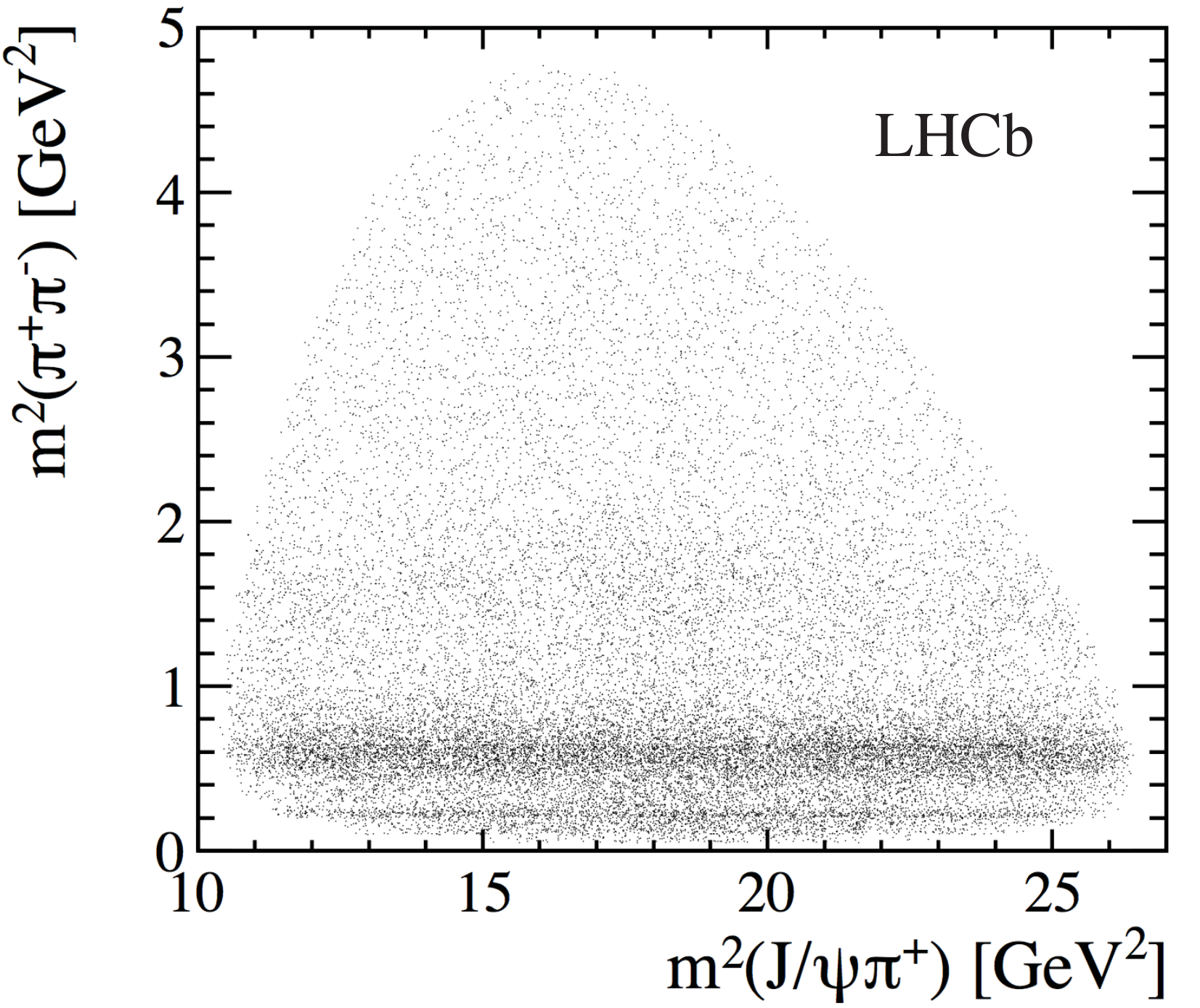}
\end{center}\label{fig:Dalitz_B0_snapshot}
\vskip -0.8cm
\caption{\small Distribution of $m^2(\pip\pim)$ versus $m^2(\jpsi\pip)$ for all events within $\pm20$ MeV of the $\Bdb$ mass.}
\end{figure}

\begin{figure}[b]
\begin{center}
\includegraphics[scale=0.7]{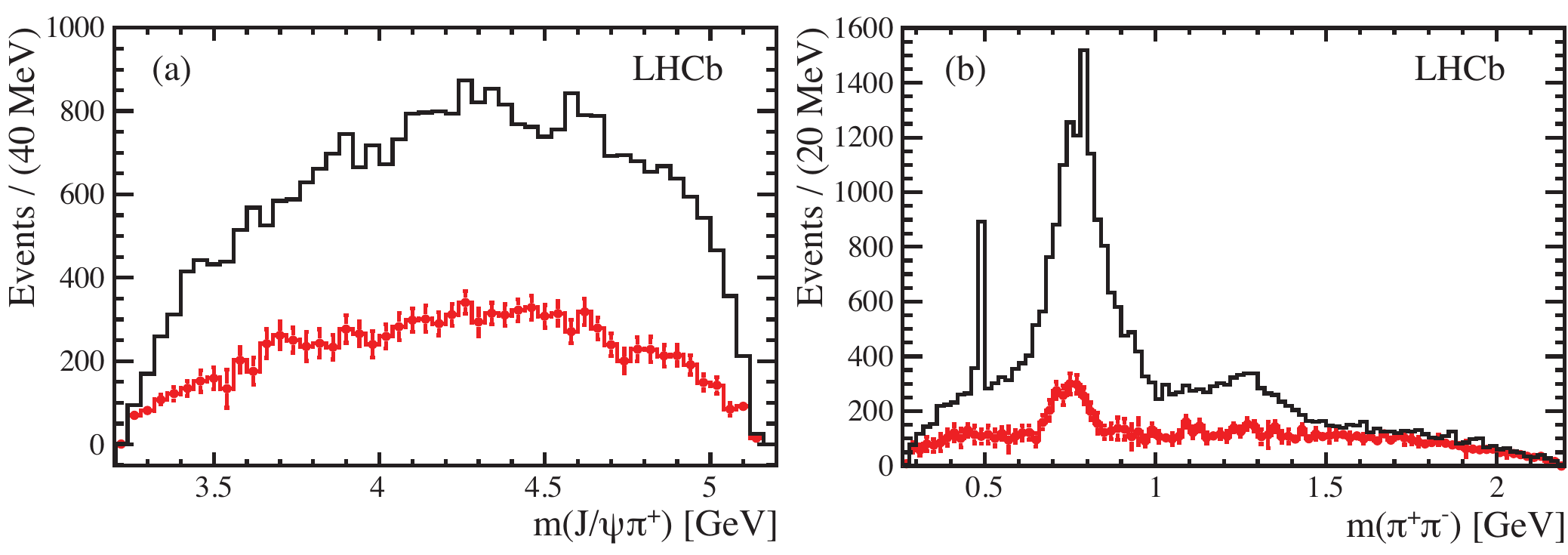}
\end{center}\label{fig:overlap_rs_data_rs_bkg}
\vskip -0.9cm
\caption{\small Distributions of (a) $m(\jpsi\pip)$ and (b) $m(\pip\pim)$ for $\Bdb\rightarrow J/\psi\pi^+\pi^-$ candidates within $\pm20$ MeV of the $\Bdb$ mass. The red points with error bars show the background contributions obtained by fitting the $m(\jpsi\pip\pim)$ distribution in bins of the plotted variables.}
\end{figure}

Examination of the event distribution for $m^2(\pip\pim)$ versus $m^2(\jpsi\pip)$ in Fig.~\ref{fig:Dalitz_B0_snapshot} shows obvious structures in $m^2(\pip\pim)$. To investigate if there are visible exotic structures in $m^2(\jpsi\pip)$, we examine the $\jpsi\pip$ invariant mass distribution as shown in Fig.~\ref{fig:overlap_rs_data_rs_bkg}(a) where we fit the $m(\jpsi\pip\pim)$ distribution to extract the background levels in bins of $m(\jpsi\pip)$ (red points).  Similarly, Fig.~\ref{fig:overlap_rs_data_rs_bkg}(b) shows the $\pip\pim$ mass distribution. Apart from a large signal peak due to the $\rho(770)$, there are visible structures at about 1270 MeV and a $\KS$ component at about 500 MeV.

\subsection{Detection efficiency}
\label{sec:mc}

The detection efficiency is determined from a sample of about four million simulated $\Bdb\rightarrow J/\psi\pi^+\pi^-$ events that are generated uniformly in phase space with unpolarized $J/\psi \rightarrow \mu^+\mu^-$ decays. 
The efficiency model can be expressed as
\begin{equation}
\varepsilon(m_{hh}, \theta_{hh}, \theta_{J/\psi}, \chi)=\varepsilon_1(s_{12}, s_{13})\times \varepsilon_2(\theta_{\jpsi} , m_{hh}) \times \varepsilon_3(\chi, m_{hh}),\label{eq:eff}
\end{equation}
where $s_{12}\equiv m^2({\jpsi\pi^+})$ and $s_{13}\equiv m^2({\jpsi\pi^-})$ are functions of $(m_{hh}, \angpi)$; such parameter transformations in $\varepsilon_1$ are implemented in order to use the Dalitz-plot based efficiency model  developed in previous publications~\cite{LHCb:2012ae,Aaij:2013zpt}. 

%To simplify the normalization of PDF's,
The efficiency dependence on $\chi$ is modeled by
\begin{equation}
\varepsilon_3(\chi , m_{hh})=\frac{1}{2\pi}(1+ p_1\cos\chi+p_2\cos2\chi),\label{eq:chi}
\end{equation}
where $p_1=p_1^0+p_1^1 \times m^2_{hh}$ and $p_2=p_2^0+p_2^1 \times m^2_{hh} + p_2^2\times m^4_{hh}$. The free parameters are determined by fitting the simulated $\chi$ distributions using Eq.~(\ref{eq:chi}) in bins of $m^2_{hh}$. The fit gives $p_1^0=-0.0065\pm0.0052$ and $p_1^1=(0.0011\pm0.0021)$\,GeV$^{-2}$; $p_2^0=-0.0006\pm0.0079$, $p_2^1=(0.0602 \pm0.0083)$\,GeV$^{-2}$ and $p_2^2=(-0.0099\pm0.0018)$\,GeV$^{-4}$.

The acceptance in $\cos \theta_{\jpsi}$ depends on $m_{hh}$. We disentangle this correlation by fitting the $\cos \theta_{\jpsi}$ distribution in 24 bins of $m^2_{hh}$ using the parameterization
\begin{equation}
\varepsilon_2(\theta_{\jpsi} , m_{hh})=\frac{1+ a\cos^2\theta_{J/\psi}}{2+2a/3}.\label{eq:cosHacc}
\end{equation}
The fitted values of $a$  are modeled by a second order polynomial function
\begin{equation}
a(m^2_{hh}) = a_0 + a_1 m^2_{hh} + a_2 m_{hh}^4,
\end{equation}
with $a_0=0.189\pm0.021$, $a_1=-0.116\pm0.021$\,GeV$^{-2}$ and $a_2=0.017\pm0.004$\,GeV$^{-4}$.

We model the detection efficiency, $\varepsilon_1(s_{12}, s_{13})$, by using the symmetric  observables
\begin{equation}
x= s_{12}/{\rm GeV}^2-18.4~,~~~~{\rm and}~~~~  y=s_{13}/{\rm GeV}^2-18.4~.
\end{equation}
These variables are related to $s_{23}$ by
\begin{eqnarray}
\label{Eq:sums}
s_{12} + s_{13} + s_{23} =m^2_B + m^2_{\jpsi} + m^2_{\pip} + m^2_{\pim}.
\end{eqnarray}

Thus, $\varepsilon_1(s_{12}, s_{13})$ can be modeled by a two-dimensional fifth order polynomial function as
\begin{eqnarray}
\varepsilon_1(s_{12}, s_{13})  &=&1+\epsilon_1(x+y)+\epsilon_2(x+y)^2+\epsilon_3 xy + \epsilon_4 (x+y)^3 + \epsilon_5 xy (x+y) \nonumber \\
&&+ \epsilon_6 (x+y)^4 + \epsilon_7 xy(x+y)^2 + \epsilon_8 x^2y^2 \nonumber \\
&&+ \epsilon_9 (x+y)^5 +\epsilon_{10} xy(x+y)^3 + \epsilon_{11} x^2y^2(x+y)
\end{eqnarray}
where all the $\epsilon_i$ are the fit parameters. The $\chi^2/ \rm ndf$ is 313/299. The values of the parameters are given in Table~\ref{tab:eps1}.

\begin{table}[b]
\centering
\caption{\small Efficiency parameters. There are substantial correlations.}\label{tab:eps1}
\begin{tabular}{cc}\hline
$\epsilon_1$ & 0.1220$\pm$0.0097\\
$\epsilon_2$ & 0.1163$\pm$0.0182\\
$\epsilon_3$ & 0.0051$\pm$0.0004\\
$\epsilon_4$ & 0.0399$\pm$0.0101\\
$\epsilon_5$ &$\!\!$ -0.0012$\pm$0.0007\\
$\epsilon_6$ & 0.10051$\pm$0.0023\\
$\epsilon_7$ & $\,\,\,$0.0002$\pm$0.0005\\
$\epsilon_8$ & $\,\,$-0.000150$\pm$0.000007\\
$\epsilon_9$ &$\,\!$ -0.000011$\pm$0.000261\\
$\epsilon_{10}$ &$\,\,\,$ 0.000350$\pm$0.000146\\
$\epsilon_{11}$ &$\,\,$ -0.000113$\pm$0.000011\\
 \hline
\end{tabular}
\end{table}

\begin{figure}[htb]
\begin{center}
    \includegraphics[width=0.48\textwidth]{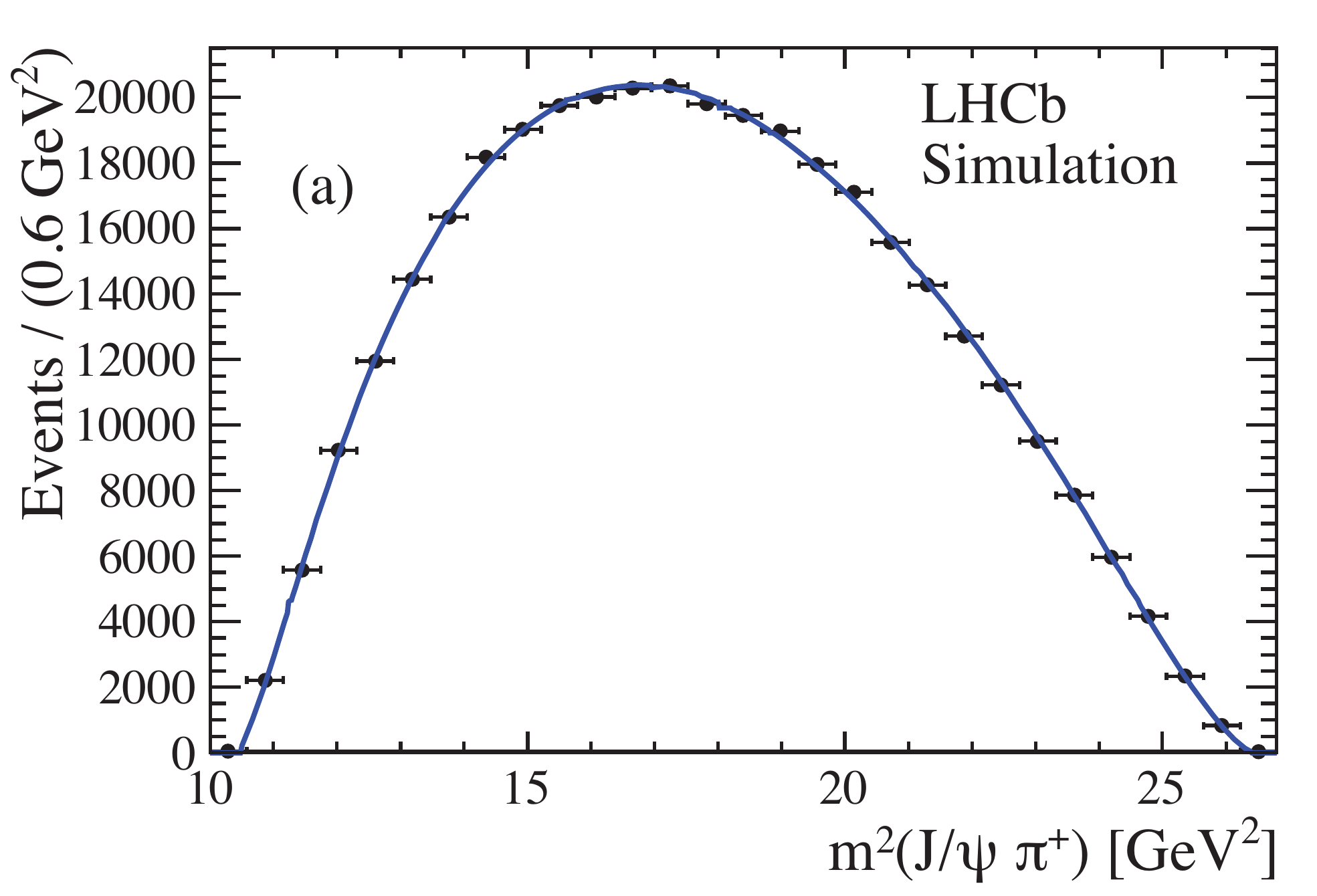}%
    \includegraphics[width =0.468\textwidth]{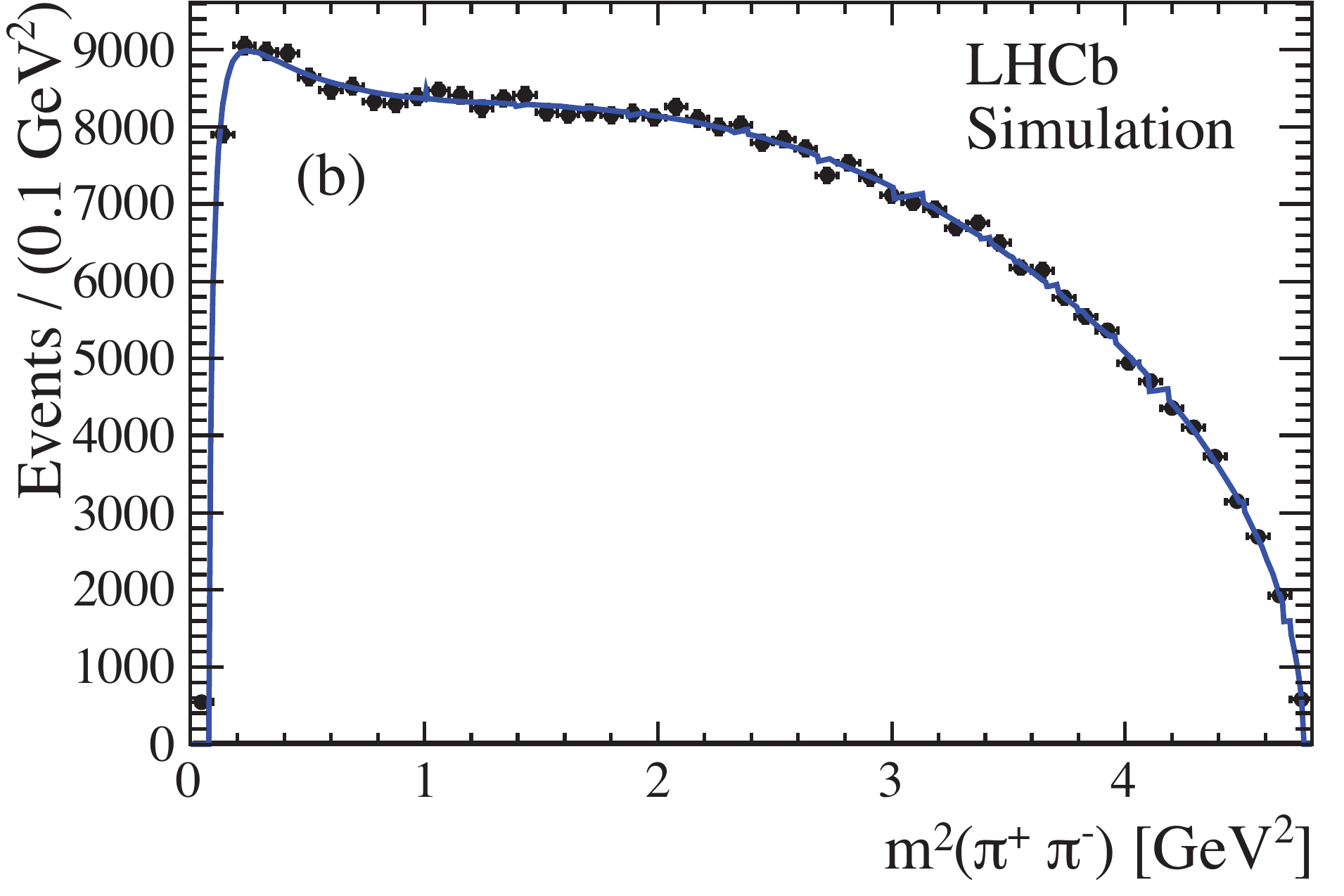}
\end{center}\label{eff2}
\vskip -0.5cm
\caption{\small Projections of invariant mass squared  (a) $s_{12}\equiv m^2(J/\psi \pi^+)$ and (b) $s_{23}\equiv m^2(\pi^+\pi^-)$ of the simulated Dalitz plot used to measure the efficiency parameters. The points represent the simulated event distributions and the curves the polynomial fit.}
\end{figure}

\begin{figure}[t]
  \begin{center}
     \includegraphics[width= 5 in]{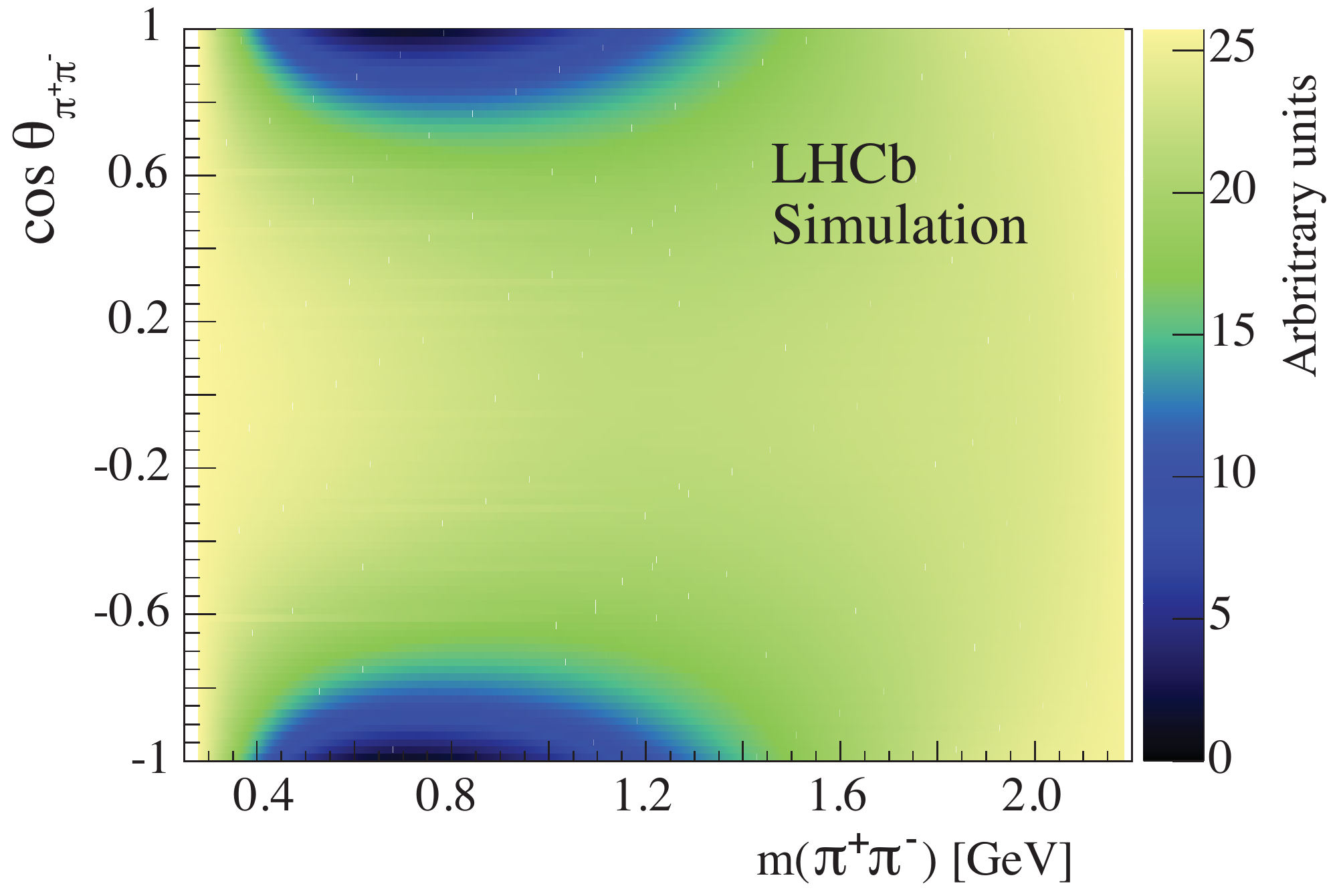}
    \caption{\small The variation of $\varepsilon_1$ is shown as a function of $m(\pi^+\pi^-)$ and $\cos \theta_{\pi^+\pi^-}$.}  \label{fig:eff_vs_mhh_coshh}
  \end{center}
\end{figure}

The projections of the fit used to measure the efficiency parameters are shown in Fig.~\ref{eff2}. The efficiency shapes are well described by the parametrization.
%[htb]
The parameterized efficiency as a function of $m(\pi^+\pi^-)$ versus $\cos\theta_{\pi^+\pi^-}$ is shown in Fig.~\ref{fig:eff_vs_mhh_coshh}.

\subsection{Background composition}
\label{sec:background modeling}

\begin{figure}[htb]
\begin{center}
    \includegraphics[width=6 in]{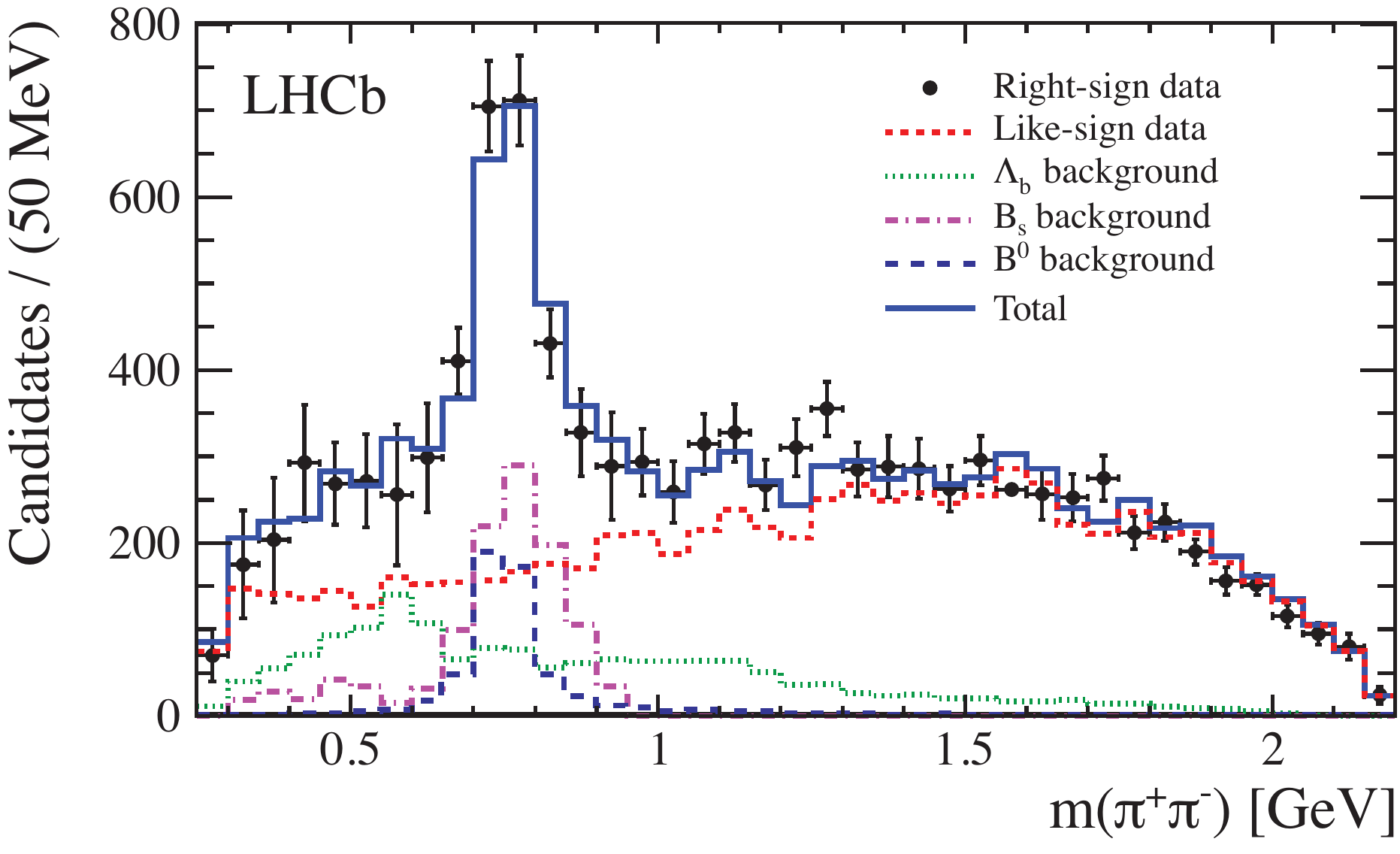}
\end{center}
\vskip -0.9cm
\caption{\small Distributions of $m(\pi^+\pi^-)$ of background components. The (blue) histogram shows the like-sign combinations added with additional backgrounds using simulations. The (black) points with error bars show the background obtained from the fits to the $m(\jpsi \pip \pim)$ mass spectrum in each bin of $\pi^+\pi^-$ mass.\label{bkgcmp}}
\end{figure}

The main background source in the $\Bzb$ signal region is combinatorial and can be taken from
the like-sign combinations within $\pm20$\,MeV of the $\Bdb$ mass peak. In addition, there is background arising from partially reconstructed $\Bsb$ decays ($\Bsb\rightarrow J/\psi\eta(^\prime) $, with $\eta(^\prime)\rightarrow \pi^+\pi^- \gamma$ and $\Bsb\to\jpsi\phi$ with $\phi\to\pip\pim\pi^0$), reflections from misidentified $\Lb \to \jpsi \Km \proton$ and $\Bdb\rightarrow J/\psi K^- \pi^+$ decays, which cannot be present in the like-sign combinations. We use simulated samples of these decays to model their contributions. The \Lb normalizations are determined from a previous analysis \cite{Aaij:2014zyy}. The background level in the opposite-sign combination ($\Bzb \to \jpsi \pip \pim$) is studied by fitting the $m(\jpsi \pip \pim)$ distributions in bins of $m(\pi^+\pi^-)$. The resulting background distribution in the $\pm20$\,MeV $\Bdb$ signal region is shown in Fig.~\ref{bkgcmp} by points with error bars.  A fit to this distribution gives a partially reconstructed $\Bsb$ background fraction of 10.7\%, the reflection from $\Bzb$ of 5.3\%, and the reflection from the $\Lb$ baryon of 15.5\% of the total background. The like-sign combinations summed with the additional backgrounds modeled by simulation are shown in Fig.~\ref{bkgcmp}.

When this data-simulation hybrid sample is used to extract the background parameters, a further re-weighting procedure is applied based on comparison of $m(\pi^+\pi^-)$ distributions between the overall fit and the background data points in 
Fig. \ref{fig:overlap_rs_data_rs_bkg}(b).

To better model the angular distributions in the $\rho(770)$ mass region, the background is separated into the  $\Kstarzb$ reflection from $\Bzb$, the $\rho$, and other backgrounds. The total background PDF is sum of these three components:
\begin{align}
B(m_{hh}, \theta_{hh}, \theta_{J/\psi}, \chi)=&\frac{f_{\Kstarzb}}{{\cal N}_{\Kstarzb}}B_{\Kstarzb}(m_{hh}, \theta_{hh}, \theta_{J/\psi}, \chi)
+\frac{f_{\rho}}{{\cal N}_{\rho}}B_{\rho}(m_{hh}, \theta_{hh}, \theta_{J/\psi}, \chi)\nonumber\\
&+ \frac{1-f_{\Kstarzb}-f_{\rho}}{{\cal N}_{\rm other}}B_{\rm other}(m_{hh}, \theta_{hh}, \theta_{J/\psi}, \chi), 
\end{align}
where the $\cal{N}$'s are normalizations, the contributing fractions having values of $f_{\Kstarzb}=(5.3\pm0.2)\%$ and $f_{\rho}=(9.5\pm0.6)\%$; the other background is normalized as $1-f_{\Kstarzb}-f_{\rho}$.

The $\Kstarzb$ background is modeled by the function
\begin{align}
B_{ \Kstarzb}(m_{hh}, \theta_{hh}, \theta_{J/\psi}, \chi)=&\left(\frac{p_R}{m_{hh}}\right)^2\frac{m_{hh}e^{-a\cdot(1-|\cos\theta_{hh}|)}}{(m^2_0-m_{hh}^2)^2+m_0^2\Gamma_0^2}
\times (1-|\cos\theta_{hh}|)^b\nonumber\\ & \times \left(1+\alpha_0\cos^2\theta_{J/\psi}\right) \times (1+p_{b1}\cos \chi + p_{b2} \cos2 \chi),
\end{align}
where $m_0$, $\Gamma_0$, $a$, $b$, $\alpha_0$ are free parameters determined by fitting to the $\Bzb\to\jpsi\Kstarzb$ simulation.
The last part $(1+p_{b1}\cos \chi + p_{b2} \cos2 \chi)$ is a function of the $\chi$ angle. We have verified that the three backgrounds have consistent $\chi$ distributions, thus the parameters $p_{b1}$ and $p_{b2}$ are determined by fitting all backgrounds simultaneously.

The $\rho$ background is described by the function
\begin{align}
B_{\rm \rho}(m_{hh}, \theta_{hh}, \theta_{J/\psi}, \chi)=&\left(\frac{p_R}{m_{hh}}\right)^2\frac{m_{hh}}{(m^2_{\rho}-m_{hh}^2)^2+m_{\rho}^2\Gamma_{\rho}^2}
\times \sin^2\theta_{hh}\nonumber\\ & \times \sin^2\theta_{J/\psi} \times (1+p_{b1}\cos \chi + p_{b2} \cos2 \chi),
\end{align}
where $m_{\rho}$, $\Gamma_{\rho}$ are free parameters. The parameters are obtained by fitting to  simulated $\Bsb\to\jpsi\eta^\prime(\to\rho\gamma)$ events.

The model for the remaining backgrounds is 
\begin{align}
B_{\rm other}(m_{hh}, \theta_{hh}, \theta_{J/\psi}, \chi)=&m_{hh}B_1(m^2_{hh},\cos \theta_{hh})\times \left(1+\alpha_1\cos^2\theta_{J/\psi}\right)
\nonumber\\
& \times (1+p_{b1}\cos \chi + p_{b2} \cos2 \chi),
\end{align}
with the function
\begin{equation}
B_1(m_{hh}^2,\cos \theta_{hh})=\left[B_2(\zeta)\frac{p_B}{m_B}+\frac{b_0}{(m^2_1-m^2_{hh})^2+m_1^2\Gamma_1^2}\right]\times\frac{1+q(\zeta)|\cos\theta_{hh}| +p(\zeta)\cos^2 \theta_{hh}}{2[1+q(\zeta)/2+p(\zeta)/3]}.
\end{equation}
Here the variable $\zeta=2(\m^2 -m^2_{\rm min})/(m^2_{\rm max}-m^2_{\rm min})-1$, where $m_{\rm min}$ and $m_{\rm max}$ are the fit boundaries, $B_2(\zeta)$ is a fifth order Chebychev polynomial with coefficients $b_i$ ($i=1$-5), and $q(\zeta)$ and $p(\zeta)$ are two first order Chebychev polynomials with parameters $c_j$ ($j=1$-4).

\begin{figure}[b]
\begin{center}
    \includegraphics[width=3 in]{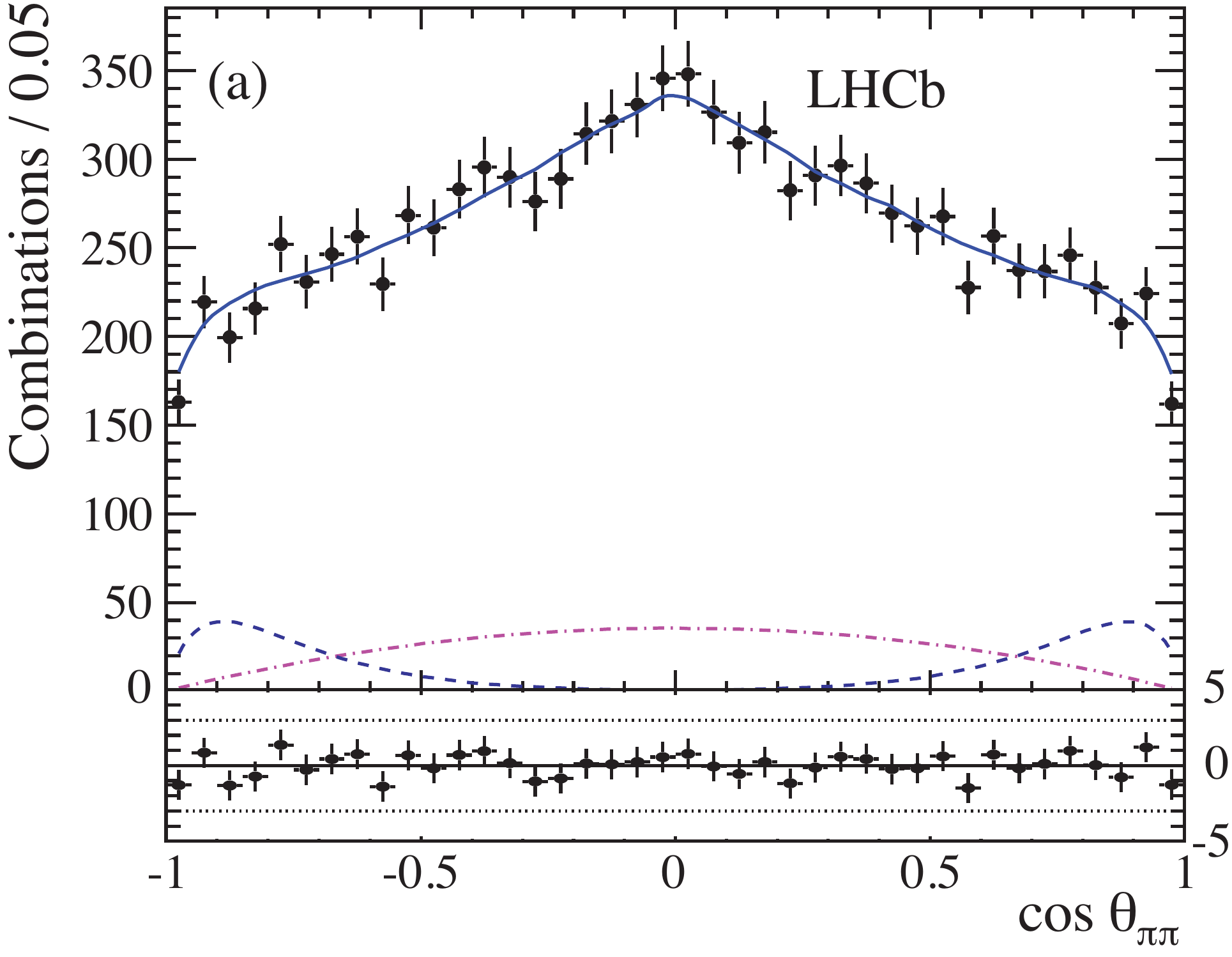}%
    \includegraphics[width=3 in]{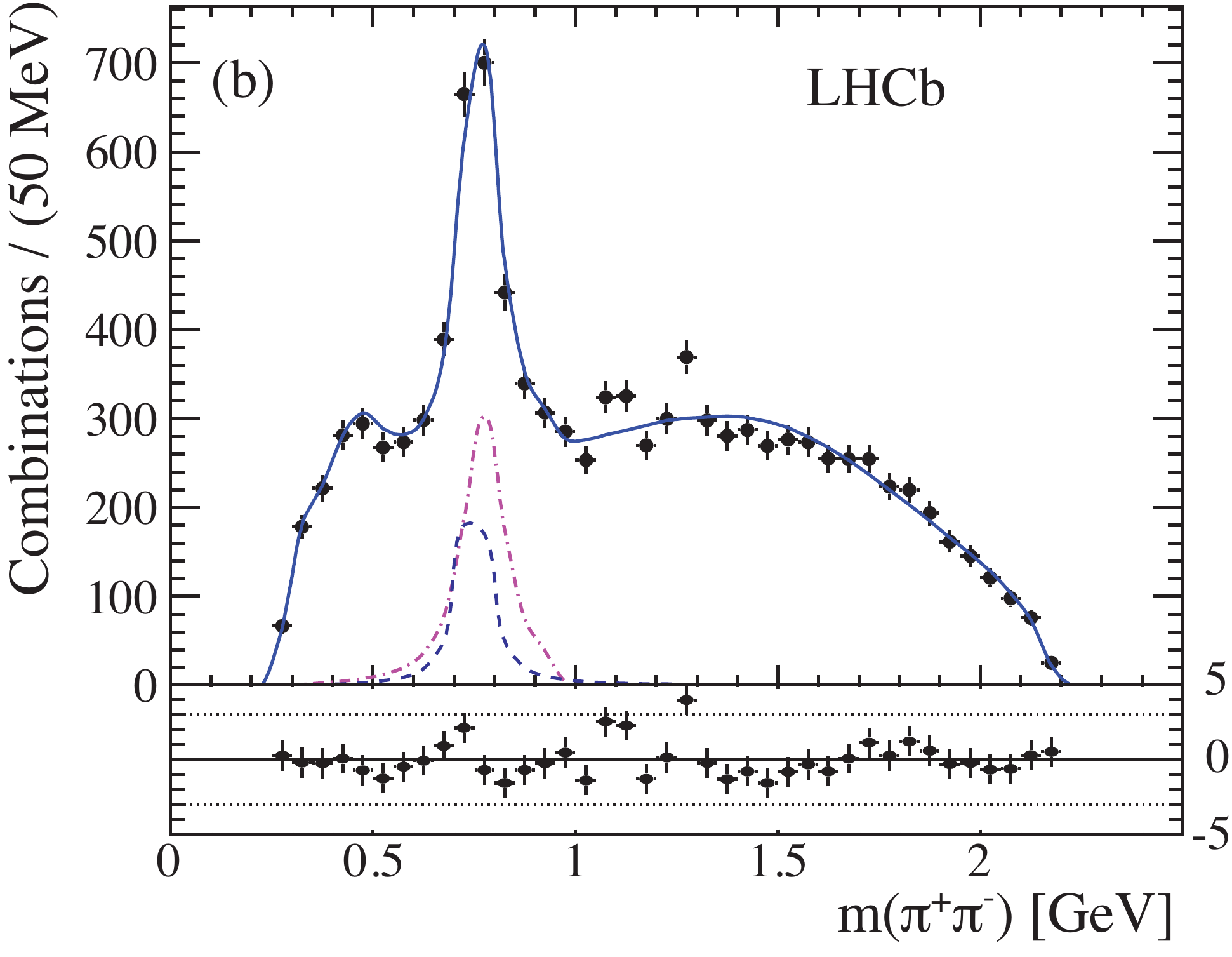}
\end{center}\label{bkg2}
\vskip -0.8cm
\caption{\small Projections of  (a) $\cos \theta_{\pi\pi}$ and (b) $m(\pip\pim)$ of the background. The points with error bars show the like-sign data combinations added with the \Lb background and additional simulated backgrounds. The (magenta) dot-dashed line shows the $\eta^{(')}\to\rho\gamma$ background, the (dark-blue) dashed line the $\Kstarb$ reflection background, and the (blue) solid line the total. The points at the bottom show the difference between the data points and the total fit divided by the statistical uncertainty on the data.}
\end{figure}

\begin{figure}[htb]
\begin{center}
\includegraphics[scale=0.5]{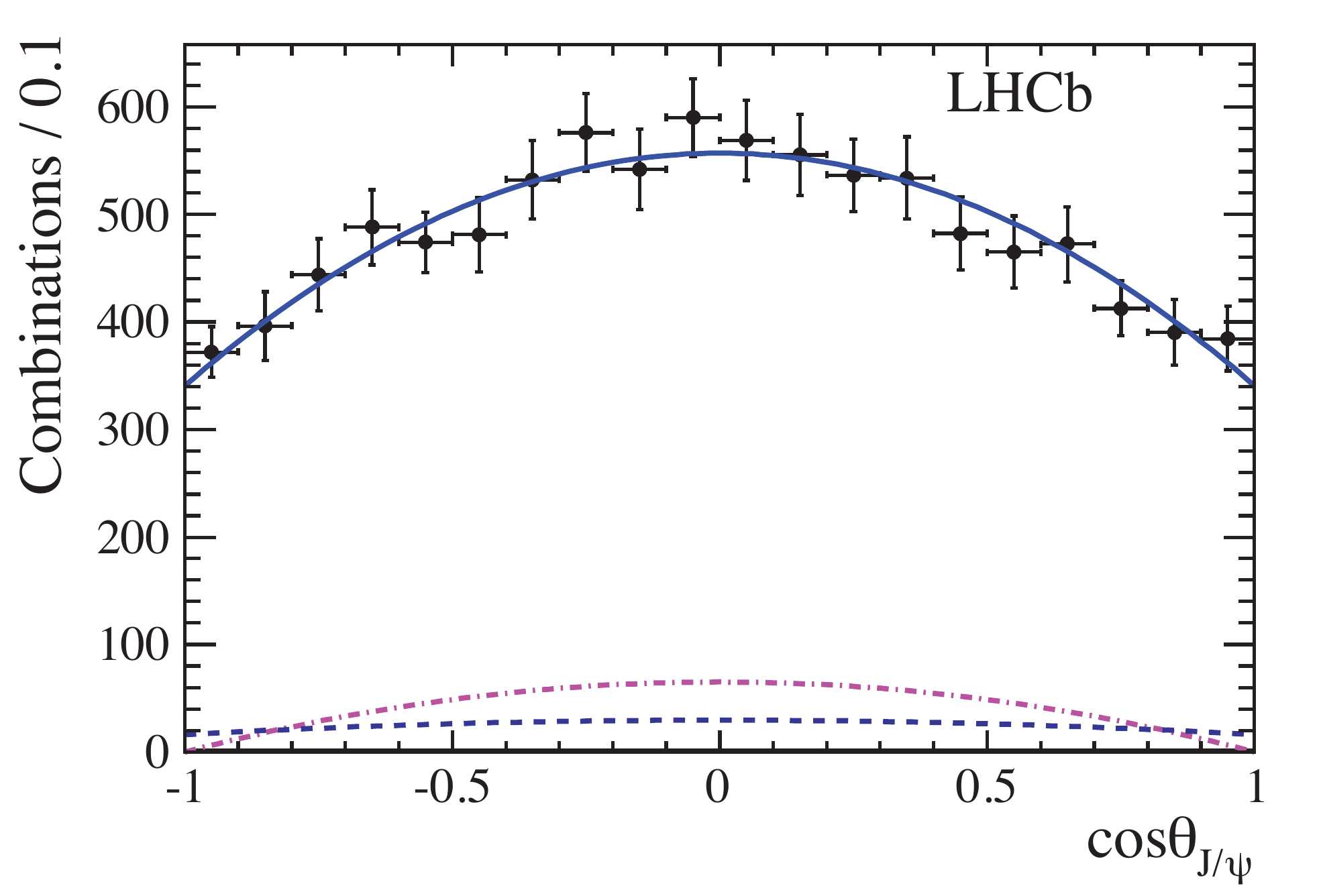}
\end{center}\label{bkgcosjpsi}
\vskip -1cm
\caption{\small The $\cos\theta_{J/\psi}$ distribution of the data-simulated hybrid background sample. The points with error bars show the like-sign data combinations added with the \Lb background and additional simulated backgrounds.  The (magenta) dot-dashed line shows the $\rho$ background, the (dark-blue) dashed the $\Kstarb$ reflection background and the (blue) solid line the total.}
\end{figure}

\begin{figure}[htb]
\begin{center}
\includegraphics[scale=0.5]{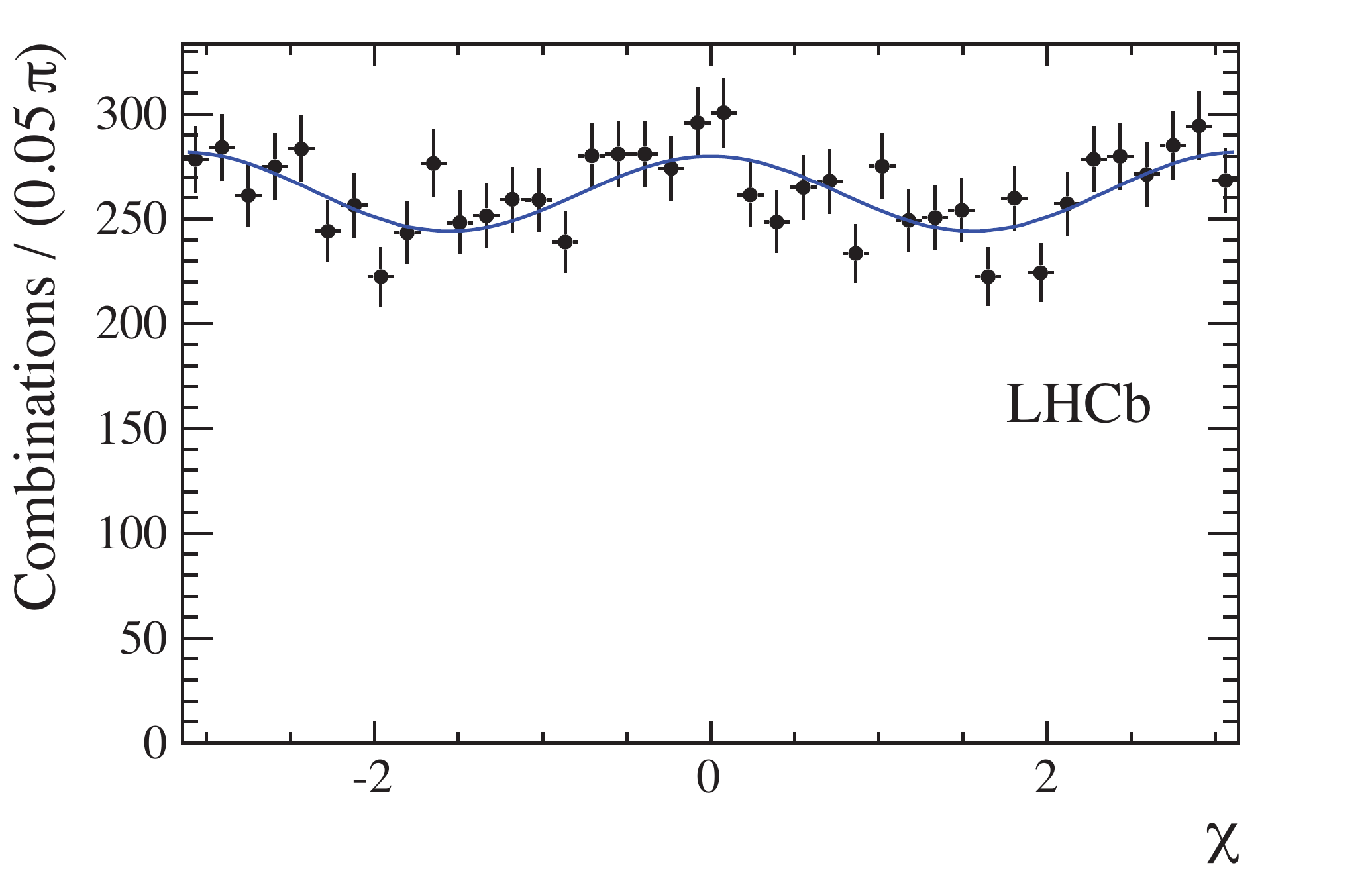}
\end{center}\label{bkg3}
\vskip -1cm
\caption{\small The $\chi$ distribution of the data-simulated hybrid background sample including the \Lb background and the fitted function $1+p_{b1}\cos \chi + p_{b2} \cos2 \chi$. The $p$-value of this fit is $40\%$.}
\label{fig:bkg_chi}
\end{figure}

Figure~\ref{bkg2} shows the projections of $\cos \theta_{\pi\pi}$ and $m(\pip\pim)$ from the like-sign data combinations added with all the additional simulated backgrounds. The other background includes the \Lb background and the combinatorial background which is described by the like-sign combinations. The fitted background parameters are given in Table \ref{tab:bkgparameter}.  The $\cos\theta_{J/\psi}$ background distribution is shown in Fig.~\ref{bkgcosjpsi}. Lastly, the $\chi$ background distribution, shown in Fig.~\ref{fig:bkg_chi} fit with the function $1+p_{b1}\cos \chi + p_{b2} \cos2 \chi$, determines the parameters $p_{b1}=-0.004\pm0.013$ and $p_{b2}=0.070\pm0.013$.


\begin{table}[h!t!p!]
\centering
\caption{\small Parameters for the background model.}
\vspace{0.2cm}
\begin{tabular}{lcrcl}
\hline
&$m_0$&$0.7473$\!\!\!\!\!&$\pm$&\!\!\!\!\!$0.0009$\gev\\
&$\Gamma_0$&$0.071$\!\!\!\!\!&$\pm$&\!\!\!\!\!$0.02$\gev\\
&$m_1$ &$~0.45$\!\!\!\!\!&$\pm$&\!\!\!\!\!$0.05$\gev\\
&$\Gamma_1$ &$~0.18$\!\!\!\!\!&$\pm$&\!\!\!\!\!$0.05$\gev\\
&$m_{\rho}$ &$~0.770$\!\!\!\!\!&$\pm$&\!\!\!\!\!$0.002$\gev\\
&$\Gamma_{\rho}$ &$~0.110$\!\!\!\!\!&$\pm$&\!\!\!\!\!$0.004$\gev\\
&$a$&$6.94$\!\!\!\!\!&$\pm$&\!\!\!\!\!$0.20$\\
&$b$&$0.76$\!\!\!\!\!&$\pm$&\!\!\!\!\!$0.04$\\
&$b_0$ &$0.0019$\!\!\!\!\!&$\pm$&\!\!\!\!\!$0.0004\,$GeV$^4$\\
&$b_1$ &$-0.536$\!\!\!\!\!&$\pm$&\!\!\!\!\!$0.053$ \\
&$b_2$ &$~0.100$\!\!\!\!\!&$\pm$&\!\!\!\!\!$0.043$ \\
&$b_3$ &$-0.100$\!\!\!\!\!&$\pm$&\!\!\!\!\!$0.042$\\
&$b_4$ &$~0.080$\!\!\!\!\!&$\pm$&\!\!\!\!\!$0.026$\\
&$b_5$ &$-0.051$\!\!\!\!\!&$\pm$&\!\!\!\!\!$0.025$\\
&$c_1$ &$-0.048$\!\!\!\!\!&$\pm$&\!\!\!\!\!$0.017$ \\
&$c_2$ &$-0.172$\!\!\!\!\!&$\pm$&\!\!\!\!\!$0.263$ \\
&$c_3$ &$-0.142$\!\!\!\!\!&$\pm$&\!\!\!\!\!$0.170$\\
&$c_4$ &$~0.855$\!\!\!\!\!&$\pm$&\!\!\!\!\!$0.259$\\
&$\alpha_0$&$0.45$\!\!\!\!\!&$\pm$&\!\!\!\!\!$0.04$\\
&$\alpha_1$&$0.30$\!\!\!\!\!&$\pm$&\!\!\!\!\!$0.03$\\\hline
\end{tabular}
\label{tab:bkgparameter}
\end{table}

%===============================================================================

\subsection{Resonance models}

%The $\KS$ is vetoed.
\begin{table}[htb]
\begin{center}
\caption{\small Possible resonance candidates in the $\Bdb\rightarrow J/\psi \pi^+\pi^-$ decay mode.}
\begin{tabular}{ccccllc}
\hline
Resonance & Spin & Helicity & Resonance   & Mass (MeV) & Width (MeV) & Source\\
& & & formalism & & &\\
\hline
$\rho(770)$ & 1 & $0,\pm 1$ & BW & $775.49\pm0.34$ &$149.1\pm0.8$&PDG \cite{PDG}\\
$f_0(500)$  & 0 & 0 & BW & ~~~~$513\pm32$ & ~~$335\,\pm67$ & CLEO \cite{Muramatsu:2002jp}\\
$f_2(1270)$ & 2 &  $0,\pm 1$ & BW & $1275.1\pm1.2$ & ~$185.1^{+2.9}_{-2.4}$&PDG \cite{PDG}\\
$\omega(782)$& 1 & $0,\pm 1$ & BW & $782.65\pm 0.12$ & ~\,$8.49\pm 0.08$ &PDG \cite{PDG} \\
$f_0(980)$ & 0 & 0 & Flatt\'e &~~~~~~~$-$ & ~~~~~$-$ & See text \\
$\rho(1450)$ & 1 & $0,\pm 1$ & BW & ~\,\,\,$1465\pm 25$ & ~\,\,\,$400\pm 60$ &PDG \cite{PDG} \\
$\rho(1700)$ & 1 & $0,\pm 1$ & BW & ~\,\,\,$1720\pm 20$ & ~\,\,\,$250\pm 100$ &PDG \cite{PDG}\\
$f_0(1500)$ & 0 & 0 & BW & ~\,\,\,$1461\pm3$ & ~\,\,\,$124\pm7$& LHCb \cite{Aaij:2014emv} \\
$f_0(1710)$ & 0 & 0 & BW &~\,\,\,$1720\pm 6$ & ~\,\,\,$135\pm 8$ &PDG \cite{PDG}\\
\hline
\end{tabular}\label{reso1}
\end{center}
\end{table}

To study the resonant structures of the decay $\Bdb\rightarrow J/\psi \pi^+\pi^-$ we use 29\,047 event candidates with invariant mass within $\pm20$ MeV of the $\Bdb$ mass peak which include $10\,207\pm178$ background candidates. The background yield is fixed in the fit.
Apart from non-resonant (NR) decays, the possible resonance candidates in the decay $\Bdb\rightarrow J/\psi \pi^+\pi^-$ are listed in Table \ref{reso1}. 
We use Breit-Wigner (BW) functions for most of the resonances except $f_0(980)$. The masses and widths of the BW resonances are listed in Table~\ref{reso1}. When used in the fit, they are fixed to these values except for the parameters of $f_0(500)$ which are allowed to vary by their uncertainties. For the $f_0(980)$ we use a Flatt\'e shape \cite{Flatte:1976xv}. 
Besides the mass, this shape has two additional parameters $g_{\pi\pi}$ and $g_{KK}$, which are fixed in the fit to the ones obtained from an amplitude analysis of $\Bsb \to J/\psi\pip\pim$ \cite{Aaij:2014emv}, where a large signal is evident. These parameters are $m_0=945.4\pm 2.2$ MeV, $g_{\pi\pi}=167\pm 7$ MeV and $g_{KK}/g_{\pi\pi}=3.47\pm 0.12$.  All background and efficiency parameters are fixed in the fit.

To determine the complex amplitudes in a specific model, the data are fitted maximizing the unbinned likelihood given as
 \begin{equation}
\mathcal{L}=\prod_{i=1}^{N}F(\m^i,\angpi^i,\theta^i_{J/\psi},\chi^i),
\end{equation}
where $N$ is the total number of candidates, and $F$ is the total PDF defined in Eq.~(\ref{eq:pdf}).

\section{Fit results}
\subsection{Final state composition}
\label{sec:Results1}

In order to compare the different models quantitatively, an estimate of the goodness of fit is calculated from 4D partitions of the fitting variables. To distinguish between models, we use the Poisson likelihood $\chi^2$ \cite{Baker:1983tu} defined as
\begin{equation}
\chi^2=2\sum_{i=1}^{N_{\rm bin}}\left[  x_i-n_i+n_i \text{ln}\left(\frac{n_i}{x_i}\right)\right],
\end{equation}
where $n_i$ is the number of events in the four-dimensional bin $i$ and $x_i$ is the expected number of events in that bin according to the fitted likelihood function. The $\chi^2/\text{ndf}$ and the negative of the logarithm of the likelihood, $\rm -ln\mathcal{L}$, of the fits are given in Table~\ref{RMchi2} for various fitting models, where ndf, the number of degrees of freedom, is equal to $N_{\rm bin}$  minus the number of fit parameters minus one. Here the five-resonance model (5R)  contains the resonances: $\rho(770)$, $f_0(500)$, $f_2(1270)$, $\rho(1450)$ and $\omega(782)$, the ``Best Model" adds a $\rho(1700)$ resonance to the 5R model, the 7R model adds a $f_0(980)$ resonance to the Best Model, and the 7R+NR model adds a non-resonant component. We also give the change of $\rm ln\mathcal{L}$ for various fits with respect to the 5R model in Table~\ref{RMchi2}.

\begin{figure}[!t]
  \begin{center}
  \includegraphics[width=0.48\textwidth]{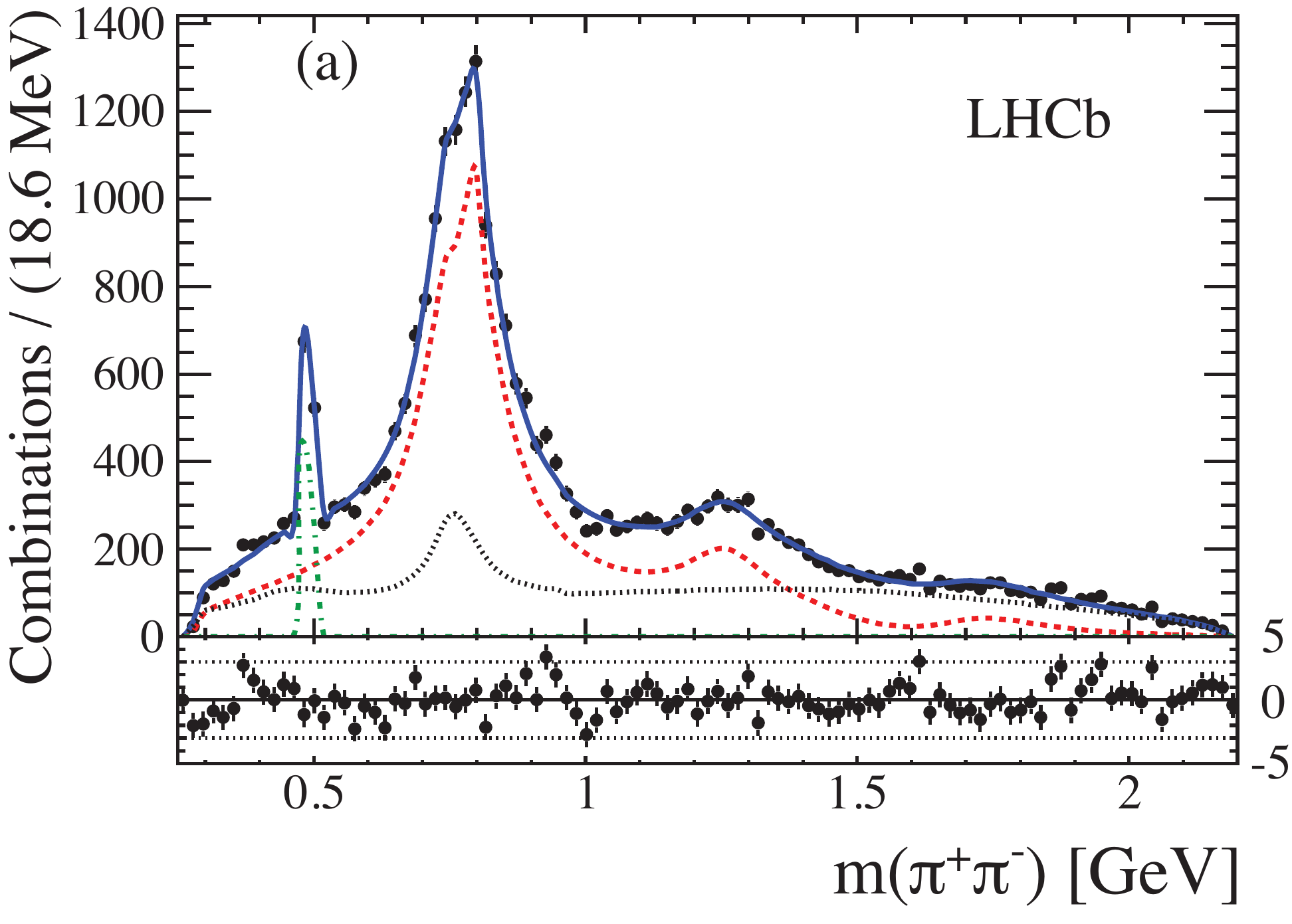}
    \includegraphics[width=0.48\textwidth]{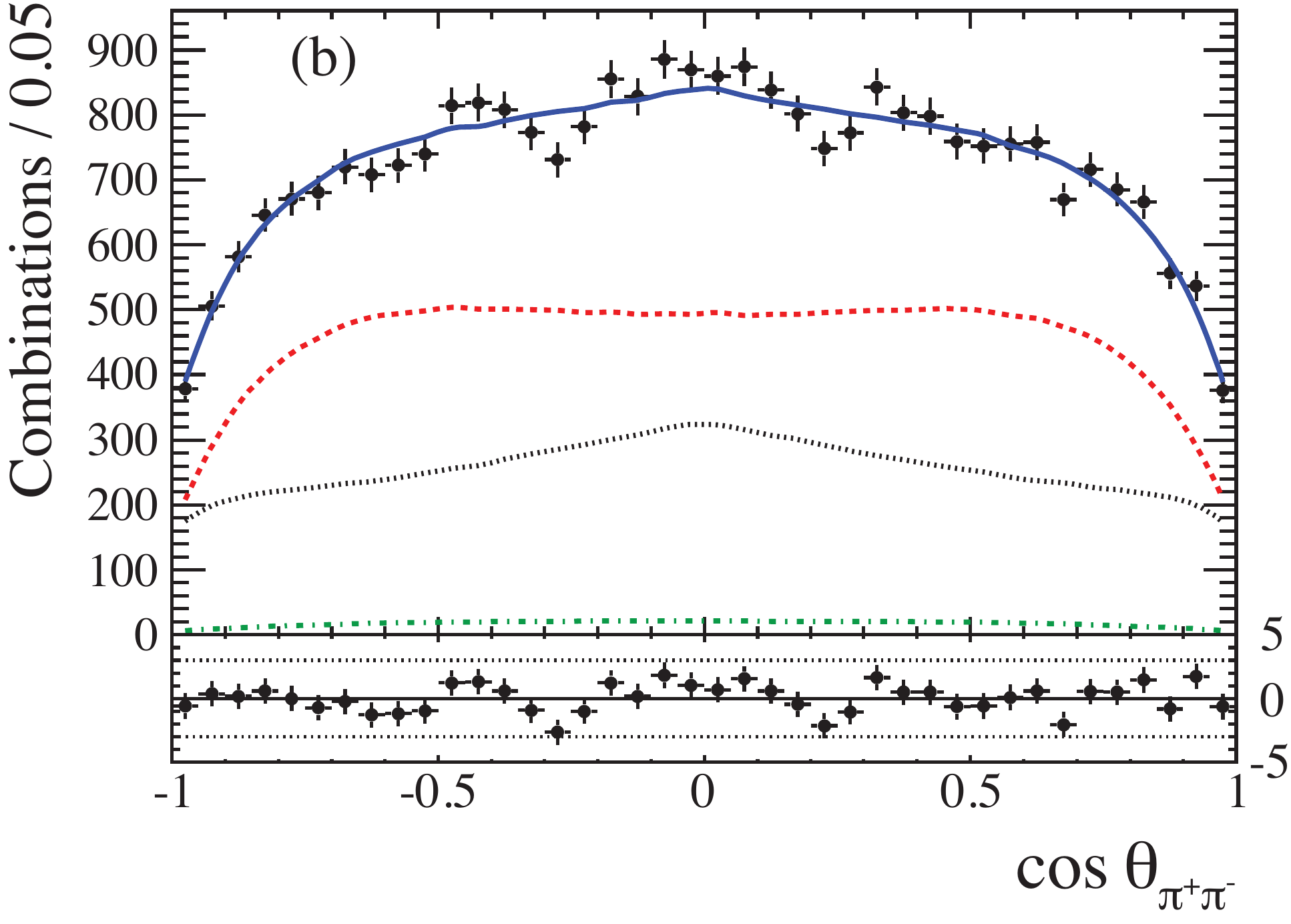}
  \includegraphics[width=0.48\textwidth]{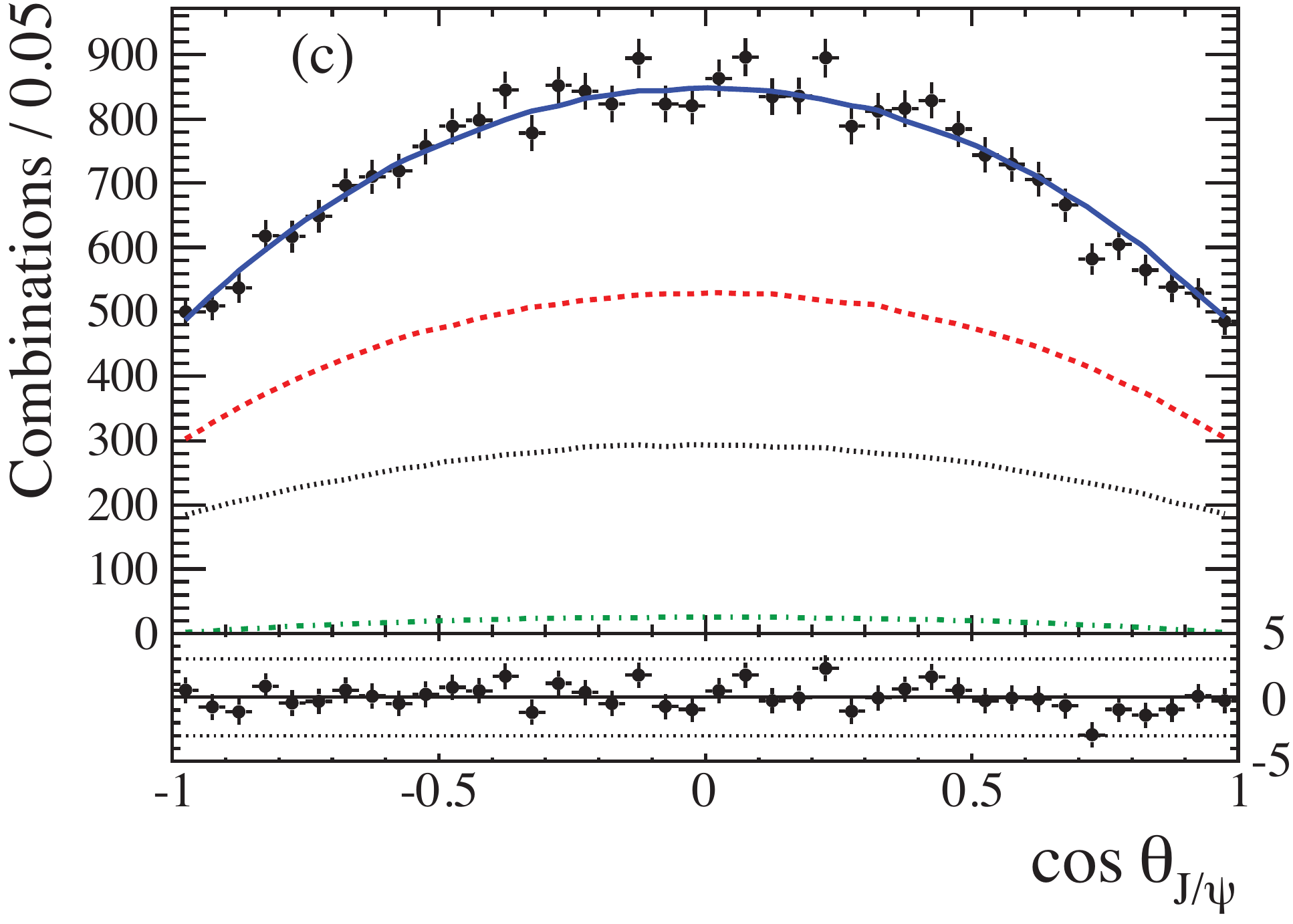}
 \includegraphics[width=0.48\textwidth]{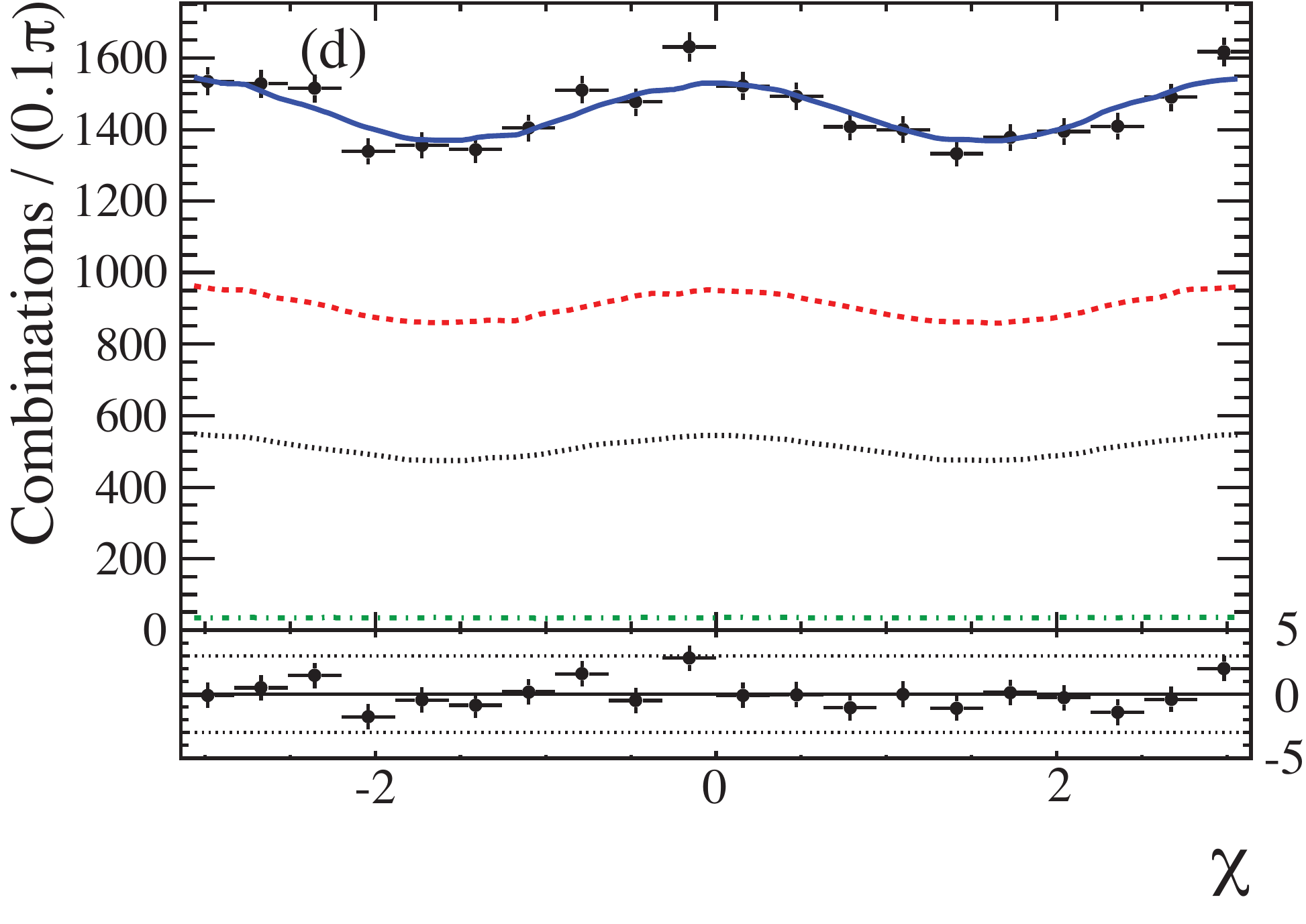}
\vspace{-2mm}
    \caption{\small Dalitz fit projections of (a) $m(\pi^+\pi^-)$, (b) $\cos(\theta_{\pip\pim})$, (c) $\cos \theta_{J/\psi}$ and (d) $\chi$ for the 5R Model + $\rho(1700)$ (Best Model). The points with error bars are data compared with the overall fit, shown by the (blue) solid line. The individual fit components are signal, shown with a (red) dashed line, background, shown with a (black) dotted line, and  $\KS$, shown with a (green) dashed line.
    }  \label{5Rrho}
  \end{center}
\end{figure}

The 7R model gives a slightly better likelihood compared to the Best Model, however, the decrease of the $\rm -ln\mathcal{L}$ due to adding $f_0(980)$ is less than the expected $\Delta \rm ln\mathcal{L}$ at 3$\sigma$ significance.
Thus, we use the Best Model, which maintains a significance larger than 3$\sigma$ for each resonance component, as our baseline fit, while the 7R model is only used to establish an upper limit on the presence of the $f_0(980)$.
The Dalitz fit projections on the four observables: $m(\pi^+\pi^-)$, $\cos(\theta_{\pip\pim})$, $\cos \theta_{J/\psi}$ and $\chi$ are shown in Fig.~\ref{5Rrho} for the Best Model. 

\begin{table}[b]
\begin{center}
\caption{\small The $\chi^2/\text{ndf}$ and the $\rm -ln\mathcal{L}$ of different resonance models. The decrease of $\rm -ln\mathcal{L}$ is with respect to the 5R model. }
\begin{tabular}{lccc}
\hline
Resonance model & $\rm -ln\mathcal{L}$& $\chi^2/\text{ndf}$ & Decrease of $\rm -ln\mathcal{L}$ \\\hline
5R Model & $-$169271  &   2396/2041  & \\
5R Model + $\rho(1700)$ (Best Model)  &  $-$169327  &   2293/2035 & $56$  \\
Best Model + $f_0(980)$ (7R Model)  &$-$169329  &2295/2033 & $58$ \\
7R + $f_0(1500)$ &  $-$169333 &  2293/2031 & $ 60$  \\
7R + $f_0(1710)$ &   $-$169329 & 2295/2031 & $56$ \\
7R + NR &  $-$169342 & 2292/2031 & $69$ \\
\hline
\end{tabular}
\label{RMchi2}
\end{center}
\end{table}

Table \ref{tab:ff1} shows the summary of fit fractions of different components for various models.  The fit fractions of the interference terms in the Best Model are computed using Eq.~(\ref{eq:inter}) and listed in Table \ref{tab:best fit fractions}. Table~\ref{tab:phase} shows the resonant phases from the Best Model.
In the Best Model the \CP-even components sum to (56.0$\pm$1.4)\%, including the interference terms, so that the \CP-odd fraction is (44.0$\pm$1.4)\%. The structure near the peak of the $\rho(770)$ is due to $\rho-\omega$ interference. The fit fraction ratio is found to be
\begin{equation}
 \frac{\Gamma(\Bzb\to \jpsi \omega(782),~\omega\to\pi^+\pi^-)}{\Gamma(\Bzb\to \jpsi \rho(770),~\rho\to\pi^+\pi^-)}= (1.07_{-0.22-0.22}^{+0.32+0.29})\times 10^{-2},\nonumber
\end{equation}
where the uncertainties are statistical and systematic, respectively; wherever two uncertainties are quoted in this paper, they will be of this form. The systematic uncertainties will be discussed in detail in Sec.~\ref{sec:syst}.

 The 7R model fit gives the ratio of observed decays into $\pi^+\pi^-$  for $f_0(980)/f_0(500)$ equal to $(0.6 _{-0.4-2.6}^{+0.7+3.3})\times 10^{-2}$.  To determine the statistical uncertainty, the full error matrix and parameter values from the fit are used to generate 500 data-size sample parameter sets. For each set, the fit fractions are calculated. The distributions of the obtained fit fractions are described by bifurcated Gaussian
functions. The widths of the Gaussians are taken as the statistical errors on the corresponding parameters.  We will discuss the implications of this measurement in Sec.~\ref{sec:mixing angle}.

\begin{sidewaystable}
%\scriptsize
\begin{center}
\caption{\small Fit fractions (\%) of contributing components for the various models. The uncertainties are statistical only. Sums can differ from 100\% due to interference (see Table~\ref{tab:best fit fractions}).}
\begin{tabular}{lccccccccc}
\hline
             	&	5R                 	&	 Best Model            	&	 7R 	&	 7R+$f_0(1500)$ 	&	 7R+$f_0(1710)$ 	&	 7R+NR  \\\hline
$\rho(770)_0$  	&	$35.5\pm1.6$    	&	 $36.2\pm1.8$   	&	 $36.1\pm1.8$ 	&	 $36.1\pm1.9$ 	&	 $36.1\pm1.8$	&	$36.0\pm1.9$\\
$\rho(770)_{\parallel}$  	&	$13.4\pm1.1$    	&	 $14.7\pm1.2$   	&	  $14.8\pm1.2$ 	&	  $14.7\pm1.2$ 	&	$14.8\pm1.2$	&	$14.9\pm1.1$\\
$\rho(770)_{\perp}$  	&	$11.7\pm0.9$     	&	  $12.1\pm1.1$  	&	 $11.9\pm1.1$  	&	 $12.0\pm1.1$ 	&	$12.0\pm1.1$	&	$15.0\pm1.3$ \\
$f_0(500)$   	&	$24.9\pm1.4$    	&	  $22.2\pm1.2$    	&	 $21.4\pm1.7$ 	&	  $20.8\pm1.9$ 	&	$21.1\pm1.8$	&	$18.7\pm3.1$ \\
$f_2(1270)_0$  	&	$4.6\pm0.4$     	&	 $4.7\pm0.4$     	&	  $5.0\pm0.4$ 	&	 $4.8	\pm0.4$ 	&	$4.9\pm0.4$	&	$4.5\pm0.4$ \\
$f_2(1270)_{\parallel}$  	&	$1.0\pm0.4$     	&	  $0.9	\pm0.4$     	&	  $1.0\pm0.5$ 	&	 $1.0\pm0.5$ 	&	$1.0\pm0.4$	&	$0.8\pm0.4$ \\
$f_2(1270)_{\perp}$  	&	$2.1\pm0.4$    	&	    $2.0\pm0.4$    	&	  $2.0\pm0.4$ 	&	 $1.9\pm0.4$	&	$2.0\pm0.5$	&	$2.2\pm0.4$\\
$\omega(782)_0$ 	&	 $0.29\pm0.11$  	&	 $0.26\pm0.10$  	&	 $0.26\pm0.11$ 	&	 $0.26\pm0.11$	&	$0.26\pm0.11$	&	$0.25\pm0.11$ \\
$\omega(782)_{\parallel}$ 	&	 $0.41\pm0.15$  	&	 $0.41\pm0.16$  	&	 $0.41\pm0.16$ 	&	 $0.42\pm0.16$	&	$0.41\pm0.15$	&	$0.39\pm0.15$ \\
$\omega(782)_{\perp}$ 	&	 $0.01_{-0.01}^{+0.06}$  	&	 $0.01_{-0.01}^{+0.06}$  	&	 $0.01_{-0.01}^{+0.05}$ 	&	 $0.01_{-0.01}^{+0.05}$	&	$0.01_{-0.01}^{+0.05}$	&	$0.01_{-0.01}^{+0.05}$\\
$f_0(980)$   	&	--	&	 --  	&	 $0.13\pm0.11$ 	&	 $0.16\pm0.12$	&	$0.14\pm0.11$	&	$0.5\pm0.3$\\
$\rho(1450)_0$ 	&	$2.5\pm0.6$     	&	  $6.8\pm2.0$  	&	 $6.2\pm2.4$ 	&	 $5.3\pm3.5$	&	$6.3\pm2.3$	&	$5.0\pm1.9$\\
$\rho(1450)_{\parallel}$ 	&	$1.8\pm0.8$    	&	 $3.1\pm1.9$   	&	 $3.2\pm1.9$ 	&	 $2.4	\pm0.8$	&	$3.4\pm2.1$	&	$	2.7\pm1.7$\\
$\rho(1450)_{\perp}$ 	&	$1.6\pm0.4$    	&	   $1.7\pm0.7$ 	&	  $1.8\pm0.7$ 	&	 $1.5\pm0.7$	&	$1.9\pm0.8$	&	$	5.8\pm2.6$\\
$f_0(1500)$  	&	--                     	&	 --              	&	-- 	&	 $0.33_{-0.18}^{+0.31}$ 	&	 -- 	&	 --\\
$f_0(1710)$  	&	--                       	&	 --            	&	--  	&	 -- 	&	$0.01_{-0.01}^{+0.12}$ 	&	 -- \\
$\rho(1700)_0$	&	--             	&	  $2.0\pm0.9$           	&	 $1.9\pm1.0$   	&	$1.4_{-0.8}^{+1.8}$	&	$2.0\pm1.0$ 	&	$1.1\pm0.7$                          \\
$\rho(1700)_{\parallel}$	&	--             	&	   $1.2_{-0.6}^{+1.2}$          	&	 $1.3_{-0.6}^{+1.1}$      	&	$1.3_{-0.7}^{+1.3}$	&	$1.3\pm1.0$ 	&	$1.0\pm0.9$\\
$\rho(1700)_{\perp}$	&	--             	&	   $1.8\pm0.7$         	&	 $1.7\pm0.6$           	&	$1.7\pm0.6$	&	$1.8\pm0.7$    	&	$3.5\pm1.2$              \\
NR           	&	--                     	&	 --                  	&	 -- 	&	 -- 	&	 -- 	&	$3.2\pm1.1$\\\hline
Sum          	&	99.8	&	110.2	&	108.8	&	105.9	&	109.3	&	 115.5\\

\hline
\end{tabular}\label{tab:ff1}
\end{center}
\end{sidewaystable}

\begin{table}[htb]
\centering
\caption{\small Non-zero interference fractions($\%$) obtained from the fit using the Best Model. The uncertainties are statistical only.}
\label{tab:best fit fractions}
\begin{tabular}{lrcl}
\hline
\!\!\!\!\!\!\!Interfering components & \multicolumn{3}{c}{\!\!\!Intererence fraction (\%) }\\
\hline
$\rho(770)	_0 + \omega(782)_0$	&	$	-0.36	$\!\!\!\!\!&$\pm$&\!\!\!\!\!$0.55$	\\
$\rho(770)	_{\parallel} + \omega(782)	_{\parallel}$	&	$	0.65$\!\!\!\!\!&$\pm$&\!\!\!\!\!$0.43	$	\\
$\rho(770)	_{\perp}\! + \omega(782)	_{\perp}$	&	$	-0.21$\!\!\!\!\!&$\pm$&\!\!\!\!\!$0.37$	\\
$\rho(770)	_0 + \rho(1450)	_0 $	&	$	-3.34	$\!\!\!\!\!&$\pm$&\!\!\!\!\!$2.60$	\\
$\rho(770)	_{\parallel} + \rho(1450)	_{\parallel} $	&	$	-4.38$\!\!\!\!\!&$\pm$&\!\!\!\!\!$1.64	$	\\
$\rho(770)	_{\perp}\! + \rho(1450)	_{\perp} $	&	$	-0.18$\!\!\!\!\!&$\pm$&\!\!\!\!\!$1.21 $	\\
$\rho(770)	_0 + \rho(1700)	_0 $	&	$	3.34$\!\!\!\!\!&$\pm$&\!\!\!\!\!$0.93	$	\\
$\rho(770)	_{\parallel} + \rho(1700)	_{\parallel} $	&	$	0.63$\!\!\!\!\!&$\pm$&\!\!\!\!\!$0.88 $	\\
$\rho(770)	_{\perp}\! + \rho(1700)	_{\perp}$	&	$	2.10$\!\!\!\!\!&$\pm$&\!\!\!\!\!$0.43 $	\\
$\omega(782)_0 + \rho(1450)	_0 $	&	$	-0.24$\!\!\!\!\!&$\pm$&\!\!\!\!\!$0.06$	\\
$\omega(782)	_{\parallel} + \rho(1450)	_{\parallel} $	&	$	-0.17$\!\!\!\!\!&$\pm$&\!\!\!\!\!$0.06	$	\\
$\omega(782)	_{\perp}\! + \rho(1450)	_{\perp} $	&	$	-0.02$\!\!\!\!\!&$\pm$&\!\!\!\!\!$0.03$	\\
$\omega(782)_0 + \rho(1700)	_0 $	&	$	0.05$\!\!\!\!\!&$\pm$&\!\!\!\!\!$0.03$	\\
$\omega(782)	_{\parallel} + \rho(1700)	_{\parallel} $	&	$	-0.05$\!\!\!\!\!&$\pm$&\!\!\!\!\!$0.03$	\\
$\omega(782)	_{\perp}\! + \rho(1700)	_{\perp} $	&	$	-0.01$\!\!\!\!\!&$\pm$&\!\!\!\!\!$0.02$	\\
$\rho(1450)	_0+ \rho(1700)	_0 $	&	$	-5.57$\!\!\!\!\!&$\pm$&\!\!\!\!\!$1.98	$	\\
$\rho(1450)	_{\parallel}+ \rho(1700)	_{\parallel} $	&	\multicolumn{3}{c}{$\!\!\!\!\!\!\!\!\!\!\!\!\!\!\!\!\!\!\!\!\!\!\!\!\!\!\!\!-1.31_{-2.89}^{+1.10}$}	\\
$\rho(1450)	_{\perp}\!+ \rho(1700)	_{\perp} $	&	$	-1.09$\!\!\!\!\!&$\pm$&\!\!\!\!\!$1.02$	\\

\hline

\end{tabular}\label{ff3}
\end{table}

\begin{table}[h!t!p!]
\begin{center}
\caption{\small The fitted resonant phases from the Best Model. The uncertainties are statistical only.}%and the parameters for $f_0(500)$ }
\begin{tabular}{lc}
\hline
Components & phase ($^\circ$) \\
\hline
$\rho(770)_{0}$ & 0 (fixed)\\
$\rho(770)_{\perp}$ & 0 (fixed)\\
$\rho(770)_{\parallel}$ & $189.8\pm~7.3$ \\
$f_0(500)$ & $336.9\pm~5.0$ \\
$f_2(1270)_{0}$ & $210.1\pm~6.9$ \\
$f_2(1270)_{\perp}$ & $165.0\pm13.3$ \\
$f_2(1270)_{\parallel}$ & $334.4\pm21.9$ \\
$\omega(782)_{0}$ & $268.8\pm11.9$ \\
$\omega(782)_{\perp}$ & $227.4\pm84.9$ \\
$\omega(782)_{\parallel}$ & $123.5\pm13.7$ \\

$\rho(1450)_{0}$ & $196.7 \pm12.1$ \\
$\rho(1450)_{\perp}$ & $182.6 \pm 22.4$ \\
$\rho(1450)_{\parallel}$ & ~\,$74.9 \pm 12.6$ \\

$\rho(1700)_{0}$ & ~\,$71.1 \pm 19.9$ \\
$\rho(1700)_{\perp}$ & $113.4 \pm 20.3$ \\
$\rho(1700)_{\parallel}$ & $~~~3.4 \pm 24.5$ \\
\hline
\end{tabular}\label{tab:phase}
\end{center}
\end{table}

\begin{figure}[b]
%\vskip -1cm
\begin{center}
    \includegraphics[width=0.82\textwidth]{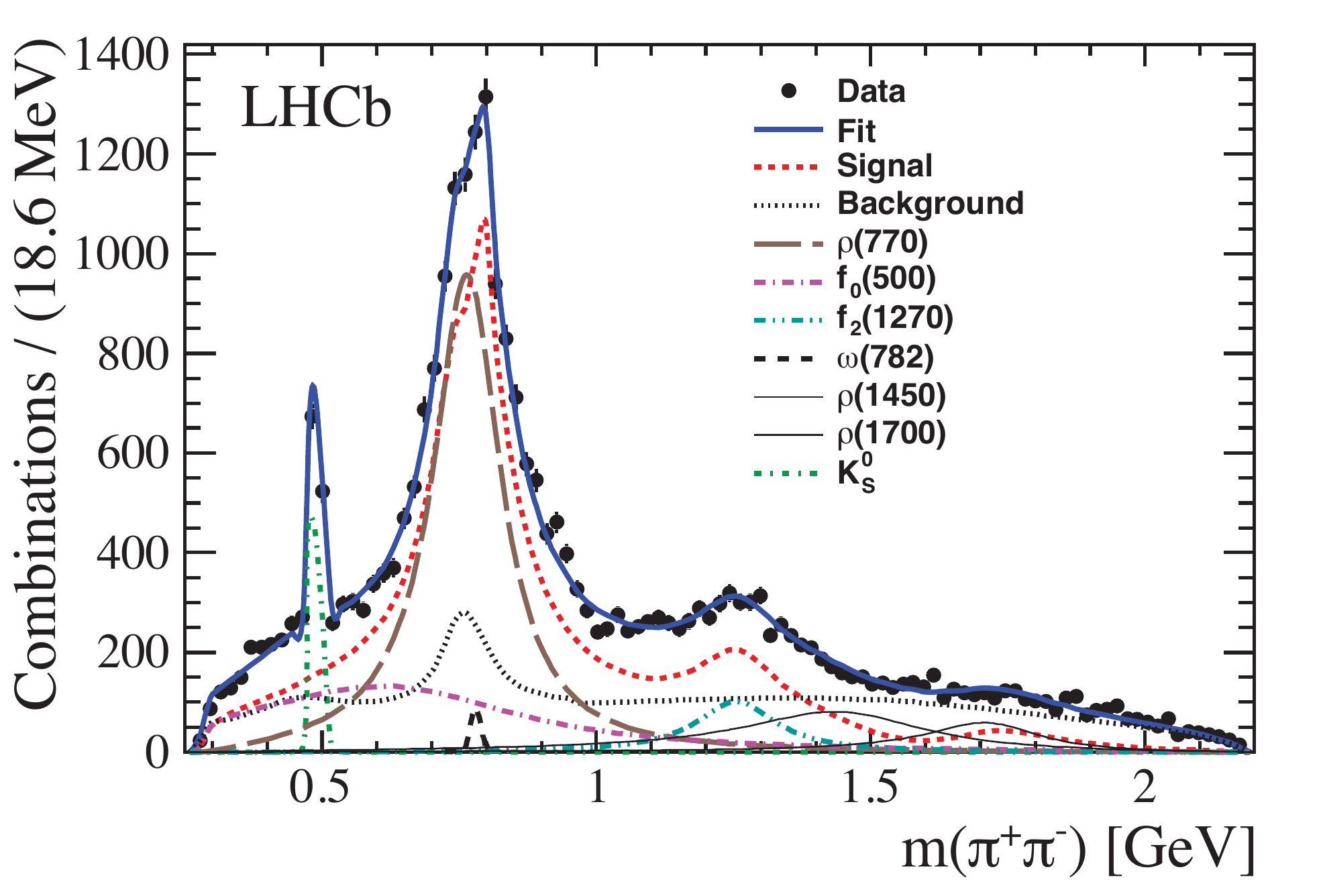}%
\end{center}
\vskip -0.6cm
\caption{\small Fit projection of $m(\pi^+\pi^-$) showing the different resonant contributions in the Best Model.}
\label{B0-mpp}
\end{figure}

In Fig.~\ref{B0-mpp} we show the fit fractions of the different resonant components in the Best Model. 
Table~\ref{tab:ff} lists the fit fractions and the transversity fractions of each contributing 
resonance. For a $P$- or $D$-wave resonance, we report its total fit fraction by summing all the three components.
\begin{table}[t]
\centering
\caption{\small Fit fractions and transversity fractions of contributing resonances in the Best Model. The first uncertainty is
statistical and the second the total systematic.}
\def\arraystretch{1.2}
\begin{tabular}{lcccc}\hline
 & & \multicolumn{3}{c}{Transversity fractions (\%)}\\
\raisebox{2.0ex}[0pt]{Component} & \raisebox{2.0ex}[0pt]{Fit fraction (\%)}& $\tau=0$ & $\tau=\|$ & $\tau=\perp$ \\\hline
$\rho(770)$ & $63.1\pm2.2_{-2.2}^{+3.4}$ & $57.4\pm2.0_{-3.1}^{+1.3}$& $23.4\pm1.7_{-1.3}^{+1.0}$& $19.2\pm1.7_{-1.2}^{+3.8}$\\
$f_0(500)$ & $22.2\pm1.2_{-3.5}^{+2.6}$ & 1 & 0 & 0\\
$f_2(1270)$ &~$7.5\pm0.6_{-0.6}^{+0.4}$&$62\pm4_{-4}^{+2}$&~$11\pm5\pm2$&\!$26\pm5_{-2}^{+4}$\\
$\omega(782)$&$0.68_{-0.14-0.13}^{+0.20+0.17}$& $39_{-13-3}^{+15+4}$&$60_{-15-4}^{+12+3}$&$1_{-1}^{+9}\pm1$\\
$\rho(1450)$&~~~$11.6\pm2.8\pm4.7$& ~~~$58\pm10_{-23}^{+14}$&~~$27\pm13_{-11}^{+7}$&~$15\pm7^{+28}_{-10}$\\
$\rho(1700)$&~~~~$5.1\pm1.2\pm3.0$&~~~$40\pm11_{-23}^{+13}$&~~$24\pm14_{-10}^{+7}$&~~~$36\pm14_{-9}^{+28}$\\\hline
\end{tabular}\label{tab:ff}
\def\arraystretch{1.0}
\end{table}

Table~\ref{tab:br} shows the branching fractions of the resonant modes calculated by multiplying the fit fraction listed in Table~\ref{tab:ff} with ${\cal B}(\Bzb\to\jpsi \pi^+\pi^-)=(3.97\pm0.09\pm0.11\pm0.16)\times10^{-5}$, obtained from our previous study~\cite{Aaij:2013zpt}, where the uncertainties are statistical, systematic, and due to normalization, respectively.  These branching fractions are proportional to the squares of the individual resonant amplitudes.

\begin{table}[t]
\centering \caption{\small Branching fractions for each channel. The first uncertainty is
statistical and the second the total systematic.}
\def\arraystretch{1.2}
\begin{tabular}{lrl}
\hline
$R$ &\multicolumn{2}{c}{${\cal B}(\Bzb \to \jpsi R, R \to \pi^+\pi^-)$}  \\
\hline
$\rho(770)$& \,\,\,\,$(2.50\pm0.10_{-0.15}^{+0.18})$&\!\!\!\!\!\!$\times10^{-5}$\\
$f_0(500)$&$(8.8\pm0.5^{+1.1}_{-1.5})$&\!\!\!\!\!\!$\times10^{-6}$\\
%$f_0(980)$& $(6.1^{+3.1+1.7}_{-2.0-1.4})\times10^{-7}$\\
$f_2(1270)$& $(3.0\pm0.3_{-0.3}^{+0.2})$&\!\!\!\!\!\!$\times10^{-6}$\\
$\omega(782)$& $(2.7_{-0.6-0.5}^{+0.8+0.7})$&\!\!\!\!\!\!$\times10^{-7}$\\
$\rho(1450)$&$(4.6\pm1.1\pm1.9)$&\!\!\!\!\!\!$\times10^{-6}$\\
$\rho(1700)$&$(2.0\pm0.5\pm1.2)$&\!\!\!\!\!\!$\times10^{-6}$\\
\hline
\end{tabular}\label{tab:br}
\def\arraystretch{1.0}
\end{table}

\subsection{Angular moments}

\begin{figure}[!htb]
\vskip -1cm
\begin{center}
    \includegraphics[width=0.9\textwidth]{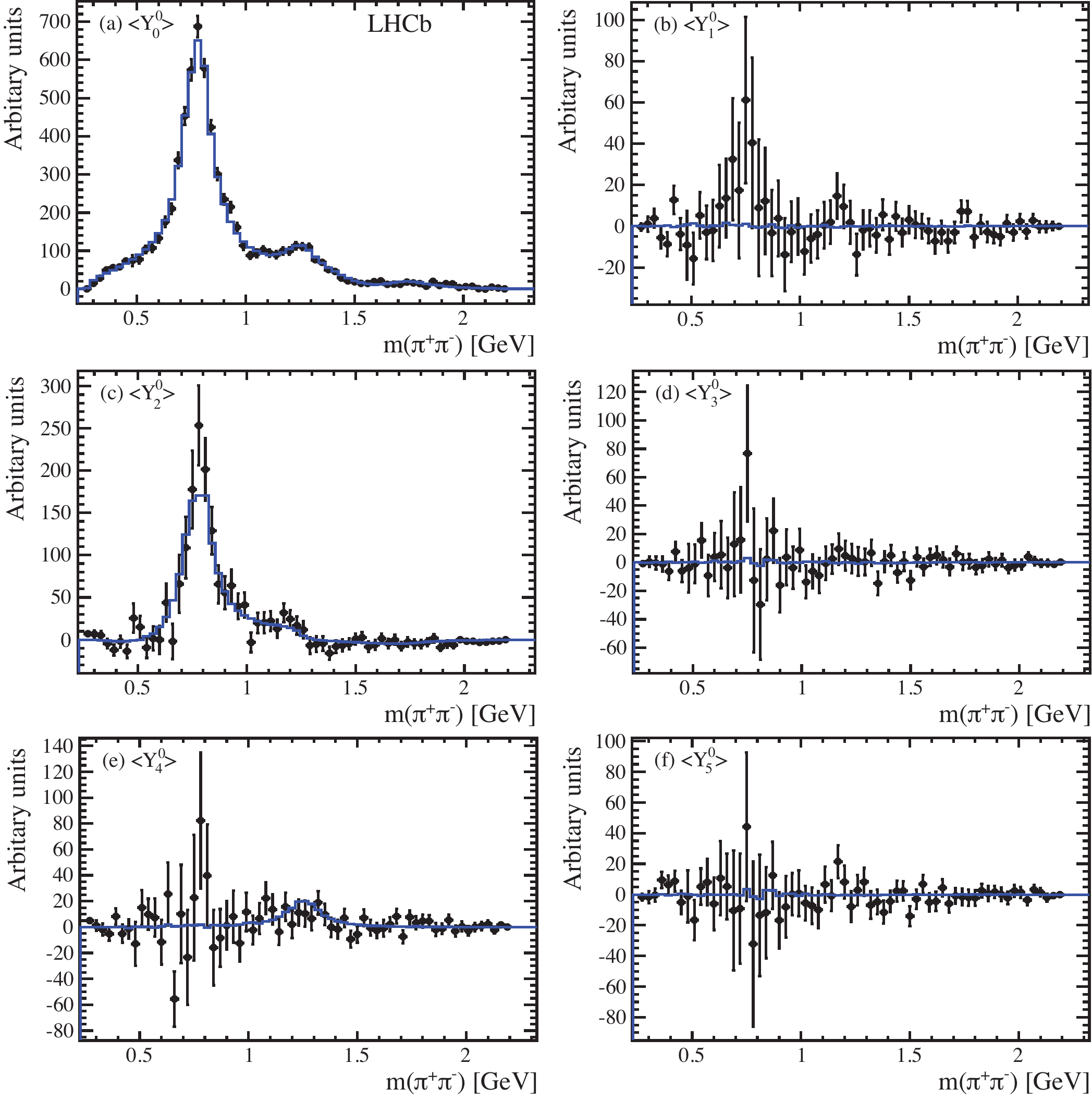}%
\end{center}
\vskip -0.5cm
\caption{\small The $\pi^+\pi^-$ mass dependence of the spherical harmonic moments of $\cos \theta_{\pi\pi}$ after efficiency corrections and background subtraction:
(a) $\langle Y^0_0\rangle$,  (b) $\langle Y^0_1\rangle$, (c) $\langle Y^0_2\rangle$, (d) $\langle Y^0_3\rangle$, (e) $\langle Y^0_4\rangle$, (f) $\langle Y^0_5\rangle$.
The errors on the black data points are statistical. The (blue) curves show the fit projections.}
\label{SPH2}
\end{figure}

Angular moments are defined as an average of the spherical harmonics, $\langle Y^0_{l}(\cos \theta_{\pi\pi})\rangle$, in each efficiency-corrected and background-subtracted $\pi^+\pi^-$ invariant mass interval. 
The moment distributions provide an additional way of visualizing the effects of different resonances and their interferences, similar to a partial wave analysis.
Figure~\ref{SPH2} shows the distributions of the angular moments for the Best Model. In general the interpretation of these moments is that $\langle Y^0_0\rangle$ is the efficiency corrected and background subtracted event distribution, $\langle Y^0_1\rangle$  the sum of the interference between S-wave and P-wave and between P-wave and D-wave amplitudes, $\langle Y^0_2\rangle$  the sum of the  P-wave, D-wave and the interference of S-wave and D-wave amplitudes, $\langle Y^0_3\rangle$  the interference between P-wave and D-wave, $\langle Y^0_4\rangle$ the D-wave, and $\langle Y^0_5\rangle$ results from an F-wave \cite{LHCb:2012ae,delAmoSanchez:2010yp}. For the moments with odd-$l$, one will always find that $\Bzb$ and $\Bz$ have opposite sign; thus the sum of their contributions is expected to be small.

\subsection{Systematic uncertainties}\label{sec:syst}

The sources of  systematic uncertainties on the results of the amplitude analysis are summarized in Table \ref{sys:dlz}. Uncertainties due to particle identification and tracking are taken from Ref.~\cite{Aaij:2013zpt} and are taken into account in the branching fraction results, but do not appear in the fit fractions as they are independent of pion kinematics. For the uncertainties due to the acceptance or background modeling, we repeat the data fit 100 times where the parameters of acceptance or background modeling are varied according to the corresponding error matrix. For the acceptance function, the error matrix is obtained by fitting the simulated acceptance as described in Sec.~\ref{sec:mc}. For the background function, the error matrix is obtained by fitting the hybrid data-simulated sample as described in Sec.~\ref{sec:background modeling}.

There is uncertainty on the fractions of sources in the hybrid MC-data sample for background modeling. Instead of using the fits to the $\pi^+\pi^-$ mass distribution to determine the background fractions, we use the fractions found from the $\jpsi\pi^+\pi^-$ mass fit
shown in Fig.~\ref{fitmass} that finds the $\Lb$ reflection is $9.6\%$, the $\Bzb$ reflection is $4.2\%$, the $\Bsb$ background is $11.5\%$ and the combinatorial part is $74.7\%$,  instead of the ones found in Sec.~\ref{sec:background modeling}. We then fit the new hybrid sample to get the background parameters. The data fit is repeated with the new background parameters; the changes on the fit results are added in quadrature with the uncertainties of the background modeling discussed above. The two background uncertainties have similar sizes.

We neglect the mass resolution in the fit where the typical resolution is 3\mev. A previous study shows that the resolution effects are negligible except for the $\omega(782)$ resonance
whose total fit fraction is underestimated by $(0.09\pm0.08)\%$. We take the quadrature of  0.09\% and 0.08\%, equal to 0.12\%, as the systematic uncertainty of the total fit fraction of the $\omega(782)$. These uncertainties are included in the ``Acceptance" category.

The uncertainties due to the fit model include adding each resonance that is listed in Table \ref{reso1} but not used in the 7R model, varying 
%hadron scale $r$'s for $B$ meson and $R$ resonance to both $3.0\gev^{-1}$,
the centrifugal barrier factors defined in Eq.~(\ref{eq:DP}) substantially, 
replacing $f_0(500)$ model by a Bugg function \cite{Bugg:2006gc} and using the alternative Gounaris and Sakurai Model \cite{GS} for the various $\rho$ mesons. The largest variation among those changes is assigned as the systematic uncertainty for modeling. We also find that increasing the default angular momentum $L_{B}$ for the P and D-wave cases gives negligible differences.

Finally, we repeat the amplitude fit by varying the mass and width of all the resonances except for the $f_0(980)$, in the 7R model within their errors one at a time, and add the changes in quadrature. For the $f_0(980)$ resonance, we change the resonance parameters $m_0$, $g_{\pi\pi}$ and $g_{KK}/g_{\pi\pi}$ to the values obtained from Solution II in \cite{Aaij:2014emv} instead of using the ones obtained from Solution I.

\begin{table}[t!]
\centering
\caption{\small Absolute systematic uncertainties on the results of the amplitude analysis estimated using the Best Model except for the
$f_0(980)$ where we use the 7R model.}
\def\arraystretch{1.25}
\begin{tabular}{lccccc}
\hline
Item& Acceptance& Background& Fit model & Resonance & Total \\
         &                      & model          &                   &parameters &\\\hline
\multicolumn{6}{c}{Fit fractions (\%)}\\\hline
$\rho(770)$	&	$	\pm	0.3	$	&	$	\pm	0.6	$	&	$_{	-1.8	}^{+	3.2	}$	&	$	\pm	1.1	$	&	$_{	-2.2	}^{+	3.4	}$	\\
%$f_0(500)$	&	$	\pm	0.3	$	&	$	\pm	0.7	$	&	$_{	-2.7	}^{+	1.2	}$	&	$	\pm	0.3	$	&	$_{	-2.8	}^{+	1.5	}$	\\
$f_0(500)	$	&	$	\pm	0.3	$	&	$	\pm	0.7	$	&	$^{+	1.2	}_{-	2.7	}$	&	$	\pm	2.2	$	&	$^{+	2.6	}_{	-3.5	}$	\\
$f_2(1270)$	&	$	\pm	0.1	$	&	$	\pm	0.2	$	&	$_{	-0.5	}^{+	0.1	}$	&	$	\pm	0.3	$	&	$_{	-0.6	}^{+	0.4	}$	\\
$\omega(782)$	&	$	\pm	0.12	$	&	$	\pm	0.02	$	&	$_{	-0.03	}^{+	0.11	}$	&	$	\pm	0.03$	&	$_{	-0.13	}^{+	0.17	}$	\\
$f_0(980)$	&	$	\pm	0.01	$	&	$_{-0.02}	^{+0.03}		$	&	$_{	-0.04	}^{+	0.37	}$	&	$	\pm	0.03$	&	$_{	-0.05	}^{+	0.37	}$	\\
$\rho(1450)$	&	$	\pm	0.15	$	&	$	\pm	1.3	$	&	$_{	-1.9	}^{+	2.3	}$	&	$	\pm	4.0	$	&	$	\pm		4.7	$	\\
$\rho(1700)$	&	$	\pm	0.13	$	&	$	\pm	0.7	$	&	$_{	-0.9	}^{+	0.7	}$	&	$	\pm	2.9	$	&	$	\pm		3.0	$	\\\hline
\multicolumn{6}{c}{Transversity $0$ fractions (\%)}\\\hline
$\rho(770)$	&	$	\pm	0.5	$	&	$	\pm	0.5	$	&	$_{	-3.0	}^{+	1.0	}$	&	$	\pm	0.5	$	&	$_{	-3.1	}^{+	1.3	}$	\\
																												
$f_2(1270)$	&	$	\pm	0.5	$	&	$	\pm	1.7	$	&	$_{	-2.9	}^{+	0.8	}$	&	$	\pm	1.0	$	&	$_{	-4	}^{+	2	}$	\\
$\omega(782)$	&	$	\pm	0.4	$	&	$	\pm	2.1	$	&	$_{	-0.6	}^{+	3.5	}$	&	$	\pm	1.5	$	&	$_{	-3	}^{+	4	}$	\\
																												
$\rho(1450)$	&	$	\pm	0.7	$	&	$	\pm	8.2	$	&	$_{	-18.4	}^{+	~2.0	}$	&	$	\pm	11.1	$	&	$_{	-23	}^{+	14	}$	\\
$\rho(1700)$	&	$	\pm	0.6	$	&	$	\pm	9.9	$	&	$_{	-18.3	}^{+	~0.4	}$	&	$	\pm	8.7	$	&	$_{	-23	}^{+	13	}$	\\\hline
\multicolumn{6}{c}{Transversity $\|$ fractions (\%)}\\\hline
$\rho(770)$	&	$	\pm	0.3	$	&	$	\pm	0.5	$	&	$_{	-0.8	}^{+	0.1	}$	&	$	\pm	0.8	$	&	$_{	-1.3	}^{+	1.0	}$	\\
																												
$f_2(1270)$	&	$	\pm	0.4	$	&	$	\pm	1.0	$	&	$_{	-1.4	}^{+	1.3	}$	&	$	\pm	1.3	$	&	$_{	-2	}^{+	2	}$	\\
$\omega(782)$	&	$	\pm	0.4	$	&	$	\pm	2.0	$	&	$_{	-3.3	}^{+	0.6	}$	&	$	\pm	1.5	$	&	$_{	-4	}^{+	3	}$	\\
																												
$\rho(1450)$	&	$	\pm	0.6	$	&	$	\pm	5.1	$	&	$_{	-8.4	}^{+	3.0	}$	&	$	\pm	4.4	$	&	$_{	-11	}^{+	~7	}$	\\
$\rho(1700)$	&	$	\pm	1.0	$	&	$	\pm	4.2	$	&	$_{	-7.9	}^{+	3.3	}$	&	$	\pm	4.0	$	&	$_{	-10	}^{+	~7	}$	\\\hline
\multicolumn{6}{c}{Transversity $\perp$ fractions (\%)}\\\hline
$\rho(770)$	&	$	\pm	0.3	$	&	$	\pm	0.3	$	&	$_{	-0.9	}^{+	3.8	}$	&	$	\pm	0.6	$	&	$_{	-1.2	}^{+	3.8	}$	\\
																												
$f_2(1270)$	&	$	\pm	0.3	$	&	$	\pm	0.9	$	&	$_{	-1.9	}^{+	4.3	}$	&	$	\pm	0.7	$	&	$_{	-2	}^{+	4	}$	\\
$\omega(782)$	&	$	\pm	0.1	$	&	$	\pm	0.2	$	&	$_{	-0.2	}^{+	0.1	}$	&	$	\pm	0.4	$	&	$	\pm		0.5	$	\\
																												
$\rho(1450)$	&	$	\pm	0.6	$	&	$	\pm	3.4	$	&	$_{	-~0.0	}^{+	26.8	}$	&	$	\pm	9.3	$	&	$_{	-10	}^{+	29	}$	\\
$\rho(1700)$	&	$	\pm	0.7	$	&	$	\pm	6.2	$	&	$_{	-~0.0	}^{+	26.2	}$	&	$	\pm	5.9	$	&	$_{	-9	}^{+	28	}$	\\
\hline
\multicolumn{6}{c}{Ratio of fit fractions (\%)}\\\hline
$f_0(980)/f_0(500)$		&	$	\pm	0.09	$	&	$	\pm	0.17	$	&	$^{+	2.1	}_{-	0.1	}$	&	$	\pm	2.6	$	&	$^{+	3.3	}_{	-2.6	}$	\\
$\omega(782)/\rho(770)$		&	$	\pm	0.19	$	&	$	\pm	0.04	$	&	$^{+	0.21	}_{-	0.10	}$	&	$	\pm	0.05	$	&	$^{+	0.29	}_{	-0.22	}$	\\

\hline
\end{tabular}
\def\arraystretch{1.0}
\label{sys:dlz}
\end{table}

\section{\boldmath Substructure of the $f_0(980)$ and $f_0(500)$ mesons}
\label{sec:mixing angle}
The substructure of mesons belonging to the scalar nonet is controversial. Most mesons are thought to be formed from a combination of a $q$ and a $\overline{q}$. Some authors introduce the concept of $q\overline{q}q\overline{q}$ states or superpositions of the tetraquark state with the $q\overline{q}$ state~\cite{Fleischer:2011au}. In either case,  the $I=0$ $f_0(500)$ and the $f_0(980)$ are thought to be mixtures of the underlying states whose mixing angle has been estimated previously.
In the $q\overline{q}$ model, the mixing is parameterized by a normal 2$\times$2 rotation matrix characterized by the angle $\varphi_m$, so that the observed states are given in terms of the base states as
\begin{eqnarray}
  \label{eq:fmix}
 \ket{f_0(980)}&=&\;\;\;\cos\varphi_m\ket{s\overline{s}}+\sin\varphi_m\ket{n\overline{n}}\nonumber\\
  \ket{f_0(500)}&=&-\sin\varphi_m\ket{s\overline{s}}+\cos\varphi_m\ket{n\overline{n}},\nonumber\\
  {\rm where~} \ket{n\overline{n}}&\equiv&\frac{1}{\sqrt{2}}\left(\ket{u\overline{u}}+\ket{d\overline{d}}\right).
\end{eqnarray}
In this case only the $\ket{d\overline{d}}$ part of the $\ket{n\overline{n}}$ wave function contributes (see Fig.~\ref{feyn3}).  Thus we have
\begin{equation}
\tan^2\varphi_m\equiv r^f_\sigma
=\frac{{\cal{B}}\left(\Bzb\to\jpsi f_0(980)\right)}{{\cal{B}}\left(\Bzb\to\jpsi f_0(500)\right)}\frac{\Phi(500)}{\Phi(980)},
\end{equation}
where the $\Phi$'s are phase space factors \cite{Stone:2013eaa,Ochs:2013gi,Fleischer:2011au}. The phase space in this pseudoscalar to vector-pseudoscalar decay is proportional to the cube of the $f_0$ momenta. Taking the average of the momentum dependent phase space over the resonant line shapes results in the ratio of phase space factors $\frac{\Phi(500)}{\Phi(980)}=1.25$.

The  7R model fit gives the ratio of branching fractions 
\begin{equation}
\frac{{\cal{B}}\left(\Bzb\to\jpsi f_0(980),~f_0(980)\to\pi^+\pi^-\right)} {{\cal{B}}\left(\Bzb\to\jpsi f_0(500),~f_0(500)\to\pi^+\pi^-\right)}
= (0.6 _{-0.4-2.6}^{+0.7+3.3})\times10^{-2}  \nonumber.
\end{equation}
We  need to correct for the individual branching fractions of the $f_0$ resonances decaying into $\pi^+\pi^-$. BaBar measures the relative branching ratios of $f_0(980) \to K^+ K^-/\pi^+\pi^-$ of $0.69\pm0.32$ using $B^{\pm}\to K^{\pm}K^{\pm}K^{\mp}$ and $B^{\pm}\to K^{\pm}\pi^{\pm}\pi^{\mp}$ decays \cite{Aubert:2006nu}. BES has extracted relative branching ratios using $\psi(2S)\to\gamma \chi_{c0}$ decays where the $\chi_{c0}\to f_0(980)f_0(980)$, and either both $f_0(980)$'s decay into $\pi^+\pi^-$ or one into $\pi^+\pi^-$ and the other into $K^+K^-$. Their results  \cite{Ablikim:2004cg,*Ablikim:2005kp} are that  the relative branching ratio of $f_0(980)\to K^+K^-/\pip\pim$  is $0.25^{+0.17}_{-0.11} $\cite{Ecklund:2009aa}. Averaging the two measurements gives
\begin{equation}
\frac{{\cal{B}}\left(f_0(980)\to K^+ K^-\right)}{{\cal{B}}\left(f_0(980)\to\pi^+\pi^-\right)}=0.35_{-0.14}^{+0.15}.
\end{equation}
Assuming that the $\pi\pi$ and $KK$ decays are dominant we can also extract
\begin{equation}
\label{eq:f02pipi}
{\cal{B}}\left(f_0(980)\to\pi^+\pi^-\right)
=\left(0.46\pm 0.06\right),
\end{equation}
where we have assumed that the only other decays are to $\pi^0\pi^0$ (one-half of the $\pi^+\pi^-$ rate), and to neutral kaons (equal to charged kaons).
We use ${\cal{B}}\left(f_0(500)\to\pi^+\pi^-\right)
=\frac{2}{3}$, which follows from isospin Clebsch-Gordan coefficients, and assuming that the only decays are into two pions. Since we have only an upper limit on the $\jpsi f_0(980)$ final state, we will only find an upper limit on the mixing angle, so if any other decay modes of the $f_0(500)$ exist, they would make the limit more stringent.

In order to set an upper limit on $|\varphi_m|$,  we  simulate the final $\varphi_m$ measurement using as input the central value of the measured ratio, the full statistical error matrix  obtained from the 7R model fit, and asymmetric Gaussian random variables different for the positive, {+3.3}\%, and negative, ${-2.6}$\%,  systematic uncertainties (see Table~\ref{sys:dlz}). The resulting rate ratios of $f_0(980)$ to $f_0(500)$ are then multiplied by a factor of ${\cal{B}}\left(f_0(500)\to\pi^+\pi^-\right)/{\cal{B}}\left(f_0(980)\to\pi^+\pi^-\right) \times \frac{\Phi(500)}{\Phi(980)}$ where a Gaussian random variable is used for ${\cal{B}}\left(f_0(980)\to\pi^+\pi^-\right)$ to take into account the uncertainty in the measurement shown in Eq.~(\ref{eq:f02pipi}).
The upper limit at 90\% confidence level is determined when 10\% of the simulations exceed the limit value. 
We find
\begin{equation}
\tan^2\varphi_m\equiv r^f_{\sigma}=\left(1.1^{+1.2+6.0}_{-0.7-0.7}\right)\times 10^{-2} <0.098 \text{~ at $90\%$ C.L \nonumber} 
\end{equation}
which translates into a limit of
\begin{equation}
|\varphi_m| < 17^{\circ} \text{~ at $90\%$ CL },\nonumber
\end{equation}
where we neglect the effect caused by the small systematic uncertainty on the ratio of phase space factors.

If the scalar meson substructure is tetraquark, the  wave functions are:
\begin{eqnarray}
\ket{f_{0}(980)} &=& \frac{1}{\sqrt{2}}\left(\ket{[su][\overline{s}\,\overline{u}]}+\ket{[sd][\overline{s}\overline{d}]}\right) \\
\ket{f_{0}(500)} &=&\ket{ [ud][\overline{u}\overline{d}]}.
\end{eqnarray}
The ratio $r^f_{\sigma}$ was predicted to be $1/2$ for pure tetraquark states in Ref.~\cite{Stone:2013eaa}. The measured upper limit on $r^f_{\sigma}$ of 0.098 at 90\% CL deviates from the tetraquark prediction by 8 standard deviations.

\section{Conclusions}

We have studied the resonance structure of $\Bzb\rightarrow J/\psi \pi^+\pi^-$ decays using a modified amplitude analysis. The decay distributions are formed by a series of final states described by individual $\pi^+\pi^-$ interfering decay amplitudes.  The data are best described by adding coherently the $\rho(770)$, $f_2(1270)$, $f_0(500)$, $\omega(782)$, $\rho(1450)$ and $\rho(1700)$ resonances, with the largest component being the $\rho(770)$. The final state is  $56.0$\% \CP-even, where we have taken into account both the fit fractions and the interference terms of the different components. Our understanding of the final state composition allows future measurements of \CP violation in these resonant final states. These results supersede those obtained in Ref.~\cite{Aaij:2013zpt}.
% (The 1.4\% uncertainty is statistical only.)

There is no evidence for $f_0(980)$ resonance production.
We limit the absolute value of the mixing angle between the lightest two scalar states, the $f_0(500)$ and the $f_0(980)$, in the $q\overline{q}$ model to be less than an absolute value of $17^{\circ}$ at 90\% confidence level. We find that $f_0(980)$ production is much smaller than predicted for tetraquarks, which we rule out at the 8 standard deviation level using the model of Ref.~\cite{Stone:2013eaa}.  Concern has been expressed \cite{Fleischer:2011au} that if the $f_0(980)$ were a tetraquark state the measurement of the mixing-dependent \CP-violating phase in the decay $\Bsb\to\jpsi f_0(980)$ could be affected due to additional decay mechanisms.
Our result here alleviates this potential source of error.

\section*{Acknowledgements}
\noindent We express our gratitude to our colleagues in the CERN
accelerator departments for the excellent performance of the LHC. We
thank the technical and administrative staff at the LHCb
institutes. We acknowledge support from CERN and from the national
agencies: CAPES, CNPq, FAPERJ and FINEP (Brazil); NSFC (China);
CNRS/IN2P3 and Region Auvergne (France); BMBF, DFG, HGF and MPG
(Germany); SFI (Ireland); INFN (Italy); FOM and NWO (The Netherlands);
SCSR (Poland); MEN/IFA (Romania); MinES, Rosatom, RFBR and NRC
``Kurchatov Institute'' (Russia); MinECo, XuntaGal and GENCAT (Spain);
SNSF and SER (Switzerland); NASU (Ukraine); STFC and the Royal Society (United
Kingdom); NSF (USA). We also acknowledge the support received from EPLANET,
Marie Curie Actions and the ERC under FP7.
The Tier1 computing centres are supported by IN2P3 (France), KIT and BMBF (Germany),
INFN (Italy), NWO and SURF (The Netherlands), PIC (Spain), GridPP (United Kingdom).
We are indebted to the communities behind the multiple open source software packages on which we depend.
We are also thankful for the computing resources and the access to software R\&D tools provided by Yandex LLC (Russia).
 
\newpage
\ifx\mcitethebibliography\mciteundefinedmacro
\PackageError{LHCb.bst}{mciteplus.sty has not been loaded}
{This bibstyle requires the use of the mciteplus package.}\fi
\providecommand{\href}[2]{#2}

%\bibliographystyle{LHCb}
%\bibliography{Dalitz-pipi}

\begin{mcitethebibliography}{10}
\mciteSetBstSublistMode{n}
\mciteSetBstMaxWidthForm{subitem}{\alph{mcitesubitemcount})}
\mciteSetBstSublistLabelBeginEnd{\mcitemaxwidthsubitemform\space}
{\relax}{\relax}

\bibitem{Fleischer:1999sj}
R.~Fleischer, \ifthenelse{\boolean{articletitles}}{{\it {Recent theoretical
  developments in \CP violation in the $B$ system}},
  }{}\href{http://dx.doi.org/10.1016/S0168-9002(00)00003-6}{Nucl.\ Instrum.\
  Meth.\  {\bf A446} (2000) 1}, \href{http://arxiv.org/abs/hep-ph/9908340}{{\tt
  arXiv:hep-ph/9908340}}\relax
\mciteBstWouldAddEndPuncttrue
\mciteSetBstMidEndSepPunct{\mcitedefaultmidpunct}
{\mcitedefaultendpunct}{\mcitedefaultseppunct}\relax
\EndOfBibitem
\bibitem{Faller:2008zc}
S.~Faller, M.~Jung, R.~Fleischer, and T.~Mannel,
  \ifthenelse{\boolean{articletitles}}{{\it {The golden modes $\Bz\to\jpsi
  K_{S,L}$ in the era of precision flavour physics}},
  }{}\href{http://dx.doi.org/10.1103/PhysRevD.79.014030}{Phys.\ Rev.\  {\bf
  D79} (2009) 014030}, \href{http://arxiv.org/abs/0809.0842}{{\tt
  arXiv:0809.0842}}\relax
\mciteBstWouldAddEndPuncttrue
\mciteSetBstMidEndSepPunct{\mcitedefaultmidpunct}
{\mcitedefaultendpunct}{\mcitedefaultseppunct}\relax
\EndOfBibitem
\bibitem{Stone:2013eaa}
S.~Stone and L.~Zhang, \ifthenelse{\boolean{articletitles}}{{\it {Use of
  $B\to\jpsi f_0$ decays to discern the $q\bar{q}$ or tetraquark nature of
  scalar mesons}},
  }{}\href{http://dx.doi.org/10.1103/PhysRevLett.111.062001}{Phys.\ Rev.\
  Lett.\  {\bf 111} (2013) 062001}, \href{http://arxiv.org/abs/1305.6554}{{\tt
  arXiv:1305.6554}}\relax
\mciteBstWouldAddEndPuncttrue
\mciteSetBstMidEndSepPunct{\mcitedefaultmidpunct}
{\mcitedefaultendpunct}{\mcitedefaultseppunct}\relax
\EndOfBibitem
\bibitem{Aubert:2007xw}
BaBar collaboration, B.~Aubert {\em et~al.},
  \ifthenelse{\boolean{articletitles}}{{\it {Branching fraction and charge
  asymmetry measurements in $B \to J/\psi \pi \pi$ decays}},
  }{}\href{http://dx.doi.org/10.1103/PhysRevD.76.031101}{Phys.\ Rev.\  {\bf
  D76} (2007) 031101}, \href{http://arxiv.org/abs/0704.1266}{{\tt
  arXiv:0704.1266}}\relax
\mciteBstWouldAddEndPuncttrue
\mciteSetBstMidEndSepPunct{\mcitedefaultmidpunct}
{\mcitedefaultendpunct}{\mcitedefaultseppunct}\relax
\EndOfBibitem
\bibitem{Aaij:2013zpt}
LHCb collaboration, R.~Aaij {\em et~al.},
  \ifthenelse{\boolean{articletitles}}{{\it {Analysis of the resonant
  components in $\Bzb\to\jpsi \pi^+ \pi^-$}},
  }{}\href{http://dx.doi.org/10.1103/PhysRevD.87.052001}{Phys.\ Rev.\  {\bf
  D87} (2013) 052001}, \href{http://arxiv.org/abs/1301.5347}{{\tt
  arXiv:1301.5347}}\relax
\mciteBstWouldAddEndPuncttrue
\mciteSetBstMidEndSepPunct{\mcitedefaultmidpunct}
{\mcitedefaultendpunct}{\mcitedefaultseppunct}\relax
\EndOfBibitem
\bibitem{Zhang:2012zk}
L.~Zhang and S.~Stone, \ifthenelse{\boolean{articletitles}}{{\it
  {Time-dependent Dalitz-plot formalism for $B_q \to J/\psi\ h^+ h^-$}},
  }{}\href{http://dx.doi.org/10.1016/j.physletb.2013.01.035}{Phys.\ Lett.\
  {\bf B719} (2013) 383}, \href{http://arxiv.org/abs/1212.6434}{{\tt
  arXiv:1212.6434}}\relax
\mciteBstWouldAddEndPuncttrue
\mciteSetBstMidEndSepPunct{\mcitedefaultmidpunct}
{\mcitedefaultendpunct}{\mcitedefaultseppunct}\relax
\EndOfBibitem
\bibitem{LHCb-det}
LHCb collaboration, A.~A. Alves~Jr. {\em et~al.},
  \ifthenelse{\boolean{articletitles}}{{\it {The LHCb detector at the LHC}},
  }{}\href{http://dx.doi.org/10.1088/1748-0221/3/08/S08005}{JINST {\bf 3}
  (2008) S08005}\relax
\mciteBstWouldAddEndPuncttrue
\mciteSetBstMidEndSepPunct{\mcitedefaultmidpunct}
{\mcitedefaultendpunct}{\mcitedefaultseppunct}\relax
\EndOfBibitem
\bibitem{LHCb-DP-2013-003}
R.~Arink {\em et~al.}, \ifthenelse{\boolean{articletitles}}{{\it {Performance
  of the LHCb outer tracker}},
  }{}\href{http://dx.doi.org/10.1088/1748-0221/9/01/P01002}{JINST {\bf 9}
  (2014) P01002}, \href{http://arxiv.org/abs/1311.3893}{{\tt
  arXiv:1311.3893}}\relax
\mciteBstWouldAddEndPuncttrue
\mciteSetBstMidEndSepPunct{\mcitedefaultmidpunct}
{\mcitedefaultendpunct}{\mcitedefaultseppunct}\relax
\EndOfBibitem
\bibitem{LHCb-DP-2012-003}
M.~Adinolfi {\em et~al.}, \ifthenelse{\boolean{articletitles}}{{\it
  {Performance of the \lhcb RICH detector at the LHC}},
  }{}\href{http://dx.doi.org/10.1140/epjc/s10052-013-2431-9}{Eur.\ Phys.\ J.\
  {\bf C73} (2013) 2431}, \href{http://arxiv.org/abs/1211.6759}{{\tt
  arXiv:1211.6759}}\relax
\mciteBstWouldAddEndPuncttrue
\mciteSetBstMidEndSepPunct{\mcitedefaultmidpunct}
{\mcitedefaultendpunct}{\mcitedefaultseppunct}\relax
\EndOfBibitem
\bibitem{LHCb-DP-2012-002}
A.~A. Alves~Jr. {\em et~al.}, \ifthenelse{\boolean{articletitles}}{{\it
  {Performance of the LHCb muon system}},
  }{}\href{http://dx.doi.org/10.1088/1748-0221/8/02/P02022}{JINST {\bf 8}
  (2013) P02022}, \href{http://arxiv.org/abs/1211.1346}{{\tt
  arXiv:1211.1346}}\relax
\mciteBstWouldAddEndPuncttrue
\mciteSetBstMidEndSepPunct{\mcitedefaultmidpunct}
{\mcitedefaultendpunct}{\mcitedefaultseppunct}\relax
\EndOfBibitem
\bibitem{Aaij:2012me}
R.~Aaij {\em et~al.}, \ifthenelse{\boolean{articletitles}}{{\it {The LHCb
  trigger and its performance in 2011}},
  }{}\href{http://dx.doi.org/10.1088/1748-0221/8/04/P04022}{JINST {\bf 8}
  (2013) P04022}, \href{http://arxiv.org/abs/1211.3055}{{\tt
  arXiv:1211.3055}}\relax
\mciteBstWouldAddEndPuncttrue
\mciteSetBstMidEndSepPunct{\mcitedefaultmidpunct}
{\mcitedefaultendpunct}{\mcitedefaultseppunct}\relax
\EndOfBibitem
\bibitem{Sjostrand:2006za}
T.~Sj{\"o}strand, S.~Mrenna, and P.~Skands,
  \ifthenelse{\boolean{articletitles}}{{\it {PYTHIA 6.4 physics and manual}},
  }{}\href{http://dx.doi.org/10.1088/1126-6708/2006/05/026}{JHEP {\bf 0605}
  (2006) 026}, \href{http://arxiv.org/abs/hep-ph/0603175}{{\tt
  arXiv:hep-ph/0603175}}\relax
\mciteBstWouldAddEndPuncttrue
\mciteSetBstMidEndSepPunct{\mcitedefaultmidpunct}
{\mcitedefaultendpunct}{\mcitedefaultseppunct}\relax
\EndOfBibitem
\bibitem{Sjostrand:2007gs}
T.~Sj\"{o}strand, S.~Mrenna, and P.~Skands,
  \ifthenelse{\boolean{articletitles}}{{\it {A brief introduction to PYTHIA
  8.1}}, }{}\href{http://dx.doi.org/10.1016/j.cpc.2008.01.036}{Comput.\ Phys.\
  Commun.\  {\bf 178} (2008) 852}, \href{http://arxiv.org/abs/0710.3820}{{\tt
  arXiv:0710.3820}}\relax
\mciteBstWouldAddEndPuncttrue
\mciteSetBstMidEndSepPunct{\mcitedefaultmidpunct}
{\mcitedefaultendpunct}{\mcitedefaultseppunct}\relax
\EndOfBibitem
\bibitem{LHCb-PROC-2010-056}
I.~Belyaev {\em et~al.}, \ifthenelse{\boolean{articletitles}}{{\it {Handling of
  the generation of primary events in \gauss, the \lhcb simulation framework}},
  }{}\href{http://dx.doi.org/10.1109/NSSMIC.2010.5873949}{Nuclear Science
  Symposium Conference Record (NSS/MIC) {\bf IEEE} (2010) 1155}\relax
\mciteBstWouldAddEndPuncttrue
\mciteSetBstMidEndSepPunct{\mcitedefaultmidpunct}
{\mcitedefaultendpunct}{\mcitedefaultseppunct}\relax
\EndOfBibitem
\bibitem{Lange:2001uf}
D.~J. Lange, \ifthenelse{\boolean{articletitles}}{{\it {The EvtGen particle
  decay simulation package}},
  }{}\href{http://dx.doi.org/10.1016/S0168-9002(01)00089-4}{Nucl.\ Instrum.\
  Meth.\  {\bf A462} (2001) 152}\relax
\mciteBstWouldAddEndPuncttrue
\mciteSetBstMidEndSepPunct{\mcitedefaultmidpunct}
{\mcitedefaultendpunct}{\mcitedefaultseppunct}\relax
\EndOfBibitem
\bibitem{Golonka:2005pn}
P.~Golonka and Z.~Was, \ifthenelse{\boolean{articletitles}}{{\it {PHOTOS Monte
  Carlo: a precision tool for QED corrections in $Z$ and $W$ decays}},
  }{}\href{http://dx.doi.org/10.1140/epjc/s2005-02396-4}{Eur.\ Phys.\ J.\  {\bf
  C45} (2006) 97}, \href{http://arxiv.org/abs/hep-ph/0506026}{{\tt
  arXiv:hep-ph/0506026}}\relax
\mciteBstWouldAddEndPuncttrue
\mciteSetBstMidEndSepPunct{\mcitedefaultmidpunct}
{\mcitedefaultendpunct}{\mcitedefaultseppunct}\relax
\EndOfBibitem
\bibitem{Allison:2006ve}
Geant4 collaboration, J.~Allison {\em et~al.},
  \ifthenelse{\boolean{articletitles}}{{\it {Geant4 developments and
  applications}}, }{}\href{http://dx.doi.org/10.1109/TNS.2006.869826}{IEEE
  Trans.\ Nucl.\ Sci.\  {\bf 53} (2006) 270}\relax
\mciteBstWouldAddEndPuncttrue
\mciteSetBstMidEndSepPunct{\mcitedefaultmidpunct}
{\mcitedefaultendpunct}{\mcitedefaultseppunct}\relax
\EndOfBibitem
\bibitem{Agostinelli:2002hh}
Geant4 collaboration, S.~Agostinelli {\em et~al.},
  \ifthenelse{\boolean{articletitles}}{{\it {Geant4: A simulation toolkit}},
  }{}\href{http://dx.doi.org/10.1016/S0168-9002(03)01368-8}{Nucl.\ Instrum.\
  Meth.\  {\bf A506} (2003) 250}\relax
\mciteBstWouldAddEndPuncttrue
\mciteSetBstMidEndSepPunct{\mcitedefaultmidpunct}
{\mcitedefaultendpunct}{\mcitedefaultseppunct}\relax
\EndOfBibitem
\bibitem{LHCb-PROC-2011-006}
M.~Clemencic {\em et~al.}, \ifthenelse{\boolean{articletitles}}{{\it {The \lhcb
  simulation application, \gauss: design, evolution and experience}},
  }{}\href{http://dx.doi.org/10.1088/1742-6596/331/3/032023}{{Journal of
  Physics: Conference Series} {\bf 331} (2011) 032023}\relax
\mciteBstWouldAddEndPuncttrue
\mciteSetBstMidEndSepPunct{\mcitedefaultmidpunct}
{\mcitedefaultendpunct}{\mcitedefaultseppunct}\relax
\EndOfBibitem
\bibitem{Nierste:2009wg}
U.~Nierste, \ifthenelse{\boolean{articletitles}}{{\it {Three lectures on meson
  mixing and CKM phenomenology}}, }{}\href{http://arxiv.org/abs/0904.1869}{{\tt
  arXiv:0904.1869}}\relax
\mciteBstWouldAddEndPuncttrue
\mciteSetBstMidEndSepPunct{\mcitedefaultmidpunct}
{\mcitedefaultendpunct}{\mcitedefaultseppunct}\relax
\EndOfBibitem
\bibitem{Bigi:2000yz}
I.~I. Bigi and A.~Sanda, \ifthenelse{\boolean{articletitles}}{{\it {CP
  violation}}, }{}Camb.\ Monogr.\ Part.\ Phys.\ Nucl.\ Phys.\ Cosmol.\  {\bf 9}
  (2000) 1\relax
\mciteBstWouldAddEndPuncttrue
\mciteSetBstMidEndSepPunct{\mcitedefaultmidpunct}
{\mcitedefaultendpunct}{\mcitedefaultseppunct}\relax
\EndOfBibitem
\bibitem{PDG}
Particle Data Group, J.~Beringer {\em et~al.},
  \ifthenelse{\boolean{articletitles}}{{\it {Review of particle physics}},
  }{}\href{http://dx.doi.org/10.1103/PhysRevD.86.010001}{Phys.\ Rev.\  {\bf
  D86} (2012) 010001}\relax
\mciteBstWouldAddEndPuncttrue
\mciteSetBstMidEndSepPunct{\mcitedefaultmidpunct}
{\mcitedefaultendpunct}{\mcitedefaultseppunct}\relax
\EndOfBibitem
\bibitem{LHCb:2012ae}
LHCb Collaboration, R.~Aaij {\em et~al.},
  \ifthenelse{\boolean{articletitles}}{{\it {Analysis of the resonant
  components in $\Bsb \to J/\psi \pi^+\pi^-$}},
  }{}\href{http://dx.doi.org/10.1103/PhysRevD.86.052006}{Phys.\ Rev.\  {\bf
  D86} (2012) 052006}, \href{http://arxiv.org/abs/1204.5643}{{\tt
  arXiv:1204.5643}}\relax
\mciteBstWouldAddEndPuncttrue
\mciteSetBstMidEndSepPunct{\mcitedefaultmidpunct}
{\mcitedefaultendpunct}{\mcitedefaultseppunct}\relax
\EndOfBibitem
\bibitem{Mizuk:2008me}
Belle collaboration, R.~Mizuk {\em et~al.},
  \ifthenelse{\boolean{articletitles}}{{\it {Observation of two resonance-like
  structures in the $\pi^+ \chi_{c1}$ mass distribution in exclusive
  $\overline{B}^0\to K^- \pi^+ \chi_{c1}$ decays}},
  }{}\href{http://dx.doi.org/10.1103/PhysRevD.78.072004}{Phys.\ Rev.\  {\bf
  D78} (2008) 072004}, \href{http://arxiv.org/abs/0806.4098}{{\tt
  arXiv:0806.4098}}\relax
\mciteBstWouldAddEndPuncttrue
\mciteSetBstMidEndSepPunct{\mcitedefaultmidpunct}
{\mcitedefaultendpunct}{\mcitedefaultseppunct}\relax
\EndOfBibitem
\bibitem{Breiman}
L.~Breiman, J.~H. Friedman, R.~A. Olshen, and C.~J. Stone, {\em Classification
  and regression trees}, Wadsworth international group, Belmont, California,
  USA, 1984\relax
\mciteBstWouldAddEndPuncttrue
\mciteSetBstMidEndSepPunct{\mcitedefaultmidpunct}
{\mcitedefaultendpunct}{\mcitedefaultseppunct}\relax
\EndOfBibitem
\bibitem{Skwarnicki:1986xj}
T.~Skwarnicki, {\em {A study of the radiative cascade transitions between the
  Upsilon-prime and Upsilon resonances}}, PhD thesis, Institute of Nuclear
  Physics, Krakow, 1986,
  {\href{http://inspirehep.net/record/230779/files/230779.pdf}{DESY-F31-86-02}}\relax
\mciteBstWouldAddEndPuncttrue
\mciteSetBstMidEndSepPunct{\mcitedefaultmidpunct}
{\mcitedefaultendpunct}{\mcitedefaultseppunct}\relax
\EndOfBibitem
\bibitem{Stone:2009hd}
S.~Stone and L.~Zhang, \ifthenelse{\boolean{articletitles}}{{\it {Measuring the
  \CP violating phase in \Bs mixing using $\Bs\to J/\psi f_0(980)$}},
  }{}\href{http://arxiv.org/abs/0909.5442}{{\tt arXiv:0909.5442}}\relax
\mciteBstWouldAddEndPuncttrue
\mciteSetBstMidEndSepPunct{\mcitedefaultmidpunct}
{\mcitedefaultendpunct}{\mcitedefaultseppunct}\relax
\EndOfBibitem
\bibitem{Hulsbergen:2005pu}
W.~D. Hulsbergen, \ifthenelse{\boolean{articletitles}}{{\it {Decay chain
  fitting with a Kalman filter}},
  }{}\href{http://dx.doi.org/10.1016/j.nima.2005.06.078}{Nucl.\ Instrum.\
  Meth.\  {\bf A552} (2005) 566},
  \href{http://arxiv.org/abs/physics/0503191}{{\tt
  arXiv:physics/0503191}}\relax
\mciteBstWouldAddEndPuncttrue
\mciteSetBstMidEndSepPunct{\mcitedefaultmidpunct}
{\mcitedefaultendpunct}{\mcitedefaultseppunct}\relax
\EndOfBibitem
\bibitem{Aaij:2014zyy}
LHCb collaboration, R.~Aaij {\em et~al.},
  \ifthenelse{\boolean{articletitles}}{{\it {Precision measurement of the ratio
  of the $\PLambda^0_b$ to $\overline{B}^0$ lifetimes}},
  }{}\href{http://arxiv.org/abs/1402.6242}{{\tt arXiv:1402.6242}}\relax
\mciteBstWouldAddEndPuncttrue
\mciteSetBstMidEndSepPunct{\mcitedefaultmidpunct}
{\mcitedefaultendpunct}{\mcitedefaultseppunct}\relax
\EndOfBibitem
\bibitem{Muramatsu:2002jp}
CLEO collaboration, H.~Muramatsu {\em et~al.},
  \ifthenelse{\boolean{articletitles}}{{\it {Dalitz analysis of $D^0 \to K^0_S
  \pi^+\pi^-$}},
  }{}\href{http://dx.doi.org/10.1103/PhysRevLett.89.251802}{Phys.\ Rev.\ Lett.\
   {\bf 89} (2002) 251802}, \href{http://arxiv.org/abs/hep-ex/0207067}{{\tt
  arXiv:hep-ex/0207067}}\relax
\mciteBstWouldAddEndPuncttrue
\mciteSetBstMidEndSepPunct{\mcitedefaultmidpunct}
{\mcitedefaultendpunct}{\mcitedefaultseppunct}\relax
\EndOfBibitem
\bibitem{Aaij:2014emv}
LHCb collaboration, R.~Aaij {\em et~al.},
  \ifthenelse{\boolean{articletitles}}{{\it {Measurement of resonant and $CP$
  components in $\overline{B}_s^0\rightarrow J/\psi \pi^+\pi^-$ decays}},
  }{}\href{http://arxiv.org/abs/1402.6248}{{\tt arXiv:1402.6248}}, to appear in
  Phys. Rev. D.\relax
\mciteBstWouldAddEndPunctfalse
\mciteSetBstMidEndSepPunct{\mcitedefaultmidpunct}
{}{\mcitedefaultseppunct}\relax
\EndOfBibitem
\bibitem{Flatte:1976xv}
S.~M. Flatt\'e, \ifthenelse{\boolean{articletitles}}{{\it {On the nature of
  $0^+$ mesons}},
  }{}\href{http://dx.doi.org/10.1016/0370-2693(76)90655-9}{Phys.\ Lett.\  {\bf
  B63} (1976) 228}\relax
\mciteBstWouldAddEndPuncttrue
\mciteSetBstMidEndSepPunct{\mcitedefaultmidpunct}
{\mcitedefaultendpunct}{\mcitedefaultseppunct}\relax
\EndOfBibitem
\bibitem{Baker:1983tu}
S.~Baker and R.~D. Cousins, \ifthenelse{\boolean{articletitles}}{{\it
  {Clarification of the use of $\chi^2$ and likelihood functions in fits to
  histograms}}, }{}\href{http://dx.doi.org/10.1016/0167-5087(84)90016-4}{Nucl.\
  Instrum.\ Meth.\  {\bf 221} (1984) 437}\relax
\mciteBstWouldAddEndPuncttrue
\mciteSetBstMidEndSepPunct{\mcitedefaultmidpunct}
{\mcitedefaultendpunct}{\mcitedefaultseppunct}\relax
\EndOfBibitem
\bibitem{delAmoSanchez:2010yp}
BaBar collaboration, P.~del Amo~Sanchez {\em et~al.},
  \ifthenelse{\boolean{articletitles}}{{\it {Dalitz plot analysis of $D_s^+ \to
  K^+ K^- \pi^+$}},
  }{}\href{http://dx.doi.org/10.1103/PhysRevD.83.052001}{Phys.\ Rev.\  {\bf
  D83} (2011) 052001}, \href{http://arxiv.org/abs/1011.4190}{{\tt
  arXiv:1011.4190}}\relax
\mciteBstWouldAddEndPuncttrue
\mciteSetBstMidEndSepPunct{\mcitedefaultmidpunct}
{\mcitedefaultendpunct}{\mcitedefaultseppunct}\relax
\EndOfBibitem
\bibitem{Bugg:2006gc}
D.~V. Bugg, \ifthenelse{\boolean{articletitles}}{{\it {The mass of the sigma
  pole}}, }{}\href{http://dx.doi.org/10.1088/0954-3899/34/1/011}{J.\ Phys.\
  {\bf G34} (2007) 151}, \href{http://arxiv.org/abs/hep-ph/0608081}{{\tt
  arXiv:hep-ph/0608081}}\relax
\mciteBstWouldAddEndPuncttrue
\mciteSetBstMidEndSepPunct{\mcitedefaultmidpunct}
{\mcitedefaultendpunct}{\mcitedefaultseppunct}\relax
\EndOfBibitem
\bibitem{GS}
G.~J. Gounaris and J.~J. Sakurai, \ifthenelse{\boolean{articletitles}}{{\it
  {Finite-width corrections to the vector-mesion-dominance prediction for
  $\rho\to e^+ e^-$}},
  }{}\href{http://dx.doi.org/10.1103/PhysRevLett.21.244}{Phys.\ Rev.\ Lett.\
  {\bf 21} (1968) 244}\relax
\mciteBstWouldAddEndPuncttrue
\mciteSetBstMidEndSepPunct{\mcitedefaultmidpunct}
{\mcitedefaultendpunct}{\mcitedefaultseppunct}\relax
\EndOfBibitem
\bibitem{Fleischer:2011au}
R.~Fleischer, R.~Knegjens, and G.~Ricciardi,
  \ifthenelse{\boolean{articletitles}}{{\it {Anatomy of $B^0_{s,d} \to J/\psi
  f_0(980)$}}, }{}\href{http://dx.doi.org/10.1140/epjc/s10052-011-1832-x}{Eur.\
  Phys.\ J.\  {\bf C71} (2011) 1832},
  \href{http://arxiv.org/abs/1109.1112}{{\tt arXiv:1109.1112}}\relax
\mciteBstWouldAddEndPuncttrue
\mciteSetBstMidEndSepPunct{\mcitedefaultmidpunct}
{\mcitedefaultendpunct}{\mcitedefaultseppunct}\relax
\EndOfBibitem
\bibitem{Ochs:2013gi}
W.~Ochs, \ifthenelse{\boolean{articletitles}}{{\it {The status of glueballs}},
  }{}\href{http://dx.doi.org/10.1088/0954-3899/40/4/043001}{J.\ Phys.\  {\bf
  G40} (2013) 043001}, \href{http://arxiv.org/abs/1301.5183}{{\tt
  arXiv:1301.5183}}\relax
\mciteBstWouldAddEndPuncttrue
\mciteSetBstMidEndSepPunct{\mcitedefaultmidpunct}
{\mcitedefaultendpunct}{\mcitedefaultseppunct}\relax
\EndOfBibitem
\bibitem{Aubert:2006nu}
BaBar collaboration, B.~Aubert {\em et~al.},
  \ifthenelse{\boolean{articletitles}}{{\it {Dalitz plot analysis of the decay
  $B^\pm \to K^\pm K^\pm K^\mp$}},
  }{}\href{http://dx.doi.org/10.1103/PhysRevD.74.032003}{Phys.\ Rev.\  {\bf
  D74} (2006) 032003}, \href{http://arxiv.org/abs/hep-ex/0605003}{{\tt
  arXiv:hep-ex/0605003}}\relax
\mciteBstWouldAddEndPuncttrue
\mciteSetBstMidEndSepPunct{\mcitedefaultmidpunct}
{\mcitedefaultendpunct}{\mcitedefaultseppunct}\relax
\EndOfBibitem
\bibitem{Ablikim:2004cg}
BES collaboration, M.~Ablikim {\em et~al.},
  \ifthenelse{\boolean{articletitles}}{{\it {Evidence for $f_0(980)f_0(980)$
  production in $\chi_{c0}$ decays}},
  }{}\href{http://dx.doi.org/10.1103/PhysRevD.70.092002}{Phys.\ Rev.\  {\bf
  D70} (2004) 092002}, \href{http://arxiv.org/abs/hep-ex/0406079}{{\tt
  arXiv:hep-ex/0406079}}\relax
\mciteBstWouldAddEndPuncttrue
\mciteSetBstMidEndSepPunct{\mcitedefaultmidpunct}
{\mcitedefaultendpunct}{\mcitedefaultseppunct}\relax
\EndOfBibitem
\bibitem{Ablikim:2005kp}
BES collaboration, M.~Ablikim {\em et~al.},
  \ifthenelse{\boolean{articletitles}}{{\it {Partial wave analysis of
  $\chi_{c0} \to \pi^+ \pi^- K^+ K^-$}},
  }{}\href{http://dx.doi.org/10.1103/PhysRevD.72.092002}{Phys.\ Rev.\  {\bf
  D72} (2005) 092002}, \href{http://arxiv.org/abs/hep-ex/0508050}{{\tt
  arXiv:hep-ex/0508050}}\relax
\mciteBstWouldAddEndPuncttrue
\mciteSetBstMidEndSepPunct{\mcitedefaultmidpunct}
{\mcitedefaultendpunct}{\mcitedefaultseppunct}\relax
\EndOfBibitem
\bibitem{Ecklund:2009aa}
CLEO collaboration, K.~M. Ecklund {\em et~al.},
  \ifthenelse{\boolean{articletitles}}{{\it {Study of the semileptonic decay
  $D_s^+\to f_0(980) e^+ \nu$ and implications for $B^0_s\to \jpsi f_0$}},
  }{}\href{http://dx.doi.org/10.1103/PhysRevD.80.052009}{Phys.\ Rev.\  {\bf
  D80} (2009) 052009}, \href{http://arxiv.org/abs/0907.3201}{{\tt
  arXiv:0907.3201}}\relax
\mciteBstWouldAddEndPuncttrue
\mciteSetBstMidEndSepPunct{\mcitedefaultmidpunct}
{\mcitedefaultendpunct}{\mcitedefaultseppunct}\relax
\EndOfBibitem
\end{mcitethebibliography}

\end{document}